\documentclass{aa}  

\usepackage{graphicx}
\usepackage{caption}
\usepackage{subcaption}
\usepackage{txfonts}
\usepackage{amsmath}
\usepackage{natbib}
\usepackage{lscape}
\usepackage{multicol}
\newcommand{\SgrB} {Sgr\,B2}
\newcommand{\mo}  {$M_\odot$} 
\newcommand{\lo}  {$L_\odot$} 
\newcommand{\phn} {$\phantom{0}$}
\newcommand{\phnn} {$\phantom{00}$}
\newcommand{\phnnn} {$\phantom{000}$}
\newcommand{\phs} {$\phantom{<}$}
\newcommand{\hii} {H{\sc ii}}
\begin{document}

   \title{The physical and chemical structure of Sagittarius\,B2}

   \subtitle{II. Continuum millimeter emission of \SgrB(M) and \SgrB(N) with ALMA}

   \author{\'A.~S\'anchez-Monge\inst{1}
          \and
          P.~Schilke\inst{1}
          \and
          A.~Schmiedeke\inst{1,2}
          \and
          A.~Ginsburg\inst{3,4}
          \and
          R.~Cesaroni\inst{5}
          \and
          D.C.~Lis\inst{6,7}
          \and
          S.-L.~Qin\inst{8}
          \and
          H.S.P.~M\"uller\inst{1}
          \and
          E.~Bergin\inst{9}
          \and
          C.~Comito\inst{1}
          \and
          Th.~M\"oller\inst{1}
          }

   \institute{I.\ Physikalisches Institut, Universit\"at zu K\"oln, Z\"ulpicher Str.\ 77, D-50937 K\"oln, Germany\\
              \email{sanchez@ph1.uni-koeln.de}
              \and
              Max-Planck Institute for Extraterrestrial Physics, Giessenbachstrasse 1, D-85748 Garching bei M\"unchen, Germany
              \and
              National Radio Astronomy Observatory, 1003 Lopezville Road, Socorro, NM 87801, USA
              \and
              European Southern Observatory, Karl-Schwarzschild-Strasse 2, D-85748 Garching bei M\"unchen, Germany
              \and
              INAF-Osservatorio Astrofisico di Arcetri, Largo E.\ Fermi 5, I-50125 Firenze, Italy
              \and
              LERMA, Observatoire de Paris, PSL Research University, CNRS, Sorbonne Universit\'es, UPMC Univ.\ Paris 06, F-75014 Paris, France
              \and
              California Institute of Technology, Pasadena, CA 91125, USA
              \and
              Department of Astronomy, Yunnan University, and Key Laboratory of Astroparticle Physics of Yunnan Province, Kumming, 650091, China
              \and
              Department of Astronomy, University of Michigan, 500 Church Street, Ann Arbor, MI 48109-1042, USA
             }

   \date{Received ; accepted }

 
  \abstract
   {The two hot molecular cores \SgrB(M) and \SgrB(N), located at the center of the giant molecular cloud complex Sagittarius\,B2, have been the targets of numerous spectral line surveys, revealing a rich and complex chemistry.}
   {We want to characterize the physical and chemical structure of the two high-mass star forming sites \SgrB(M) and \SgrB(N) using high-angular resolution observations at millimeter wavelengths, reaching spatial scales of about 4000~au.}
   {We use the Atacama Large Millimeter/submillimeter Array (ALMA) to perform an unbiased spectral line survey of both regions in the ALMA band 6, with a frequency coverage from 211~GHz to 275~GHz. The achieved angular resolution is 0\farcs4, which probes spatial scales of about 4000~au, i.e.\ able to resolve different cores and fragments. In order to determine the continuum emission in these line-rich sources, we use a new statistical method, STATCONT, that has been applied successfully to this and other ALMA datasets and to synthetic observations.}
   {We detect 27 continuum sources in \SgrB(M) and 20 sources in \SgrB(N). We study the continuum emission variation across the ALMA band~6 (i.e.\ spectral index), and compare the ALMA 1.3~mm continuum emission with previous SMA 345~GHz and VLA 40~GHz observations, to study the nature of the sources detected. The brightest sources are dominated by (partially optically thick) dust emission, while there is an important degree of contamination from ionized gas free-free emission in weaker sources. While the total mass in \SgrB(M) is distributed in many fragments, most of the mass in \SgrB(N) arises from a single object, with filamentary-like structures converging towards the center. There seems to be a lack of low-mass dense cores in both regions. We determine H$_2$ volume densities for the cores of about $10^7$--$10^9$~cm$^{-3}$ (or $10^5$--$10^7$~\mo~pc$^{-3}$), one to two orders of magnitude higher than the stellar densities of super star clusters. We perform a statistical study of the chemical content of the identified sources. In general, \SgrB(N) is chemically richer than \SgrB(M). The chemically richest sources have about 100 lines per GHz, and the fraction of luminosity contained in spectral lines at millimeter wavelengths with respect to the total luminosity is about 20\%--40\%. There seems to be a correlation between the chemical richness and the mass of the fragments, with more massive clumps being more chemically rich. Both \SgrB(N) and \SgrB(M) harbour a cluster of hot molecular cores. We compare the continuum images with predictions from a detailed 3D radiative transfer model that reproduces the structure of \SgrB\ from 45~pc down to 100~au.}
   {This ALMA dataset, together with other on-going observational projects in the range 5~GHz to 200~GHz, better constrain the 3D structure of \SgrB, and allow us to understand its physical and chemical structure.}


   \maketitle
%
\section{Introduction}\label{s:intro}

The giant molecular cloud complex Sagittarius\,B2 (hereafter \SgrB) is the most massive region with ongoing star formation in the Galaxy. It is located at a distance of $8.34\pm0.16$~kpc \citep{Reid2014} and thought to be within less than 100~pc in projected distance to the Galactic center \citep{Molinari2010}. The high densities ($>10^{5}$~cm$^{-3}$) and relatively warm temperatures ($\sim50$--$70$~K), together with its proximity to the Galactic center, make \SgrB\ an interesting environment of extreme star formation, different to the typical star forming regions in the Galactic disk but similar to the active galactic centers that dominate star formation throughout the Universe at high redshifts.

\begin{figure*}[h!]
\begin{center}
\begin{tabular}[b]{c}
        \includegraphics[width=0.9\textwidth]{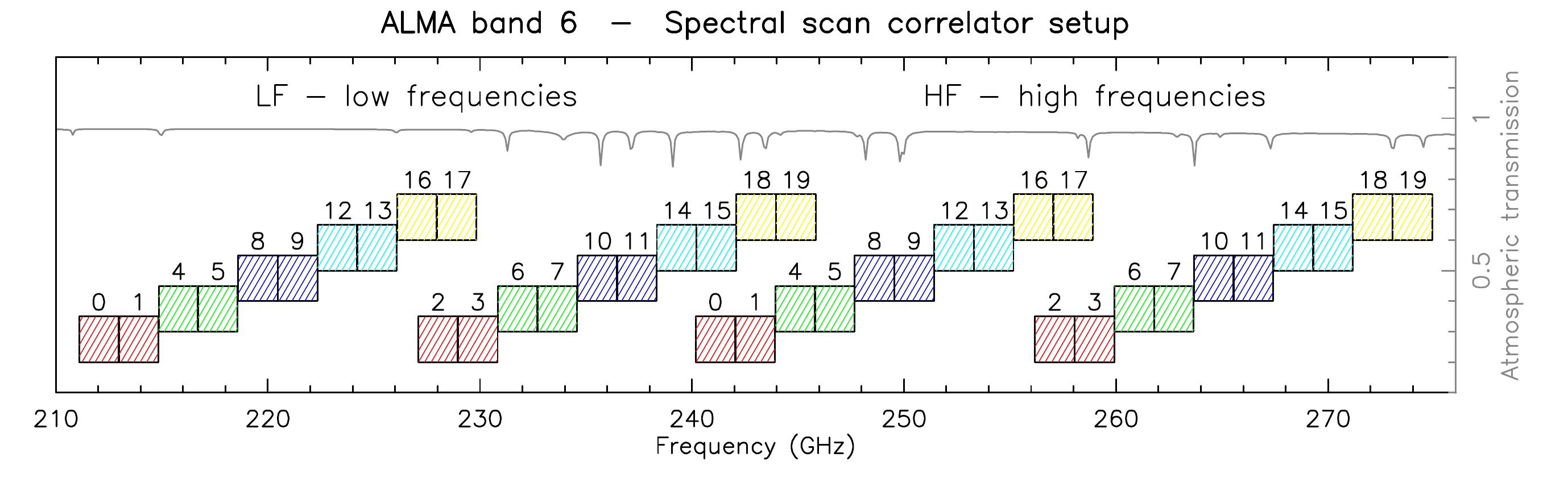} \\
\end{tabular}
\caption{Sketch of the setup of the ALMA correlator during the spectral scan observations in the ALMA band 6. The observations were divided in two frequency ranges: low frequencies from 211 to 246~GHz, and high frequencies from 240 to 275~GHz. Each frequency tuning is depicted with a different color, while the different boxes refer to the four spectral windows per tuning. The number of the spectral windows, as given in the ALMA observations, are given on top of each box. The gray line shows the transmission of the atmosphere at the ALMA site with a precipitable water vapor of 0.7~mm.}
\label{f:freqcover}
\end{center}
\end{figure*}

The whole \SgrB\ complex contains a total mass of about $10^{7}$~\mo\ (with a total luminosity of about $10^{7}$~\lo, \citealt{Goldsmith1990}) distributed in a large envelope of about 22~pc in radius (\citealt{LisGoldsmith1989}; see also \citealt{Schmiedeke2016}). Despite the large mass reservoir, star formation seems to be mainly occurring in the two hot molecular cores \SgrB(M) and \SgrB(N). These two sites of active star formation are located at the center of the envelope, occupy an area of around 2~pc in radius, contain at least 70 high-mass stars with spectral types from O5 to B0 \citep[e.g.][]{dePree1998, dePree2014, Gaume1995}, and constitute one of the best laboratories for the search of new chemical species in the Galaxy \citep[e.g.][]{Belloche2013, Belloche2014, Schilke2014}. The reader is referred to \citet{Schmiedeke2016} for a more detailed description of \SgrB\ and the two hot cores.

Due to their exceptional characteristics, \SgrB(M) and \SgrB(N) have been the targets of numerous spectral line surveys, with instruments such as the IRAM-30\,m telescope and the \textit{Herschel} space observatory \citep[e.g.][]{Nummelin1998, Bergin2010, Belloche2013, Neill2014, Corby2015}. These studies, although lacking the information on the spatial structure of the cores, have revealed a very different chemical composition of the two objects, with \SgrB(M) being very rich in sulphur-bearing species and \SgrB(N) in complex organic molecules. High-angular (sub-arcsecond) resolution observations at (sub)millimeter wavelengths were conducted for the first time by \citet{Qin2011} using the Submillimeter Array (SMA) at 345~GHz. The authors find different morphologies in the two objects, with \SgrB(M) being highly fragmented into many cores and \SgrB(N) remaining, at least at the sensitivity and image fidelity of the SMA observations, monolithic consisting of only a main core and a northern, secondary core.

With the aim of better understanding the physical and chemical structure of these two active high-mass star forming sites, we started an ALMA project in Cycle~2 to perform an unbiased spectral line survey of \SgrB(M) and \SgrB(N) in the ALMA band 6 (from 211 to 275~GHz). This is the second in a series of papers investigating the structure of \SgrB. In the Paper~I of this series \citep{Schmiedeke2016}, we present a 3D radiative transfer model of the whole \SgrB\ region. In the current paper we describe the ALMA project, apply a new method to determine the continuum emission in line-rich sources like \SgrB, and compare the obtained continuum images with previous high-angular resolution observations at different wavelengths \citep[e.g.][]{Qin2011, Rolffs2011} and with predictions from the 3D radiative transfer model presented by \citet{Schmiedeke2016}.

This paper is organized as follows. In Sect.~\ref{s:obs} we describe our ALMA observations. In Sect.~\ref{s:results} we study the continuum emission towards \SgrB(M) and \SgrB(N), and identify the compact sources and main structures. In Sect.~\ref{s:analysis} we analyze and discuss our findings focusing on the properties of the continuum sources, its possible origin, its chemical content and we compare our findings with a 3D radiative transfer model. Finally, in Sect.~\ref{s:summary} we summarize the main results and draw the conclusions.

\begin{figure*}[h!]
\begin{center}
\begin{tabular}[b]{c c c}
        \includegraphics[width=0.8\columnwidth]{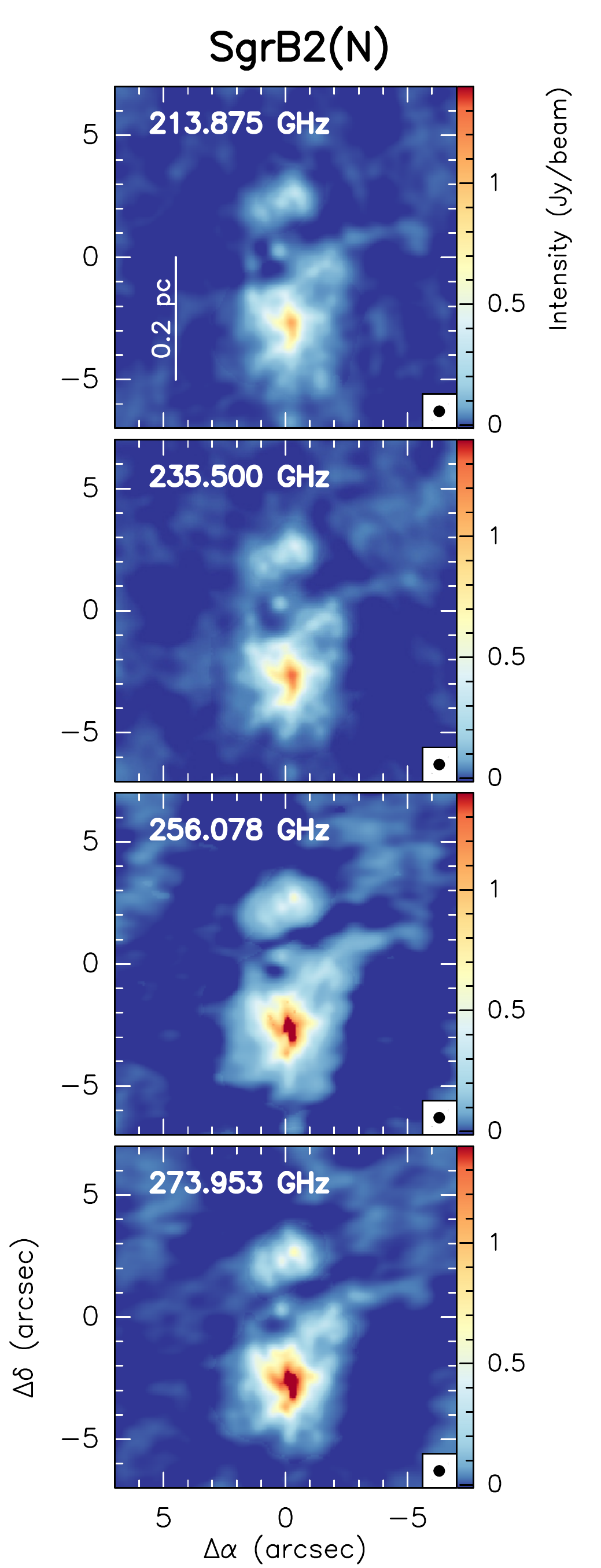} &
        \includegraphics[width=0.8\columnwidth]{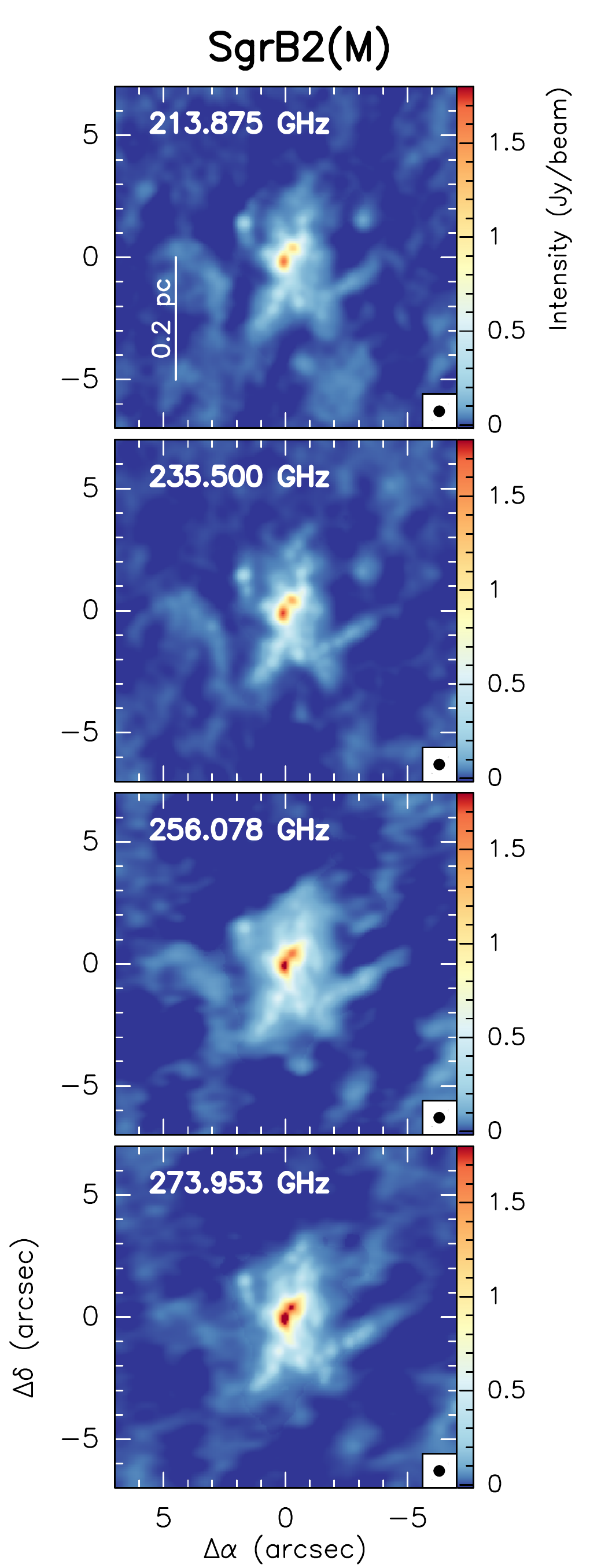} \\
\end{tabular}
\caption{Continuum emission maps for \SgrB-N (left column) and \SgrB-M (right column) produced using STATCONT as described in Sect.~\ref{s:cont1mm} for four different spectral windows, corresponding to the central frequencies 213.875~GHz, 235.500~GHz, 256.078~GHz and 273.953~GHz from top to bottom. The synthesized beam is $0\farcs4$ and is indicated in the bottom-right corner of each panel. The intensity color scale has been fixed for each source to better show the variation in intensity with frequency.}
\label{f:SgrB2contfreqs}
\end{center}
\end{figure*}

\section{Observations}\label{s:obs}

\SgrB\ was observed with ALMA (Atacama Large Millimeter/submillimeter Array; \citealt{ALMA2015}) during Cycle~2 in 2014 June and 2015 June (project number 2013.1.00332.S), using 34--36 antennas in an extended configuration with baselines in the range from 30~m to 650~m, which in the frequency range 211--275~GHz results in a sensitivity to structures in the range 0\farcs4--5\arcsec. The observations were done in the `spectral scan' mode covering the broad frequency range from 211~GHz to 275~GHz (ALMA band 6) with 10 different spectral tunings. For each tuning, the digital correlator was configured in four spectral windows of 1875~MHz and 3840~channels each (with dual polarization), providing a resolution of 0.5--0.7~km~s$^{-1}$ across the full frequency band. In Figure~\ref{f:freqcover}, we show a sketch of the frequency coverage. The two sources \SgrB(M) and \SgrB(N) were observed in track-sharing mode, with phase centers at $\alpha$(J2000)=$17^\mathrm{h}47^\mathrm{m}20\fs157$, $\delta$(J2000)=$-28\degr23\arcmin04\farcs53$ for \SgrB(M), and at $\alpha$(J2000)=$17^\mathrm{h}47^\mathrm{m}19\fs887$, $\delta$(J2000)=$-28\degr22\arcmin15\farcs76$ for \SgrB(N). Flux calibration was obtained through observations of the bright quasar J1733$-$1304 (flux 1.36~Jy at 228.265~GHz, with spectral index $-0.71$) and the satellite Titan. The phases were calibrated by interleaved observations of the quasars J1744$-$3116 (bootstrapped flux of 0.39~Jy at 228.265~GHz, with spectral index $-0.55$) and J1752$-$2956 (bootstrapped flux of 0.031~Jy at 228.265~GHz, with spectral index $-0.99$). The gain calibrators, separated from \SgrB\ on the sky by $\sim2$--3\degr, were observed every 6~minutes. The bandpass correction was obtained by observing the bright quasar J1733$-$1304. The on-source observing time per source and frequency is about 2.5~minutes. The amount of precipitable water vapor during the different observing days was about 0.7~mm.

The calibration and imaging were performed in CASA\footnote{The Common Astronomy Software Applications (CASA) software can be downloaded at http://casa.nrao.edu} version 4.4.0. The gains of the sources were determined from the interpolation of the gains derived for the nearby quasars J1744$-$3116 and J1752$-$2956 at each corresponding frequency, with the exception of spectral windows 12 to 15 (in both the low-frequency and high-frequency regimes; see Fig.~\ref{f:freqcover}) for which the phases of the gain calibrator were not properly determined. For these spectral windows we transferred the phases derived from nearby (in frequency and in observing time) spectral windows. This method results in a bit noiser data for the affected spectral windows, but consistent with the expected results for the spectral windows 12 to 15 in the low frequency (LF) regime. For the high frequency (HF) regime, the dispersion of the phases is still too large. We applied two iterations of self-calibration in phase-mode only, and a last step in both amplitude and phase to correct the phases of the more noisy spectral windows, using only the strongest component (or bright emission above 500~mJy) in the model. We note that this process is used only to recover the main structure of both sources for future analysis, but we do not consider the self-calibrated images for the analysis of the continuum properties of \SgrB(M) and (N), since the presence of extended emission can slightly modify the overall flux density scale (see e.g.\ \citealt{AntonucciUlvestad1985}). Channel maps for each spectral window were created using the task CLEAN in CASA, with the robust parameter of \citet{Briggs1995} set equal to 0.5, as a compromise between resolution and sensitivity to extended emission. The resulting images have a synthesized CLEANed beam that varies from about $0\farcs39$ to $0\farcs65$. These images were restored with a Gaussian beam to have final synthesized beams of $0\farcs7$ (medium resolution images) and $0\farcs4$ (super-resolution images). The medium resolution images are better suited to study the spectral line emission that will be presented in forthcoming papers, while the super-resolution images can be used to study in more detail the structure of the continuum emission. All the images are corrected for the primary beam response of the ALMA antennas. The rms noise level of each spectral channel (of 490~kHz, or 0.5--0.7~km~s$^{-1}$) is typically in the range 10 to 20~mJy~beam$^{-1}$. The final continuum image, produced considering the whole observed frequency range and following the procedure described in S\'anchez-Monge et al.\ (submitted, see also Sect.~\ref{s:cont1mm}), has a rms noise level of 8~mJy~beam$^{-1}$.

\section{Results}\label{s:results}

In this section we aim at identifying and characterizing the different sources and structures seen in the \SgrB(N) and \SgrB(M) continuum maps at 1.3~mm (211--275~GHz). We first present the continuum maps created following the method described in S\'anchez-Monge et al.\ submitted. Then, we identify the continuum sources and main structures in the ALMA images.

\subsection{Continuum maps at 1.3~mm with ALMA}\label{s:cont1mm}

We have used the STATCONT\footnote{\url{http://www.astro.uni-koeln.de/~sanchez/statcont}} python-based tool (see S\'anchez-Monge et al.\ submitted) to create the continuum emission images of \SgrB(M) and \SgrB(N) from the spectral scan observations presented in Sect.~\ref{s:obs}. Each spectral window, corresponding to a bandwidth of 1.87~GHz, has been processed independently in order to produce 40 different continuum emission maps throughout the frequency range 211~GHz to 275~GHz. The method used to determine the continuum level is based on the sigma-clipping algorithm (or corrected sigma-clipping method, hereafter cSCM, within the STATCONT python-based tool). Each cube from each spectral window is inspected on a pixel basis. The spectrum is analyzed iteratively: in a first step the median ($\mu$) and dispersion ($\sigma$) of the entire intensity distribution is calculated. In a second step, the algorithm removes all the points that are smaller or larger than $\mu\pm\alpha\sigma$, where $\alpha$ is set to 2. In each iteration, a number of outliers will be removed and $\sigma$ will decrease or remain the same. The process is stopped when $\sigma$ is within a certain tolerance level determined as ($\sigma_\mathrm{old}-\sigma_\mathrm{new})/\sigma_\mathrm{new}$. This method has been proved against synthetic observations to be accurate within an uncertainty level of 5--10\% (see S\'anchez-Monge et al.\ submitted). An image map of the uncertainty in the determination of the continuum level is also produced. The continuum level slightly depends on the spectral features, both in emission and absorption, contained in the spectrum. It is then important to note that each spectral window contains different molecular transitions, and therefore different spectra. Thus, the comparison of continuum images constructed from different spectral ranges permits us to confirm the real emission and to search for artifacts produced by the shape of the spectra. Furthermore, it will also allow us to study how the continuum emission changes with frequency.

\begin{figure}[t]
\begin{center}
\begin{tabular}[b]{c}
        \includegraphics[width=0.9\columnwidth]{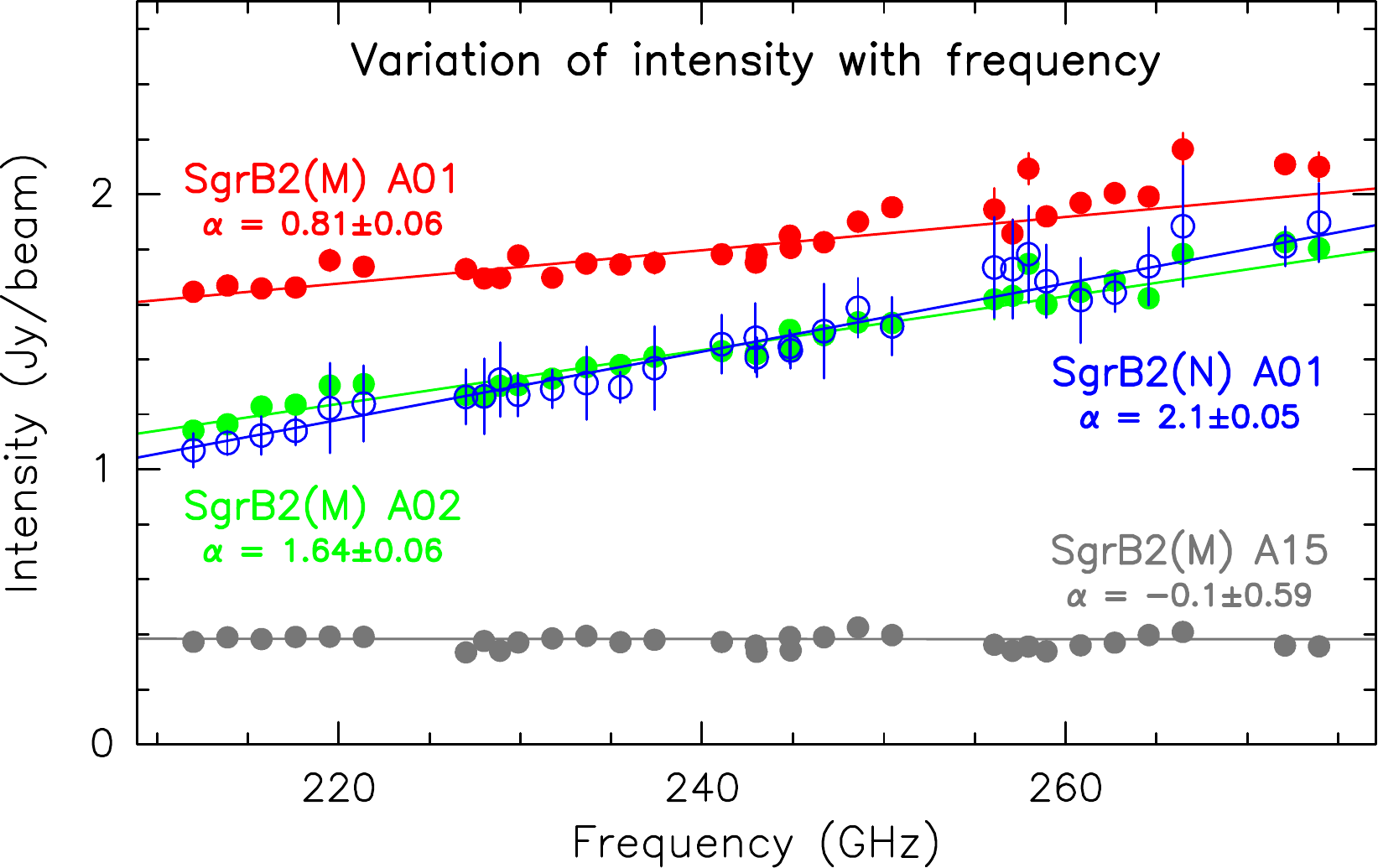} \\
\end{tabular}
\caption{Variation of the intensity (in Jy~beam$^{-1}$) with frequency (in GHz) for four pixels in the \SgrB(N) (open circles) and \SgrB(M) (filled circles) regions. The vertical lines correspond to the error in the determination of the continuum level as computed by STATCONT (see Sect.~\ref{s:cont1mm}). The intensity is extracted for each continuum map created at different frequencies (see Sect.~\ref{s:cont1mm}). The numbers in parenthesis next to the name of the sources refer to the number of the sources as listed in Table~\ref{t:SgrB2Nsources} and \ref{t:SgrB2Msources} (see also Sect.~\ref{s:sources}). The spectral index ($\alpha$, defined as $I\propto\nu^\alpha$) is indicated for each source. The intensity of source \SgrB(M)~AM15 has been manually shifted by +0.3~Jy~beam$^{-1}$.}
\label{f:fluxfreq}
\end{center}
\end{figure}

\begin{figure*}[h!]
\begin{center}
\begin{tabular}[b]{c}
        \includegraphics[width=0.82\textwidth]{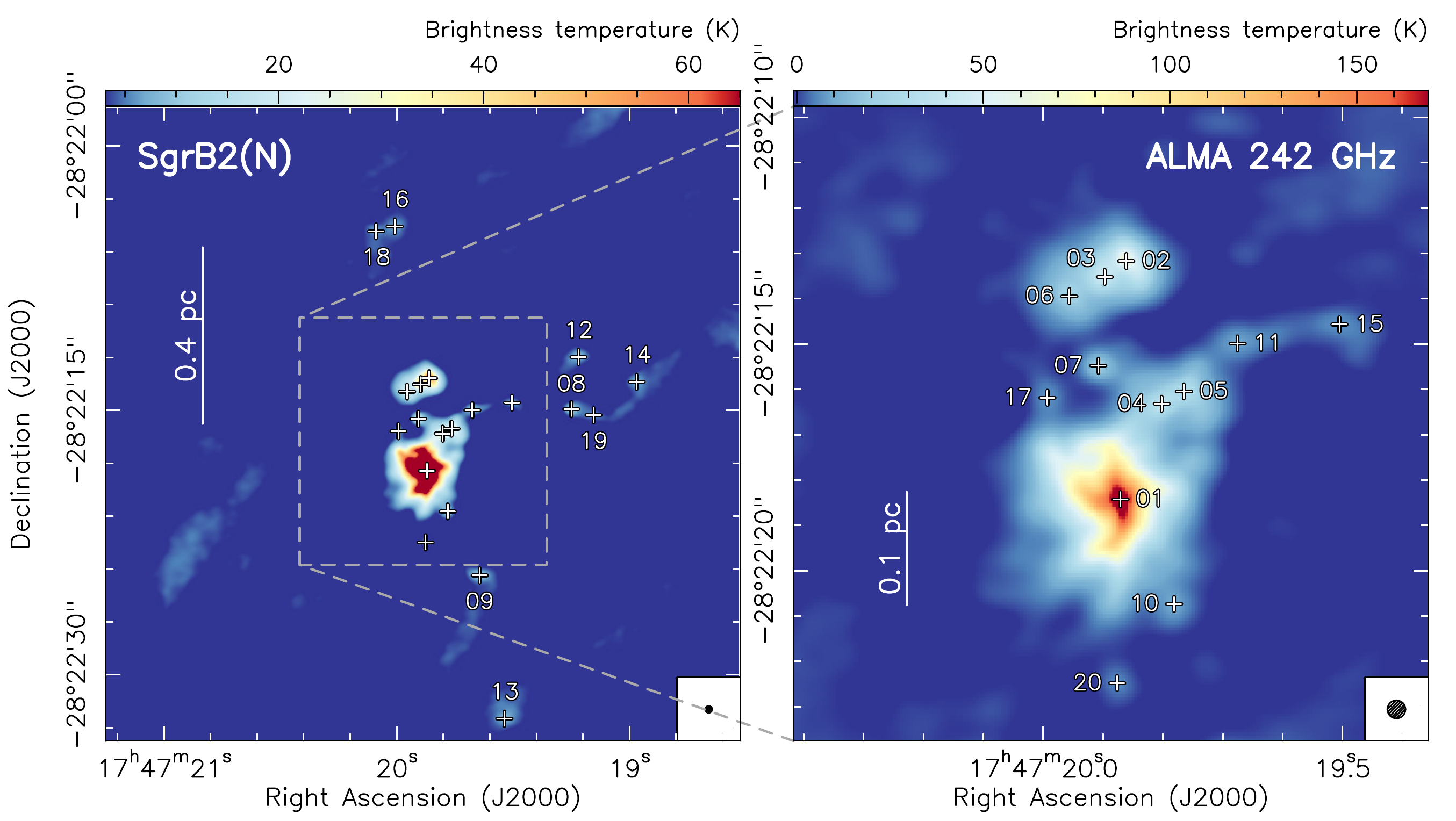} \\
\end{tabular}
\caption{ALMA 242~GHz continuum emission towards \SgrB(N). The right panel shows a close-up view of the central 0.56~pc. The identified sources are indicated with crosses. The number indicates the source ID as listed in Table~\ref{t:SgrB2Msources}. The synthesized beam is $0\farcs4$ and is shown in the bottom-right corner. The intensity color scale is shown in units of brightness temperature, the conversion factor to flux units is 130.2~Jy/K. The rms noise level is 8~mJy~beam$^{-1}$ (or about 1~K).}
\label{f:SgrB2Nsources}
\end{center}
\end{figure*}

\begin{figure*}[h!]
\begin{center}
\begin{tabular}[b]{c}
        \includegraphics[width=0.82\textwidth]{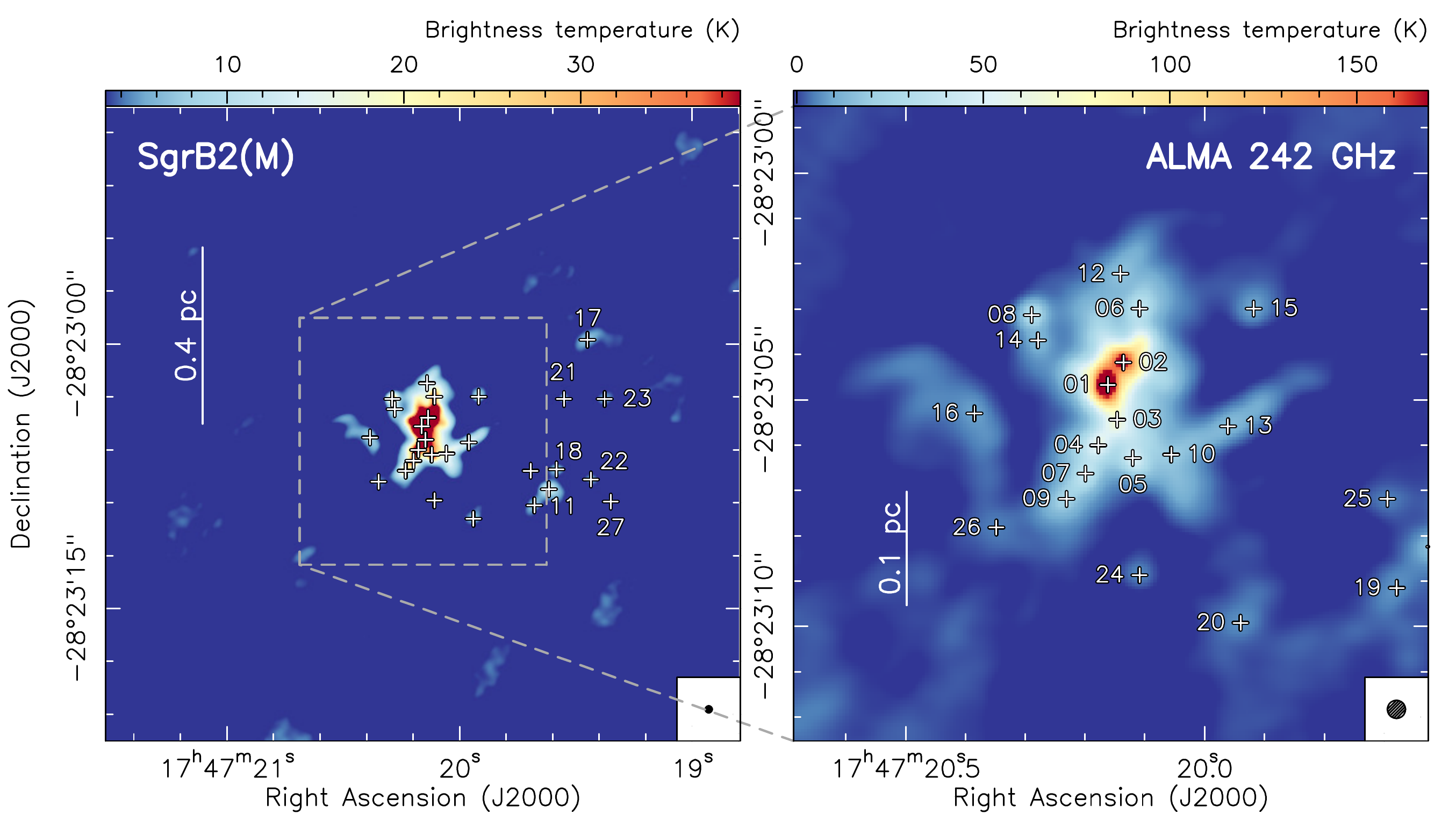} \\
\end{tabular}
\caption{ALMA 242~GHz continuum emission towards \SgrB(M). The right panel shows a close-up view of the central 0.56~pc. The identified sources are indicated with crosses. The number indicates the source ID as listed in Table~\ref{t:SgrB2Msources}. The synthesized beam is $0\farcs4$ and is shown in the bottom-right corner. The intensity color scale is shown in units of brightness temperature, the conversion factor to flux units is 130.2~Jy/K. The rms noise level is 8~mJy~beam$^{-1}$ (or about 1~K).}
\label{f:SgrB2Msources}
\end{center}
\end{figure*}

Figure~\ref{f:SgrB2contfreqs} shows the continuum emission for \SgrB(N) (left column) and \SgrB(M) (right column) at different frequencies. Each column in the figure is divided in four panels with increasing frequency from top to bottom. The color scale has been fixed in all the panels to better show the changes in intensity. In the following we describe the results obtained from the continuum images. The spatial structure is similar at all frequencies. The continuum emission of \SgrB(N) comes from two well distinguished objects, with the second brightest object located about 5~arcsec to the north of the brightest source\footnote{These two sources are also identified in the sub-arcsecond SMA observations of \citet{Qin2011}, with names SMA1 and SMA2 for the brightest core and the core located 5\arcsec to the north, respectively. These sources have also been named N1 and N2 \citep[e.g.\ ][]{Mueller2016}.}. For \SgrB(M), the map consists of a double-peak bright object surrounded by extended and weak features tracing fainter sources. The properties of different cores and its comparison with previous observations is done in Sects.~\ref{s:origin}--\ref{s:chemistry}. In contrast to the spatial distribution, the intensity of the continuum emission changes with frequency. For most of the sources in both images, the intensity increases with frequency as expected if the emission is dominated by dust with a dependence of the flux, $S_\nu$, with frequency, $\nu$, given by $S_{\nu}\propto\nu^{\alpha}$, where $\alpha=2$ for optically thick emission and $\alpha=2+\beta$ for optically thin dust emission. The coefficient $\beta$ is, in the interstellar medium, usually between 1 and 2 \citep[e.g.\ ][]{Schnee2010, Schnee2014, Juvela2015, Reach2015}, and depends on grain properties. The change of intensity with frequency is better shown in Fig.~\ref{f:fluxfreq}, where we plot the continuum intensity for each spectral window for the three brightest sources in \SgrB(N) and \SgrB(M). However, not all the sources have a flux that increases with frequency. Some sources have a constant flux that in some cases slightly decreases with frequency (see Fig.~\ref{f:fluxfreq}) suggesting a different origin for the emission of these objects than the dust core nature of the brightest sources.

\subsection{Source determination}\label{s:sources}

As shown in Fig.~\ref{f:SgrB2contfreqs}, the structure of the continuum emission at different frequencies is similar but not identical. The small differences in the continuum maps are essentially due to two main reasons: (i) differences in the sampling of the \textit{u,v} domain that may result in different cleaning artifacts. As mentioned in Sect.~\ref{s:obs}, our ALMA observations are sensitive to spatial structures $\le5$\arcsec, and (ii) different accuracy in the determination of the continuum level due to the different lines (in emission and absorption) covered in each spectral window.

\begin{sidewaystable*}
\centering
\caption{\SgrB(N) continuum sources detected at 1.3~mm with ALMA (see Fig.~\ref{f:SgrB2Nsources} and Section~\ref{s:sources}) and fluxes at different frequencies}
\label{t:SgrB2Nsources}
\begin{tabular}{l c c c c c c c c c c c c c c}
\hline\hline

&
&
&\multicolumn{3}{c}{ALMA$_\mathrm{211-275}$}
&VLA\tablefootmark{a}
&ALMA$_{220}$
&ALMA$_{235}$
&ALMA$_{250}$
&ALMA$_{265}$
&SMA\tablefootmark{b}
&\multicolumn{2}{c}{}
\\
\cline{4-6}
\noalign{\smallskip}

&R.A.
&Dec.
&$I_\nu$
&$F_\nu$
&$\theta_\mathrm{s}$
&$F_\nu$
&$F_\nu$
&$F_\nu$
&$F_\nu$
&$F_\nu$
&$F_\nu$
&\multicolumn{2}{c}{}
\\
\multicolumn{1}{c}{ID}
&(h~:~m~:~s)
&($^\circ$~:~\arcmin~:~\arcsec)
&(Jy~beam$^{-1}$)
&(Jy)
&(\arcsec)
&(Jy)
&(Jy)
&(Jy)
&(Jy)
&(Jy)
&(Jy)
&\multicolumn{2}{c}{Spectral index\tablefootmark{c}}
\\
\hline

AN01 &17:47:19.87 &$-$28:22:18.43 &$1.454\pm0.101$ &$25.74\pm2.97$\phn  &1.45 &\phs$0.24$ &$19.7$\phnnn&$22.0$\phnnn&$29.2$\phnnn&$32.2$\phnnn&$50.0$\phn   &$+2.81\pm0.19$ &\phs$2.54$ \\ 
AN02 &17:47:19.86 &$-$28:22:13.17 &$0.502\pm0.071$ &$1.836\pm0.307$     &0.64 &$<$$0.01$  &$1.45$\phn  &$1.50$\phn  &$2.20$\phn  &$2.22$\phn  &\phs$3.93$   &$+3.07\pm0.36$ &$>$$2.83$ \\
AN03 &17:47:19.90 &$-$28:22:13.52 &$0.337\pm0.055$ &$1.404\pm0.240$     &0.70 &\phs$0.02$ &$1.11$\phn  &$1.03$\phn  &$1.73$\phn  &$1.75$\phn  &\phs$4.32$   &$+3.17\pm0.48$ &\phs$2.57$ \\ 
AN04 &17:47:19.80 &$-$28:22:16.32 &$0.258\pm0.031$ &$1.236\pm0.154$     &0.77 &$<$$0.01$  &$1.03$\phn  &$0.99$\phn  &$1.57$\phn  &$1.36$\phn  &\phs$2.42$   &$+2.43\pm0.41$ &$>$$2.64$ \\
AN05 &17:47:19.77 &$-$28:22:16.04 &$0.234\pm0.020$ &$1.019\pm0.092$     &0.84 &$<$$0.01$  &$0.80$\phn  &$0.80$\phn  &$1.35$\phn  &$1.12$\phn  &\phs$1.51$   &$+2.65\pm0.44$ &$>$$2.54$ \\
AN06 &17:47:19.96 &$-$28:22:13.94 &$0.175\pm0.024$ &$1.017\pm0.112$     &0.90 &$<$$0.02$  &$0.88$\phn  &$0.70$\phn  &$1.22$\phn  &$1.26$\phn  &\phs$2.44$   &$+2.37\pm0.62$ &$>$$2.23$ \\
AN07 &17:47:19.91 &$-$28:22:15.48 &$0.151\pm0.009$ &$0.305\pm0.032$     &0.44 &$<$$0.01$  &$0.29$\phn  &$0.23$\phn  &$0.37$\phn  &$0.34$\phn  &\phs$0.75$   &$+2.18\pm0.71$ &$>$$1.93$ \\
AN08 &17:47:19.25 &$-$28:22:14.92 &$0.094\pm0.016$ &$0.308\pm0.054$     &0.59 &$<$$0.02$  &$0.26$\phn  &$0.31$\phn  &$0.32$\phn  &$0.34$\phn  &$<$$0.30$    &$+1.45\pm0.55$ &--- \\
AN09 &17:47:19.64 &$-$28:22:24.37 &$0.074\pm0.007$ &$0.416\pm0.067$     &0.98 &$<$$0.01$  &$0.37$\phn  &$0.31$\phn  &$0.54$\phn  &$0.45$\phn  &$<$$0.30$    &$+2.00\pm1.12$ &--- \\
AN10 &17:47:19.78 &$-$28:22:20.74 &$0.073\pm0.010$ &$0.213\pm0.035$     &0.64 &\phs$0.05$ &$0.31$\phn  &$0.24$\phn  &$0.15$\phn  &$0.16$\phn  &$<$$0.30$    &$-2.60\pm0.55$ &$<$$0.85$ \\
AN11 &17:47:19.68 &$-$28:22:14.99 &$0.066\pm0.007$ &$0.235\pm0.037$     &0.84 &$<$$0.01$  &$0.078$     &$0.16$\phn  &$0.47$\phn  &$0.24$\phn  &$<$$0.30$    &$+5.18\pm1.48$ &--- \\
AN12 &17:47:19.22 &$-$28:22:11.98 &$0.063\pm0.007$ &$0.413\pm0.074$     &1.09 &$<$$0.01$  &$0.49$\phn  &$0.23$\phn  &$0.46$\phn  &$0.47$\phn  &$<$$0.30$    &$-1.02\pm0.88$ &--- \\
AN13 &17:47:19.54 &$-$28:22:32.49 &$0.060\pm0.007$ &$1.039\pm0.198$     &2.01 &$<$$0.01$  &$1.08$\phn  &$0.82$\phn  &$1.50$\phn  &$0.75$\phn  &$<$$0.93$    &$+1.89\pm1.27$ &--- \\
AN14 &17:47:18.97 &$-$28:22:13.38 &$0.054\pm0.007$ &$0.493\pm0.092$     &1.48 &$<$$0.01$  &$0.38$\phn  &$0.26$\phn  &$0.61$\phn  &$0.73$\phn  &$<$$1.19$    &$+3.13\pm0.94$ &--- \\
AN15 &17:47:19.50 &$-$28:22:14.57 &$0.049\pm0.006$ &$0.147\pm0.028$     &0.80 &$<$$0.01$  &$0.088$     &$0.13$\phn  &$0.21$\phn  &$0.16$\phn  &$<$$0.30$    &$+2.29\pm0.84$ &--- \\
AN16 &17:47:20.01 &$-$28:22:04.56 &$0.047\pm0.007$ &$0.468\pm0.089$     &1.59 &$<$$0.03$  &$0.55$\phn  &$0.37$\phn  &$0.57$\phn  &$0.38$\phn  &$<$$0.30$    &$-0.74\pm1.07$ &--- \\
AN17 &17:47:19.99 &$-$28:22:16.18 &$0.047\pm0.008$ &$0.076\pm0.016$     &0.53 &$<$$0.01$  &$0.069$     &$0.059$     &$0.082$     &$0.096$     &$<$$0.30$    &$+2.08\pm0.75$ &--- \\
AN18 &17:47:20.09 &$-$28:22:04.84 &$0.047\pm0.006$ &$0.359\pm0.066$     &1.35 &$<$$0.01$  &$0.29$\phn  &$0.20$\phn  &$0.56$\phn  &$0.38$\phn  &$<$$0.30$    &$+2.10\pm1.25$ &--- \\
AN19 &17:47:19.16 &$-$28:22:15.27 &$0.040\pm0.006$ &$0.118\pm0.022$     &0.83 &$<$$0.01$  &$0.12$\phn  &$0.11$\phn  &$0.12$\phn  &$0.12$\phn  &$<$$0.30$    &$+2.01\pm0.82$ &--- \\
AN20 &17:47:19.88 &$-$28:22:22.48 &$0.039\pm0.007$ &$0.043\pm0.011$     &0.44 &\phs$0.01$ &$0.060$     &$0.062$     &$0.036$     &$0.015$     &$<$$0.30$    &$-2.71\pm1.04$ &$<$$2.39$ \\
\hline
\end{tabular}
\tablefoot{Each source is numbered with an ID according to their intensity peak, from the brightest to the weakest. The intensity ($I_\nu$) and fluxes ($F_\nu$) for the different frequencies have been computed over the polygon defined by the $3\sigma$ contour level on the ALMA$_\mathrm{211-275}$ image, with $\sigma$ being the rms noise of the map, 8~mJy~beam$^{-1}$. The deconvolved size ($\theta_\mathrm{S}$) is determined from the 50\% contour level. See Sect.~\ref{s:sources} for details.\\
\tablefoottext{a}{VLA data at 40~GHz from \citet{Rolffs2011}.}
\tablefoottext{b}{SMA data at 345~GHz from \citet{Qin2011}.}
\tablefoottext{c}{The first spectral index is computed from all the continuum measurements in the ALMA frequency coverage (from 211 to 275~GHz) listed in Table~\ref{t:SgrB2Nfluxes}. The second value is derived from the continuum measurements at 40~GHz (VLA), 242~GHz (ALMA) and 345~GHz (SMA).}
}
\end{sidewaystable*}

\begin{sidewaystable*}
\centering
\caption{\SgrB(M) continuum sources detected at 1.3~mm with ALMA (see Fig.~\ref{f:SgrB2Msources} and Section~\ref{s:sources}) and fluxes at different frequencies}
\label{t:SgrB2Msources}
\begin{tabular}{l c c c c c c c c c c c c c c}
\hline\hline

&
&
&\multicolumn{3}{c}{ALMA$_\mathrm{211-275}$}
&VLA\tablefootmark{a}
&ALMA$_{220}$
&ALMA$_{235}$
&ALMA$_{250}$
&ALMA$_{265}$
&SMA\tablefootmark{b}
&\multicolumn{2}{c}{}
\\
\cline{4-6}
\noalign{\smallskip}

&R.A.
&Dec.
&$I_\nu$
&$F_\nu$
&$\theta_\mathrm{s}$
&$F_\nu$
&$F_\nu$
&$F_\nu$
&$F_\nu$
&$F_\nu$
&$F_\nu$
&\multicolumn{2}{c}{}
\\
\multicolumn{1}{c}{ID}
&(h~:~m~:~s)
&($^\circ$~:~\arcmin~:~\arcsec)
&(Jy~beam$^{-1}$)
&(Jy)
&(\arcsec)
&(Jy)
&(Jy)
&(Jy)
&(Jy)
&(Jy)
&(Jy)
&\multicolumn{2}{c}{Spectral index\tablefootmark{c}}
\\
\hline

AM01 &17:47:20.16 &$-$28:23:04.67 &$1.815\pm0.025$ &$5.619\pm0.135$ &0.58 &\phs$1.28$ &$4.69$\phn &$5.00$\phn &$6.14$\phn &$6.65$\phn &\phs$11.26$\phn  &$+1.95\pm0.13$ &\phs$0.89$ \\ 
AM02 &17:47:20.14 &$-$28:23:04.18 &$1.437\pm0.025$ &$5.745\pm0.149$ &0.63 &\phs$0.78$ &$4.74$\phn &$5.16$\phn &$6.26$\phn &$6.82$\phn &\phs$11.85$\phn  &$+2.14\pm0.14$ &\phs$1.17$ \\ 
AM03 &17:47:20.15 &$-$28:23:05.44 &$0.716\pm0.027$ &$1.792\pm0.090$ &0.70 &$<$$0.01$  &$1.33$\phn &$1.52$\phn &$2.03$\phn &$2.29$\phn &\phs$3.96$       &$+3.09\pm0.14$ &$>$$2.87$ \\
AM04 &17:47:20.18 &$-$28:23:06.00 &$0.469\pm0.018$ &$1.199\pm0.046$ &0.60 &\phs$0.07$ &$0.90$\phn &$1.00$\phn &$1.38$\phn &$1.52$\phn &\phs$2.07$       &$+2.96\pm0.22$ &\phs$1.59$ \\ 
AM05 &17:47:20.12 &$-$28:23:06.28 &$0.373\pm0.024$ &$1.196\pm0.075$ &0.59 &$<$$0.01$  &$0.94$\phn &$0.98$\phn &$1.42$\phn &$1.44$\phn &\phs$1.94$       &$+2.59\pm0.26$ &$>$$2.63$ \\
AM06 &17:47:20.11 &$-$28:23:02.99 &$0.310\pm0.010$ &$0.937\pm0.051$ &0.59 &$<$$0.01$  &$0.74$\phn &$0.76$\phn &$1.18$\phn &$1.07$\phn &\phs$2.90$       &$+2.80\pm0.43$ &$>$$2.60$ \\
AM07 &17:47:20.20 &$-$28:23:06.63 &$0.310\pm0.012$ &$0.800\pm0.044$ &0.53 &\phs$0.01$ &$0.63$\phn &$0.65$\phn &$0.93$\phn &$0.99$\phn &\phs$1.57$       &$+2.67\pm0.29$ &\phs$2.27$ \\ 
AM08 &17:47:20.29 &$-$28:23:03.13 &$0.252\pm0.007$ &$0.401\pm0.021$ &0.35 &\phs$0.13$ &$0.41$\phn &$0.35$\phn &$0.44$\phn &$0.41$\phn &\phs$0.41$       &$+0.48\pm0.38$ &\phs$0.63$ \\ 
AM09 &17:47:20.23 &$-$28:23:07.19 &$0.208\pm0.009$ &$0.683\pm0.054$ &0.59 &$<$$0.01$  &$0.55$\phn &$0.53$\phn &$0.78$\phn &$0.87$\phn &\phs$1.59$       &$+2.81\pm0.43$ &$>$$2.37$ \\
AM10 &17:47:20.06 &$-$28:23:06.21 &$0.172\pm0.011$ &$1.002\pm0.071$ &1.09 &$<$$0.01$  &$0.94$\phn &$0.85$\phn &$1.13$\phn &$1.09$\phn &\phs$1.53$       &$+1.13\pm0.33$ &$>$$2.54$ \\
AM11 &17:47:19.62 &$-$28:23:08.24 &$0.132\pm0.007$ &$0.525\pm0.053$ &0.59 &$<$$0.01$  &$0.37$\phn &$0.38$\phn &$0.82$\phn &$0.53$\phn &\phs$0.70$       &$+3.92\pm1.19$ &$>$$2.20$ \\
AM12 &17:47:20.14 &$-$28:23:02.22 &$0.126\pm0.007$ &$0.544\pm0.042$ &0.81 &$<$$0.03$  &$0.44$\phn &$0.38$\phn &$0.70$\phn &$0.65$\phn &\phs$1.98$       &$+3.12\pm0.59$ &$>$$1.73$ \\
AM13 &17:47:19.96 &$-$28:23:05.58 &$0.109\pm0.006$ &$0.548\pm0.047$ &0.92 &$<$$0.01$  &$0.34$\phn &$0.40$\phn &$0.78$\phn &$0.68$\phn &\phs$1.24$       &$+4.37\pm0.65$ &$>$$2.25$ \\
AM14 &17:47:20.28 &$-$28:23:03.69 &$0.129\pm0.006$ &$0.297\pm0.024$ &0.63 &$<$$0.01$  &$0.24$\phn &$0.23$\phn &$0.36$\phn &$0.37$\phn &$<$$0.30$        &$+3.24\pm0.55$ &--- \\
AM15 &17:47:19.92 &$-$28:23:02.99 &$0.071\pm0.005$ &$0.151\pm0.018$ &0.55 &\phs$0.14$ &$0.15$\phn &$0.13$\phn &$0.19$\phn &$0.14$\phn &$<$$0.30$        &$+0.49\pm0.78$ &$<$$0.39$ \\
AM16 &17:47:20.39 &$-$28:23:05.30 &$0.067\pm0.005$ &$0.475\pm0.061$ &1.23 &\phs$0.28$ &$0.54$\phn &$0.36$\phn &$0.50$\phn &$0.49$\phn &$<$$0.30$        &$+0.71\pm0.78$ &$<$$0.01$ \\
AM17 &17:47:19.45 &$-$28:22:59.77 &$0.054\pm0.005$ &$0.371\pm0.052$ &1.25 &\phs$0.11$ &$0.44$\phn &$0.27$\phn &$0.41$\phn &$0.37$\phn &$<$$0.30$        &$+0.86\pm1.17$ &$<$$0.46$ \\
AM18 &17:47:19.58 &$-$28:23:07.12 &$0.052\pm0.005$ &$0.134\pm0.020$ &0.72 &$<$$0.01$  &$0.091$    &$0.079$    &$0.24$\phn &$0.13$\phn &$<$$0.30$        &$+6.32\pm2.17$ &--- \\
AM19 &17:47:19.68 &$-$28:23:09.15 &$0.050\pm0.005$ &$0.184\pm0.026$ &0.90 &$<$$0.02$  &$0.094$    &$0.027$    &$0.38$\phn &$0.24$\phn &$<$$0.30$        &$+3.98\pm2.05$ &--- \\
AM20 &17:47:19.94 &$-$28:23:09.92 &$0.048\pm0.005$ &$0.160\pm0.024$ &1.00 &$<$$0.01$  &$0.16$\phn &$0.13$\phn &$0.20$\phn &$0.16$\phn &\phs$0.35$       &$-0.48\pm0.76$ &$>$$1.56$ \\
AM21 &17:47:19.55 &$-$28:23:03.13 &$0.042\pm0.005$ &$0.064\pm0.011$ &0.59 &$<$$0.01$  &$0.052$    &$0.052$    &$0.068$    &$0.084$    &$<$$0.30$        &$+2.67\pm0.95$ &--- \\
AM22 &17:47:19.44 &$-$28:23:07.68 &$0.040\pm0.005$ &$0.073\pm0.012$ &0.65 &$<$$0.01$  &$0.063$    &$0.037$    &$0.096$    &$0.096$    &$<$$0.30$        &$+4.57\pm1.34$ &--- \\
AM23 &17:47:19.38 &$-$28:23:03.13 &$0.039\pm0.005$ &$0.140\pm0.024$ &1.02 &$<$$0.01$  &$0.13$\phn &$0.095$    &$0.14$\phn &$0.20$\phn &$<$$0.30$        &$+3.45\pm1.04$ &--- \\
AM24 &17:47:20.11 &$-$28:23:08.87 &$0.039\pm0.005$ &$0.036\pm0.007$ &0.37 &\phs$0.06$ &$0.054$    &$0.040$    &$0.038$    &$0.010$    &$<$$0.30$        &$-2.43\pm0.93$ &$<$$0.86$ \\
AM25 &17:47:19.70 &$-$28:23:07.19 &$0.033\pm0.005$ &$0.054\pm0.011$ &0.61 &$<$$0.01$  &$0.066$    &$0.033$    &$0.088$    &$0.031$    &$<$$0.30$        &$+2.22\pm1.13$ &--- \\
AM26 &17:47:20.35 &$-$28:23:07.82 &$0.033\pm0.005$ &$0.115\pm0.022$ &0.95 &$<$$0.01$  &$0.069$    &$0.11$\phn &$0.13$\phn &$0.16$\phn &$<$$0.30$        &$+5.03\pm0.91$ &--- \\
AM27 &17:47:19.35 &$-$28:23:08.94 &$0.028\pm0.005$ &$0.048\pm0.011$ &0.62 &$<$$0.01$  &$0.062$    &$0.067$    &$0.020$    &$0.042$    &$<$$0.30$        &$+0.13\pm1.27$ &--- \\
\hline
\end{tabular}
\tablefoot{Each source is numbered with an ID according to their intensity peak, from the brightest to the weakest. The intensity ($I_\nu$) and fluxes ($F_\nu$) for the different frequencies have been computed over the polygon defined by the $3\sigma$ contour level on the ALMA$_\mathrm{211-275}$ image, with $\sigma$ being the rms noise of the map, 8~mJy~beam$^{-1}$. The deconvolved size ($\theta_\mathrm{S}$) is determined from the 50\% contour level. See Sect.~\ref{s:sources} for details.\\
\tablefoottext{a}{VLA data at 40~GHz from \citet{Rolffs2011}.}
\tablefoottext{b}{SMA data at 345~GHz from \citet{Qin2011}.}
\tablefoottext{c}{The first spectral index is computed from all the continuum measurements in the ALMA frequency coverage (from 211 to 275~GHz) listed in Table~\ref{t:SgrB2Mfluxes}. The second value is derived from the continuum measurements at 40~GHz (VLA), 242~GHz (ALMA) and 345~GHz (SMA).}
}
\end{sidewaystable*}

In order to identify the continuum sources, we have created new continuum maps combining the images produced for each individual spectral window. By doing this, we downweight the artifacts that appear only in some maps and we emphasize the real sources. However, it is important to consider also the variation of the flux with frequency of some sources (see Fig.~\ref{f:fluxfreq}). We have averaged the different individual continuum maps in different ways. Four maps were produced by averaging the continuum images of the spectral windows that are plotted in each one of the four blocks shown in Fig.~\ref{f:freqcover}. They have central frequencies of 220~GHz, 235~GHz, 250~GHz and 265~GHz. Additionally, two maps were produced averaging the spectral windows that correspond to the low-frequency (LF) setup and the high-frequency (HF) setup (see Sect.~\ref{s:obs}), and are centered at the frequencies 227.5~GHz and 256.6~GHz, respectively. One final image is produced averaging all the spectral windows, and is centered at the frequency of 242~GHz. This results in a total of seven continuum maps (shown in Fig.~\ref{f:SgrB2ALLcontinuum}), complementary to the maps created for each spectral window. We have searched these new images for emission features that appear in most of them and considered them to be real emission (i.e.\ emission from a source). If some feature is only observed in one or two maps, we consider them artifacts of the cleaning process, or imperfect removal of the line contamination due to different spectral line features in each spectral window.

For each real continuum emission feature, we identify a source if there is at least one closed contour above the 3$\sigma$ level (with $\sigma$ the rms noise level of the map: 8~mJy~beam$^{-1}$). In some cases, there are extensions or elongations that suggest the presence of additional weaker, nearby objects that cannot be completely resolved in our observations (angular resolution of $0\farcs4$). The identified sources are listed in Table~\ref{t:SgrB2Nsources} for \SgrB(N) and in Table~\ref{t:SgrB2Msources} for \SgrB(M). Each source has been labeled with the letters AN or AM to indicate they are sources detected with ALMA in \SgrB(N) and \SgrB(M), respectively, followed by a number that orders the sources from the brightest to the faintest. In total, we have detected 20 sources in \SgrB(N), 12 of them being located in the central 10\arcsec\ around the main core, and 27 sources in \SgrB(M) with 18 in the central 10\arcsec. The spatial distribution of the sources is shown in Figs.~\ref{f:SgrB2Nsources} and \ref{f:SgrB2Msources}, and their coordinates are listed in Tables~\ref{t:SgrB2Nsources} and \ref{t:SgrB2Msources}. The number of detected sources in the central 10\arcsec\ (12 for \SgrB(N) and 18 for \SgrB(M)) differs from the number of sources identified by \citet{Qin2011} in SMA observations at 345~GHz (2 in \SgrB(N) and 12 in \SgrB(M)). This suggests that the ALMA observations at 1.3~mm have a better \textit{uv}-coverage leading to a higher image fidelity, and that they are sensitive to weak sources, not previously seen in the SMA images due to a low sensitivity (20--30~mJy~beam$^{-1}$, \citealt{Qin2011}), and to a different population of objects not detectable in the previous sub-millimeter images. As discussed in the following sections, some of these sources have a thermal free-free origin, i.e.\ they are \hii\ regions still bright at 1~mm but completely attenuated at 850~$\mu$m (or 345~GHz). The nature of each detected source is discussed in detail in Sect.~\ref{s:origin}.

In addition to the identified compact sources, extended structures are clearly visible in the continuum images of Figs.~\ref{f:SgrB2Nsources} and \ref{f:SgrB2Msources}. Particularly striking is the filamentary structure to the northwest of the main core \SgrB(N)-AN01, which contains the sources AN04, AN05, AN11 and AN15. These four sources seem to lie along a faint, extended filament that points towards the brightest center of source AN01. Apart from this filament, three additional filamentary structures are visible to the east and south of source AN01, all of them converging towards the center. Similar structures have been found in other high-mass star-forming regions (e.g.\ G33.92+0.11, \citealt{Liu2015}) likely tracing accretion channels fuelling material to the star-forming dense cores. The bright core \SgrB(N)-AN01 was reported to be monolithic by \citet{Qin2011} in their SMA observations (with a similar angular resolution, 0\farcs35). The ALMA images confirm that AN01 appears monolithic. In Sect.~\ref{s:properties} we discuss about its possible nature. Other elongated structures are visible in \SgrB(M), for example the one connecting sources AM10 and AM13 to the main emission, and the one associated with source AM16. This second one has an arc-shape structure, with the cometary head pointing towards the brightest sources. As will be discussed in following sections, source AM16 is associated with a cometary \hii\ region, and the emission detected at 1.3~mm is probably highly contaminated by free-free ionized gas emission.

For each identified source we have defined the polygon that delineates the 50\% contour level with respect to the intensity peak and the 3$\sigma$ contour level. We used the image at 242~GHz (average of all the spectral window continuum images) to define the polygons and cross-checked with the other images to confirm them. These polygons are used to determine the size and the flux of each source. We refrain from fitting Gaussians due to the complex and extended structure of some sources. The flux is determined as the integrated flux over the 3$\sigma$ polygon. We also determine the intensity peak as the maximum intensity within the polygon. In Tables~\ref{t:SgrB2Nsources} and \ref{t:SgrB2Msources}, we list the intensities and fluxes obtained from the 242~GHz continuum image. The uncertainties are computed from the noise map that is automatically produced while creating the continuum image (see S\'anchez-Monge et al.\ submitted, and Fig.~\ref{f:SgrB2ALLcontinuumnoise}). The polygon at the 50\% contour level is used to determine the size of the source. We used the approach described in \citet{SanchezMonge2013a} to determine the effective observed full width at half maximum, and the deconvolved size (after taking into account the contribution of the beam to the observed size). The deconvolved sizes are listed in Tables~\ref{t:SgrB2Nsources} and \ref{t:SgrB2Msources}. For \SgrB(N) the sources have sizes in the range 1700--13000~au, and a mean value of 6600~au (assuming a distance of 8.34~kpc). \SgrB(M) contains slightly more compact sources with sizes in the range 1700--9400~au, and a mean size of 5400~au.

\section{Analysis and discussion}\label{s:analysis}

In this section we study the properties of the sources identified in the ALMA continuum images, we compare them with previous VLA 40~GHz \citep{Rolffs2011} and SMA 345~GHz \citep{Qin2011} images in order to infer their nature. We then characterize their physical properties, origin of the millimeter continuum emission, and chemical content. Finally, we compare the observed structure with predictions from a 3D radiative transfer model \citep{Schmiedeke2016} that models the whole structure of \SgrB\ from 45~pc down to 100~au.
 
\subsection{Origin of the continuum emission}\label{s:origin}

The nature of the emission at 1.3~mm for each source can be inferred by studying their millimeter continuum spectrum or spectral energy distribution (hereafter `mm-SED' for simplicity). We have used the polygon at the 3$\sigma$ level to determine the intensity peak and flux density of each source at all available continuum images, i.e.\ for each individual spectral window. This permits us to sample the mm-SED of each source in the range 211~GHz to 275~GHz. In Appendix~\ref{a:fluxes} we list the fluxes for all the sources and spectral windows, while in Tables~\ref{t:SgrB2Nsources} and \ref{t:SgrB2Msources} we list the fluxes measured in the four images created after combining different spectral windows (see Sect.~\ref{s:sources}) with central frequencies 220~GHz, 235~GHz, 250~GHz and 265~GHz. For completeness, we also list the flux at 40~GHz \citep[from VLA observations,][]{Rolffs2011} and the flux at 345~GHz \citep[from SMA observations,][]{Qin2011}, see below for a more detailed description. The fluxes at the different frequencies within the ALMA band can be used to investigate the nature of the 1.3~mm emission for each source. This is done by studying the spectral index ($\alpha$, defined as $S_\nu\propto\nu^\alpha$). For dust emission, $\alpha$ must be in the range 2 (optically thick dust) to 4 (optically thin dust). Flatter spectral indices may suggest an origin different than dust. For example, values of the spectral index between 1 and 2 are usually found when observing hypercompact \hii\ regions, spectral indices $\sim0.6$ are typically associated with jets and winds, while values of $-0.1$ are usually found towards optically thin \hii\ regions \citep[see][]{SanchezMonge2013b}.

\begin{figure}[t!]
\begin{center}
\begin{tabular}[b]{c}
        \includegraphics[width=0.9\columnwidth]{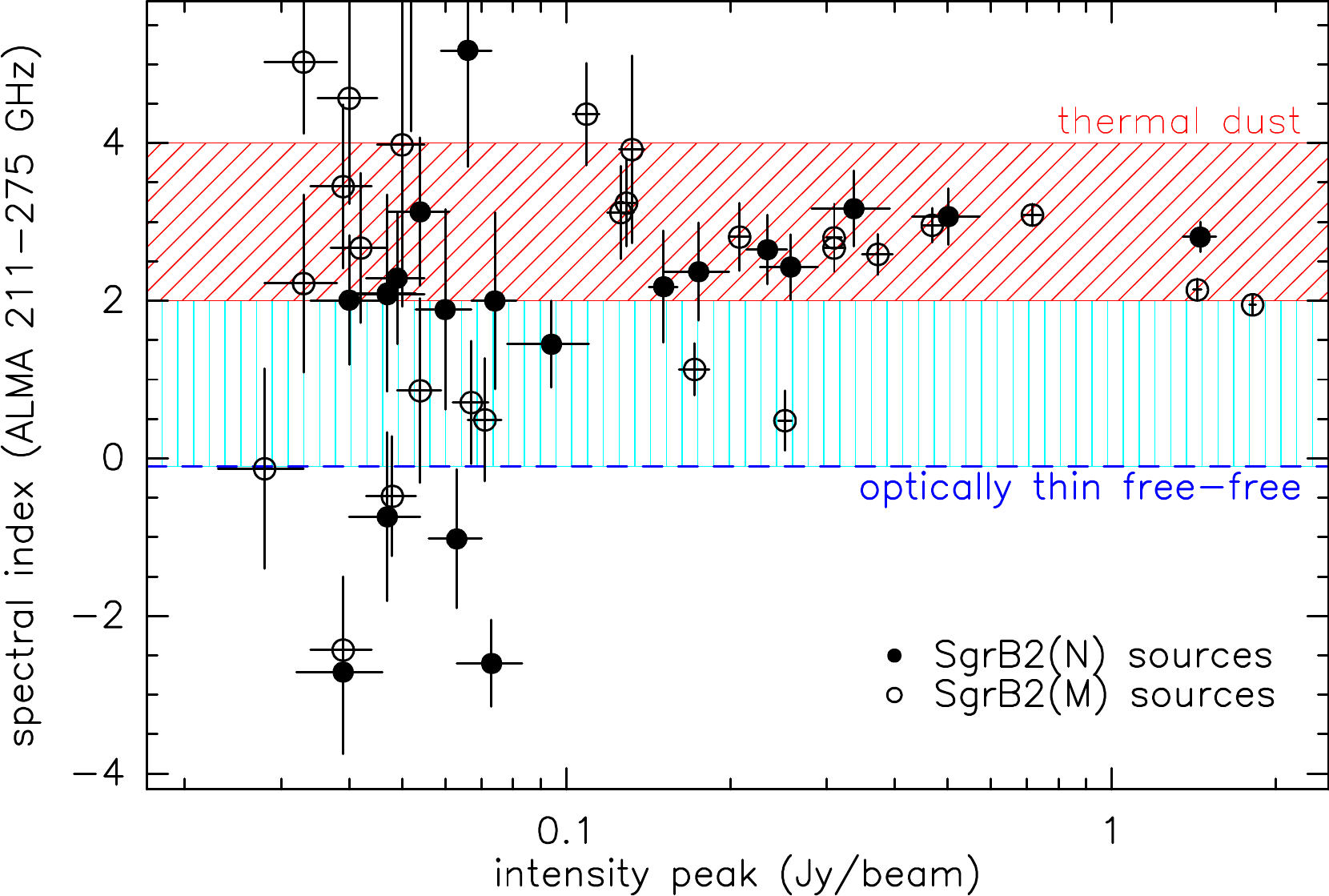} \\
\end{tabular}
\caption{Spectral index within the ALMA frequency coverage (from 211~GHz to 275~GHz) as a function of the intensity peak of the continuum emission at 242~GHz (see Tables~\ref{t:SgrB2Nsources} and \ref{t:SgrB2Msources}. Filled circles depict the continuum sources found in \SgrB(N), while open circles correspond to sources identified in \SgrB(M). The red area shows the spectral index regime expected for dust sources (from $+2$ for optically thick dust emission to $+4$ for optically thin dust emission). The light blue area shows the regime of spectral indices for ionized gas sources, ranging from $-0.1$ for optically thin, homogeneous \hii\ regions (also marked with a dark blue dashed line) to $+2$ for optically thick sources. The vertical and horizontal lines show 1$\sigma$ error bars for the spectral index and intensity peak.}
\label{f:spindex_flux}
\end{center}
\end{figure}

\begin{figure}[h!]
\begin{center}
\begin{tabular}[b]{c}
        \includegraphics[width=0.88\columnwidth]{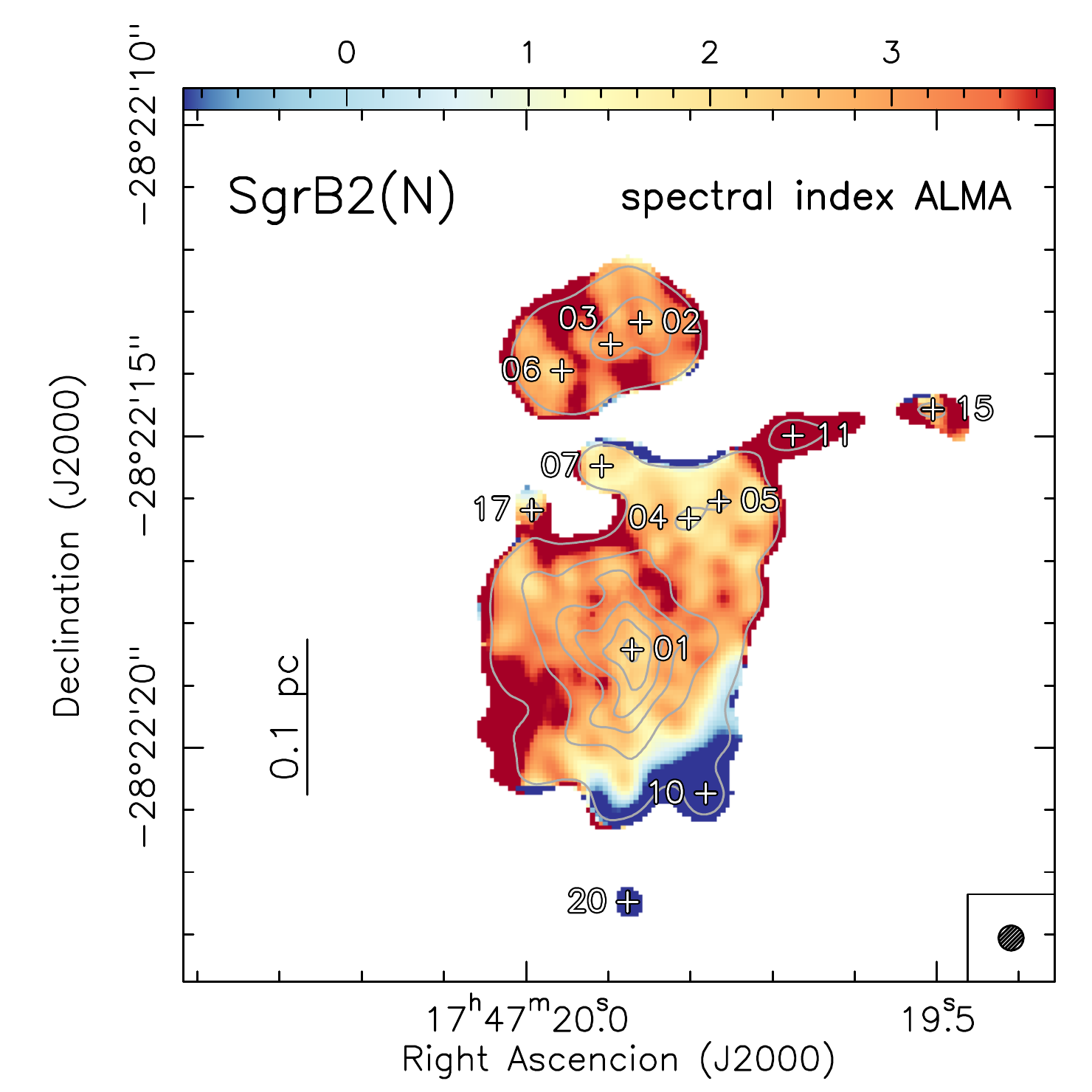} \\
        \includegraphics[width=0.88\columnwidth]{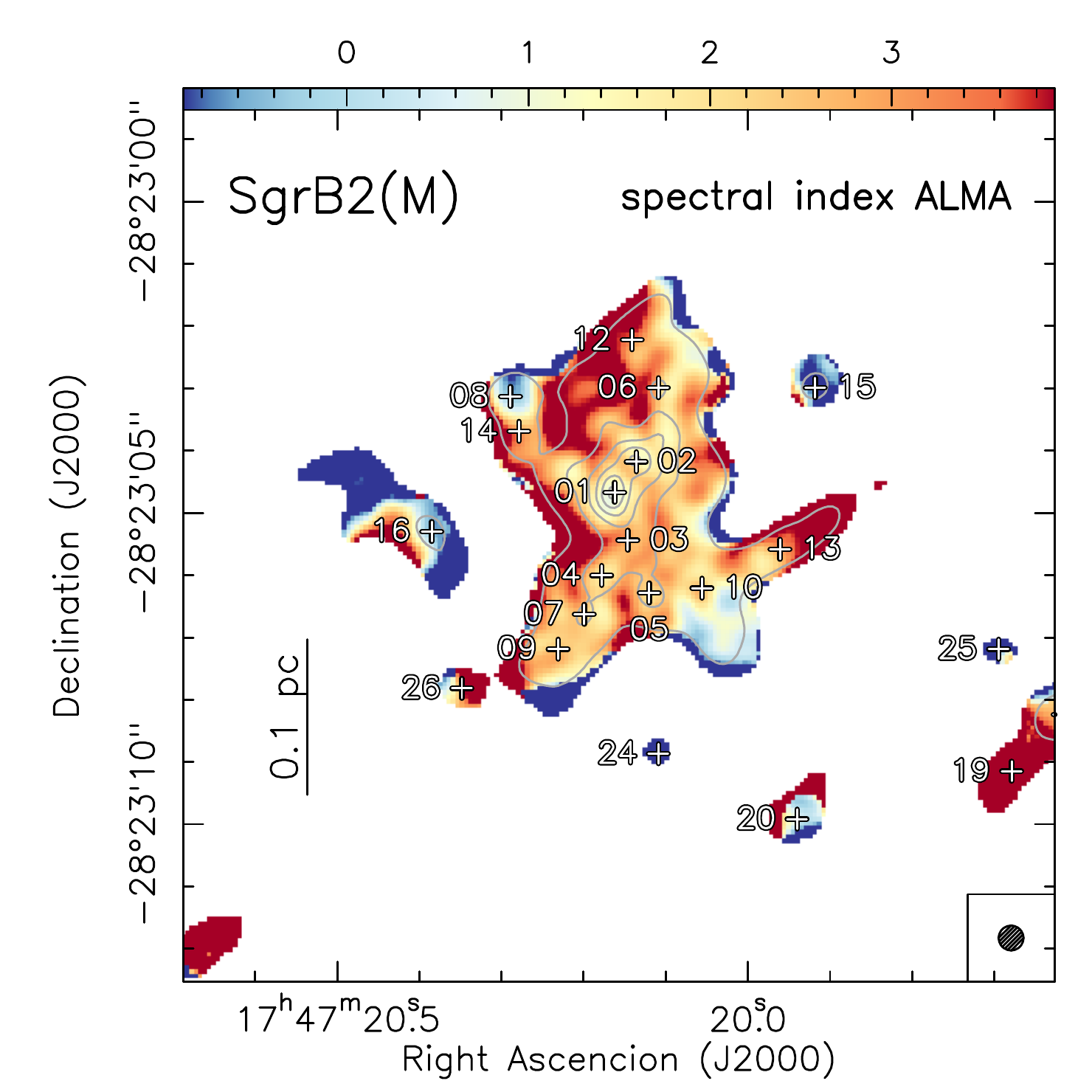} \\
\end{tabular}
\caption{Spectral index map from the ALMA continuum measurements from 211~GHz to 275~GHz (see Sect.~\ref{s:origin}) for \SgrB(N) (left panel) and \SgrB(M) (right panel). Overlaid in gray contours there is the ALMA 1.3~mm (242~GHz) continuum emission as shown in Figs.~\ref{f:SgrB2Nsources} and \ref{f:SgrB2Msources}. The synthesized beam is $0\farcs4$ and is shown in the bottom-right corner of each penal. The continuum sources listed in Tables~\ref{t:SgrB2Nsources} and \ref{t:SgrB2Msources} are indicated with crosses and numbers.}
\label{f:SgrB2spindex}
\end{center}
\end{figure}

\begin{figure*}[h!]
\begin{center}
\begin{tabular}[b]{c}
        \includegraphics[width=0.89\textwidth]{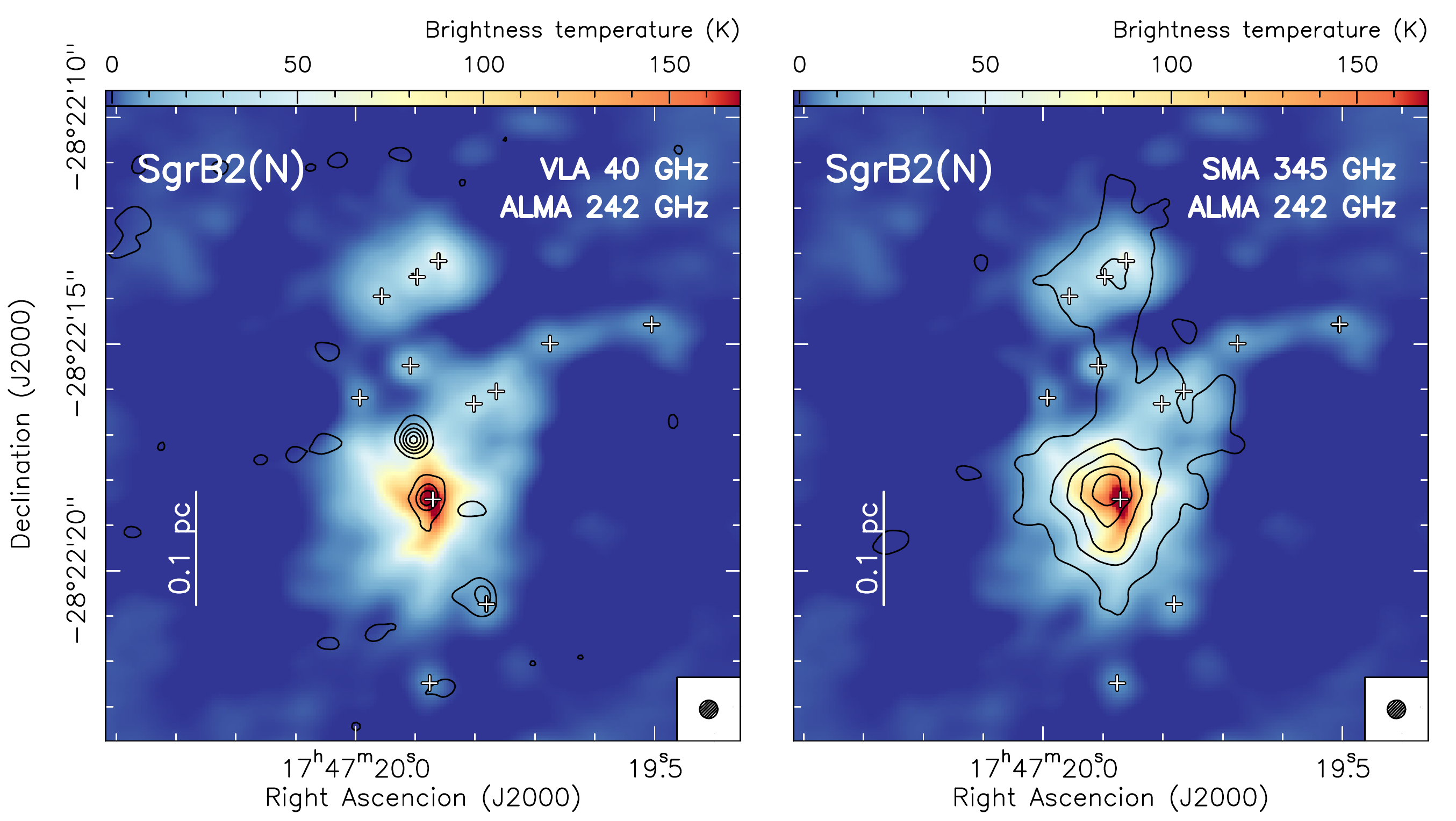} \\
        \includegraphics[width=0.89\textwidth]{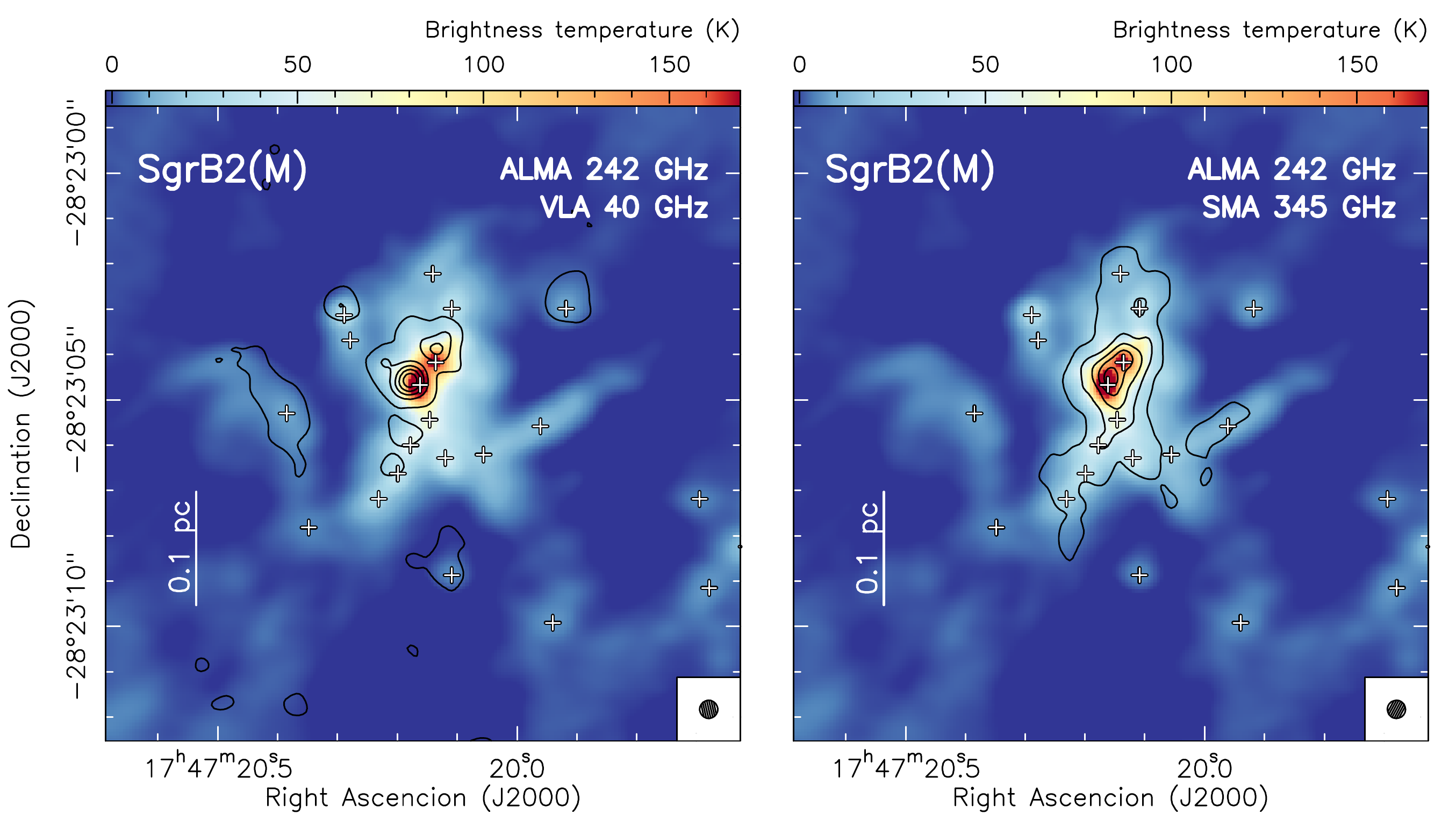} \\
\end{tabular}
\caption{(\textit{top panels}) \SgrB(N) ALMA continuum emission image at 242~GHz compared to the VLA 40~GHz continuum emission in contours (\textit{left panel}; from \citealt{Rolffs2011}, and convolved to $0\farcs$4) and to the SMA 345~GHz in contours (\textit{right panel}; from \citealt{Qin2011}; convolved to $0\farcs4$). The contour levels are from 10\% to 100\% the intensity peak 0.085~Jy~beam$^{-1}$, in steps of 20\%, and from 5\% to 100\% the intensity peak 2.82~Jy~beam$^{-1}$, in steps of 20\%, for the VLA and SMA images, respectively.
(\textit{bottom panels}) Same as in the top panels for \SgrB(M). The contour levels are from 2\% to 100\% the intensity peak 0.62~Jy~beam$^{-1}$, in steps of 20\%, and from 5\% to 100\% the intensity peak 3.41~Jy~beam$^{-1}$, in steps of 20\%, for the VLA and SMA images, respectively.
In all panels, the crosses mark the position of the continuum sources listed in Tables~\ref{t:SgrB2Nsources} and \ref{t:SgrB2Msources}.}
\label{f:SgrB2VLASMA}
\end{center}
\end{figure*}

The spectral index obtained by fitting the fluxes of all the spectral windows is listed in the last column of Tables~\ref{t:SgrB2Nsources} and \ref{t:SgrB2Msources}. In Figs.~\ref{f:SgrB2NSEDs} and \ref{f:SgrB2MSEDs}, we show the mm-SEDs of each source. We investigate the dominant nature origin for the emission at 1.3~mm by comparing the spectral index to the brightness of the sources. We find that the brightest sources usually show positive spectral indices, while the faint sources have spectral indices going from negative to positive values, with larger uncertainties. This is better shown in Fig.~\ref{f:spindex_flux}, where we plot the variation of the spectral index as a function of the intensity peak at 242~GHz. From this plot, we can infer that the brightest sources are generally tracing a dust component, while some of the faint objects might be \hii\ regions still bright in the millimeter regime. A few sources show negative spectral indices with values $<-0.1$. Most of these sources are weak with large uncertainties. We note that the ALMA observations are sensitive to angular scales of only $<5$\arcsec, and this may result in resolving out extended emission which hinders an accurate determination of the spectral index for faint sources distributed throughout the extended envelopes of \SgrB(N) and \SgrB(M). In Fig.~\ref{f:SgrB2spindex}, we show spectral index maps computed on a pixel to pixel basis from the continuum maps produced for each spectral window observed with ALMA (see Sect.~\ref{s:cont1mm}). The emission in both regions have typically positive spectral indices with values $\alpha>+2$, suggesting a major contribution from thermal dust to the continuum emission at 1.3~mm. In both regions, \SgrB(N) and \SgrB(M), there is a trend with the central pixels of the brightest sources having spectral indices close to $+2$ and the surrounding emission with values between $+3$ and $+4$. This is consistent with the dust being optically thick towards the central and most dense areas of the most massive cores, and optically thin in the outskirts. Finally, the presence of different sources with dominant flat ($\alpha\sim0$) or negative spectral index emission, commonly found toward \hii\ regions, is noticeable. For example, source AN10 in \SgrB(N) and sources AM08, AM15 and AM16 in \SgrB(M).

In order to confirm the spectral index derived from the ALMA band~6 observations and better establish the nature of the 1.3~mm continuum emission, we have compared the ALMA continuum images with VLA and SMA continuum images at 40~GHz and 345~GHz respectively. The details of these observations can be found in \citet{Rolffs2011} and \citet{Qin2011}, the angular resolution of the images is $0\farcs14\times0\farcs08$ at 40~GHz and $0\farcs4\times0\farcs3$ at 345~GHz. We have convolved the images to the same angular resolution as the ALMA maps, i.e.\ $0\farcs4$, and we have extracted the fluxes for each ALMA source using the polygons defined in Sect.~\ref{s:sources}. In Tables~\ref{t:SgrB2Nsources} and \ref{t:SgrB2Msources} we list the fluxes at 40 and 345~GHz. The second number in the last column of both tables lists the spectral index computed after taking into account the VLA and SMA fluxes. The derived values are in agreement with the spectral indices derived only with the ALMA data from 211 to 275~GHz, and, therefore, the spectral indices found to be around 0 confirm the presence of thermal ionized gas free-free emission contributing to the emission at 1.3~mm of some sources. A direct comparison of the \hii\ regions detected in the VLA map and the thermal free-free sources detected with ALMA is shown in the left panels of Fig.~\ref{f:SgrB2VLASMA}. Differently to ALMA, the SMA observations at 345~GHz seem to be essentially tracing dust continuum emission only. The distribution of dust at 345~GHz is also compared to the ALMA continuum map in the right panels of Fig.~\ref{f:SgrB2VLASMA}. The overlay of these maps clearly confirms that those sources with flat/negative spectral indices in the ALMA images are directly associated with \hii\ regions bright at 40~GHz and not detected at 345~GHz (e.g.\ sources AN10 in \SgrB(N) and sources AM08, AM15 and AM16 in \SgrB(M)). This analysis confirms a degree of contamination from free-free emission in the \SgrB\ clouds even at wavelengths as long as 1~mm, and therefore, one must be careful when analyzing longer wavelength millimeter images (e.g.\ continuum emission at 3~mm or 2~mm) and assuming that the emission comes from dust cores only. A more accurate determination of the ionized gas content at millimeter wavelengths will be possible in the near future thanks to on-going sub-arcsecond resolution observational projects studying the \SgrB\ region at different wavelengths from 5~GHz to 150~GHz (e.g.\ Ginsburg et al.\ in prep; Meng et al.\ in prep).

\begin{table*}[h!]
\caption{\label{t:properties} Dust emission, ionized gas and molecular properties of the ALMA continuum sources in \SgrB}
\centering
\begin{tabular}{l c c c c c c c c c c c c}
\hline\hline

&\multicolumn{3}{c}{dust emission}
&
&\multicolumn{3}{c}{ionized gas}
&
&\multicolumn{3}{c}{molecular content}
\\
\cline{2-4}
\cline{6-8}
\cline{10-12}

&$M_\mathrm{d+g}$
&$n_\mathrm{H_2}$
&$N_\mathrm{H_2}$
&
&$n_\mathrm{e}$
&$M_\mathrm{i}$
&$\dot{N}_\mathrm{i}$
&
&lines
&$L_\mathrm{lines}$
&$L_\mathrm{l}$/$L_\mathrm{c+l}$
\\
\multicolumn{1}{c}{ID}
&(\mo)
&($10^8$~cm$^{-3}$)
&($10^{24}$~cm$^{-2}$)
&
&(cm$^{-3}$)
&(\mo)
&(s$^{-1}$)
&
&per GHz
&($0.1$~\lo)
&(\%)
\\
\hline
\multicolumn{6}{l}{\phn \SgrB(N)} \\
\hline
AN01 &8668~--~4083  &20.94~--~9.86\phn &130.9~--~61.6\phn &&$4.8\times10^{4}$ &$0.13$\phnn &$1.9\times10^{48}$ &&\phn74.8 \phn(6.9) &92.61    &22.1 \\
AN02 &623~--~294    &7.77~--~3.66      &48.6~--~22.9      &&\ldots            &\ldots      &\ldots             &&101.6 \phn(6.8)    &11.77    &33.7 \\
AN03 &470~--~221    &4.83~--~2.27      &30.2~--~14.2      &&$4.4\times10^{4}$ &$0.013$\phn &$1.7\times10^{47}$ &&\phn92.0 \phn(4.9) &10.81    &37.8 \\
AN04 &419~--~198    &3.59~--~1.69      &22.4~--~10.6      &&\ldots            &\ldots      &\ldots             &&100.3 \phn(7.6)    &\phn7.98 &33.6 \\
AN05 &346~--~163    &2.47~--~1.16      &15.4~--~7.3\phn   &&\ldots            &\ldots      &\ldots             &&\phn99.2 (24.1)    &\phn3.12 &19.4 \\
AN06 &345~--~163    &2.16~--~1.02      &13.5~--~6.4\phn   &&\ldots            &\ldots      &\ldots             &&100.9 (29.4)       &\phn3.60 &22.0 \\
AN07 &104~--~49\phn &2.70~--~1.27      &16.8~--~7.9\phn   &&\ldots            &\ldots      &\ldots             &&\phn52.2 (24.9)    &\phn0.48 &11.3 \\
AN08 &105~--~49\phn &1.51~--~0.71      &9.5~--~4.5        &&\ldots            &\ldots      &\ldots             &&\phn65.8 (23.9)    &\phn0.90 &19.1 \\
AN09 &141~--~66\phn &0.75~--~0.35      &4.7~--~2.2        &&\ldots            &\ldots      &\ldots             &&\phn10.4 \phn(8.7) &\phn0.02 &\phn0.4 \\
AN10 &57~--~27      &0.71~--~0.34      &4.5~--~2.1        &&$7.7\times10^{4}$ &$0.017$\phn &$4.1\times10^{47}$ &&\phn73.3 (11.6)    &\phn1.39 &33.7 \\
AN11 &80~--~38      &0.57~--~0.27      &3.6~--~1.7        &&\ldots            &\ldots      &\ldots             &&\phnn8.1 (11.1)    &\phn0.01 &\phn0.3 \\
AN12 &140~--~66\phn &0.60~--~0.28      &3.8~--~1.8        &&\ldots            &\ldots      &\ldots             &&\phn32.9 (33.2)    &\phn0.07 &\phn1.4 \\
AN13 &353~--~166    &0.44~--~0.21      &2.8~--~1.3        &&\ldots            &\ldots      &\ldots             &&\phn48.2 (42.3)    &\phn0.36 &\phn3.0 \\
AN14 &167~--~79\phn &0.39~--~0.18      &2.4~--~1.1        &&\ldots            &\ldots      &\ldots             &&\phn18.1 (10.6)    &\phn0.11 &\phn1.9 \\
AN15 &50~--~23      &0.40~--~0.19      &2.5~--~1.2        &&\ldots            &\ldots      &\ldots             &&\phn10.0 \phn(9.9) &\phn0.01 &\phn0.4 \\
AN16 &159~--~75\phn &0.32~--~0.15      &\phn2.0~--~0.94   &&\ldots            &\ldots      &\ldots             &&\phn24.9 (15.4)    &\phn0.14 &\phn2.7 \\
AN17 &26~--~12      &0.47~--~0.22      &2.9~--~1.4        &&\ldots            &\ldots      &\ldots             &&\phn60.0 (26.2)    &\phn0.36 &26.8 \\
AN18 &122~--~57\phn &0.34~--~0.16      &2.1~--~1.0        &&\ldots            &\ldots      &\ldots             &&\phn22.3 (13.0)    &\phn0.11 &\phn2.7 \\
AN19 &40~--~19      &0.30~--~0.14      &\phn1.8~--~0.87   &&\ldots            &\ldots      &\ldots             &&\phn17.8 (11.9)    &\phn0.02 &\phn1.5 \\
AN20 &12~--~6\phn   &0.31~--~0.14      &\phn1.9~--~0.90   &&$5.8\times10^{4}$ &$0.0043$    &$7.6\times10^{46}$ &&\phn10.1 \phn(5.5) &\phn0.01 &\phn3.5 \\
\hline
\multicolumn{6}{l}{\phn \SgrB(M)} \\
\hline
AM01 &1544~--~727\phn &23.59~--~11.11 &147.4~--~69.4\phn &&$4.4\times10^{5}$ &$0.072$\phn &$9.8\times10^{48}$ &&\phn66.2 (32.0)    &\phn5.90 &\phn7.6 \\
AM02 &1731~--~815\phn &21.92~--~10.32 &137.0~--~64.5\phn &&$3.0\times10^{5}$ &$0.064$\phn &$5.9\times10^{48}$ &&\phn69.5 (37.0)    &\phn7.01 &\phn8.7 \\
AM03 &608~--~286      &6.33~--~2.98   &39.5~--~18.6      &&\ldots            &\ldots      &\ldots             &&\phn75.7 (22.2)    &\phn4.21 &15.5 \\
AM04 &388~--~183      &5.42~--~2.55   &33.9~--~15.9      &&$9.4\times10^{4}$ &$0.018$\phn &$5.2\times10^{47}$ &&\phn29.6 \phn(7.3) &\phn1.48 &\phn8.8 \\
AM05 &406~--~191      &5.87~--~2.77   &36.7~--~17.3      &&\ldots            &\ldots      &\ldots             &&\phn75.7 (38.9)    &\phn2.60 &14.6 \\
AM06 &318~--~150      &4.60~--~2.17   &28.8~--~13.5      &&\ldots            &\ldots      &\ldots             &&\phn40.5 \phn(9.1) &\phn2.28 &16.0 \\
AM07 &268~--~126      &4.92~--~2.32   &30.8~--~14.5      &&$5.2\times10^{4}$ &$0.0065$    &$1.1\times10^{47}$ &&\phn44.1 \phn(8.4) &\phn1.41 &12.1 \\
AM08 &99~--~46        &4.15~--~1.96   &26.0~--~12.2      &&$3.0\times10^{5}$ &$0.011$\phn &$1.0\times10^{48}$ &&\phn20.7 (10.0)    &\phn0.47 &\phn8.5 \\
AM09 &232~--~109      &3.35~--~1.58   &20.9~--~9.9\phn   &&\ldots            &\ldots      &\ldots             &&\phn40.4 (10.3)    &\phn1.29 &13.1 \\
AM10 &340~--~160      &1.46~--~0.69   &9.2~--~4.3        &&\ldots            &\ldots      &\ldots             &&\phn35.3 \phn(8.5) &\phn1.32 &\phn9.5 \\
AM11 &178~--~84\phn   &2.58~--~1.21   &16.1~--~7.6\phn   &&\ldots            &\ldots      &\ldots             &&\phnn5.6 \phn(5.3) &\phn0.05 &\phn0.7 \\
AM12 &185~--~87\phn   &1.44~--~0.68   &9.0~--~4.2        &&\ldots            &\ldots      &\ldots             &&\phn15.6 \phn(8.8) &\phn0.62 &\phn8.5 \\
AM13 &186~--~88\phn   &1.12~--~0.53   &7.0~--~3.3        &&\ldots            &\ldots      &\ldots             &&\phn26.2 (10.5)    &\phn0.55 &\phn7.5 \\
AM14 &101~--~48\phn   &1.30~--~0.61   &8.1~--~3.8        &&\ldots            &\ldots      &\ldots             &&\phn19.4 \phn(7.9) &\phn0.73 &16.1 \\
AM15 &11~--~5\phn     &0.18~--~0.08   &\phn1.1~--~0.52   &&$1.6\times10^{5}$ &$0.023$\phn &$1.1\times10^{48}$ &&\phn10.5 \phn(6.9) &\phn0.18 &\phn8.9 \\
AM16 &83~--~39        &0.28~--~0.13   &\phn1.7~--~0.82   &&$6.6\times10^{4}$ &$0.10$\phnn &$2.1\times10^{48}$ &&\phnn2.9 \phn(2.9) &\phn0.06 &\phn1.1 \\
AM17 &95~--~45        &0.31~--~0.15   &\phn1.9~--~0.91   &&$4.0\times10^{4}$ &$0.067$\phn &$8.2\times10^{47}$ &&\phnn5.4 \phn(4.6) &\phn0.04 &\phn1.0 \\
AM18 &46~--~21        &0.45~--~0.21   &2.8~--~1.3        &&\ldots            &\ldots      &\ldots             &&\phnn3.3 \phn(3.4) &\phn0.01 &\phn0.9 \\
AM19 &62~--~29        &0.39~--~0.19   &2.5~--~1.2        &&\ldots            &\ldots      &\ldots             &&\phnn3.0 \phn(4.0) &\phn0.02 &\phn0.8 \\
AM20 &54~--~26        &0.28~--~0.13   &\phn1.7~--~0.81   &&\ldots            &\ldots      &\ldots             &&\phnn3.5 \phn(2.9) &\phn0.02 &\phn1.1 \\
AM21 &22~--~10        &0.32~--~0.15   &\phn2.0~--~0.93   &&\ldots            &\ldots      &\ldots             &&\phnn1.9 \phn(2.0) &\phn0.01 &\phn0.2 \\
AM22 &25~--~12        &0.30~--~0.14   &\phn1.9~--~0.88   &&\ldots            &\ldots      &\ldots             &&\phnn3.7 \phn(3.1) &\phn0.01 &\phn0.9 \\
AM23 &48~--~22        &0.23~--~0.11   &\phn1.5~--~0.69   &&\ldots            &\ldots      &\ldots             &&\phnn3.4 \phn(4.1) &\phn0.02 &\phn1.3 \\
AM24 &\ldots          &\ldots         &\ldots            &&$1.9\times10^{5}$ &$0.0083$    &$4.8\times10^{47}$ &&\phnn6.1 \phn(3.6) &\phn0.03 &\phn6.2 \\
AM25 &18~--~9\phn     &0.25~--~0.12   &\phn1.6~--~0.75   &&\ldots            &\ldots      &\ldots             &&\phnn1.4 \phn(1.6) &\phn0.01 &\phn0.8 \\
AM26 &39~--~18        &0.22~--~0.10   &\phn1.4~--~0.65   &&\ldots            &\ldots      &\ldots             &&\phnn6.5 \phn(4.0) &\phn0.08 &\phn6.4 \\
AM27 &16~--~8\phn     &0.22~--~0.10   &\phn1.3~--~0.63   &&\ldots            &\ldots      &\ldots             &&\phnn2.7 \phn(2.5) &\phn0.01 &\phn0.9 \\
\hline
\end{tabular}
\tablefoot{The symbol `\ldots' denotes if a source is not associated with VLA 40~GHz emission and therefore the physical parameters of an \hii\ region cannot be derived. For source AM24 in \SgrB(M), all the ALMA continuum flux at 242~GHz results seems to be originated by ionized gas, the physical parameters of the dust content cannot be derived.}
\end{table*}

\begin{figure*}[t!]
\begin{center}
\begin{tabular}[b]{c}
        \includegraphics[width=0.9\textwidth]{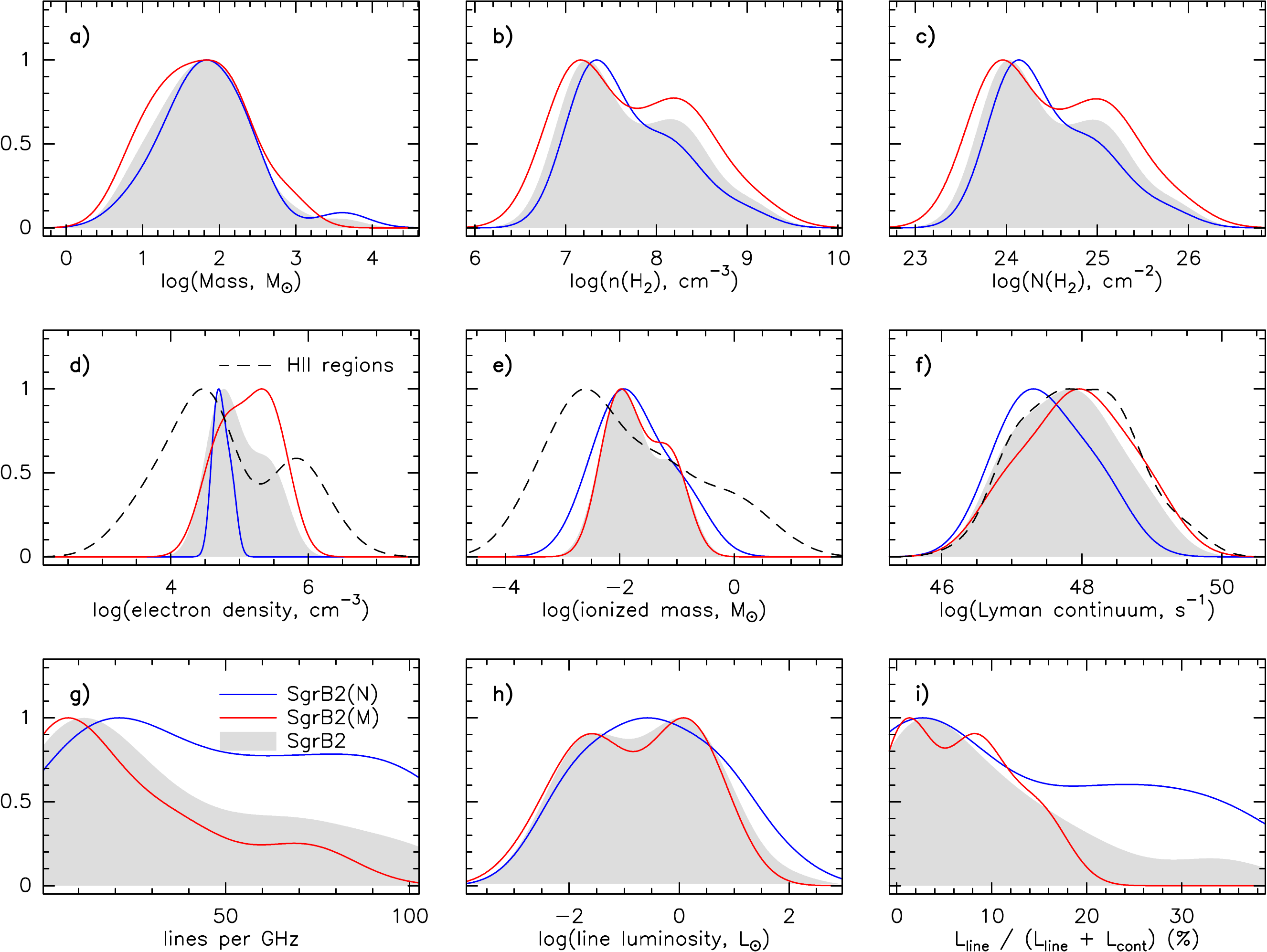} \\
\end{tabular}
\caption{Distributions of \textit{a)} dust and gas mass; \textit{b)} H$_2$ volume density; \textit{c)} H$_2$ column density; \textit{d)} electron density; \textit{e)} ionized gas mass; \textit{f)} number of ionizing photons or Lyman contiuum; \textit{g)} number of lines per GHz; \textit{h)} line luminosity; and \textit{i)} percentage of line luminosity with respect to the total luminosity (line plus continuum), for the ALMA continuum sources detected towards \SgrB\ (see Table~\ref{t:properties}). The gray filled area shows the distribution of all the sources, while the blue solid line corresponds to the sources in \SgrB(N) and the red solid line to the sources in \SgrB(M). The black dashed line in the middle panels is derived from the physical parameters of all the \hii\ regions detected in the \SgrB\ star-forming complex \citep[see Appendix~B of][]{Schmiedeke2016}. All the KDE distributions have been normalized to their maximum value to better show differences between the different regions.}
\label{f:physical_histograms}
\end{center}
\end{figure*}

\subsection{Physical properties of the continuum sources}\label{s:properties}

In Sect.~\ref{s:sources} we have presented the detection of a number of continuum sources in \SgrB(N) (20 sources) and \SgrB(M) (27 sources). The spectral index analysis shows that the emission at 1.3~mm is mainly dominated by dust, with some contribution from ionized gas. In this section we derive the physical properties of the dust and ionized gas content of all the sources. The contamination of ionized gas emission at 1.3~mm is taken into account by using the fluxes at 40~GHz \citep[VLA observations,][]{Rolffs2011} listed in Tables~\ref{t:SgrB2Nsources} and \ref{t:SgrB2Msources}. We assume a spectral index of $-0.1$ for the ionized gas, which corresponds to optically thin emission, and extrapolate the flux measured at 40~GHz to the frequency of 242~GHz of the ALMA continuum image. This procedure may result in a lower limit to the contribution of the ionized gas for optically thick \hii\ regions (with positive spectral indices ranging from $+0.6$ to $+2$; e.g.\ \citealt{Kurtz2005, SanchezMonge2013b}).

In Table~\ref{t:properties} we list the (dust and gas) masses for each source determined from the expression \citep{Hildebrand1983}
\begin{equation}
M_\mathrm{d+g}=\frac{S_\nu D^2}{B_\nu(T_\mathrm{d}) \kappa_\nu},
\end{equation}
where $S_\nu$ is the flux density at 242~GHz (listed in Tables~\ref{t:SgrB2Nsources} and \ref{t:SgrB2Msources}), $D$ is 8.34~kpc for \SgrB, $B_\nu(T_\mathrm{d})$ is the Planck function at a dust temperature $T_\mathrm{d}$, and $\kappa_\nu$ is the absorption coefficient per unit of total mass (gas and dust) density. We assume a dust mass opacity coefficient at 230~GHz of 1.11~g~cm$^{-2}$ (agglomerated grains with thin ice mantles in cores of densities $10^8$~cm$^{-3}$; \citealt{OssenkopfHenning1994}), optically thin emission, and a gas to dust conversion factor of 100. In Table~\ref{t:properties} we list the mass derived for two different dust temperatures 50~K and 100~K (see e.g.\ \citealt{Qin2008, Rolffs2011, Belloche2016, Schmiedeke2016}). The volume density, $n_\mathrm{H_2}$, is determined, assuming spherical cores, as
\begin{equation}
n_\mathrm{H_2} = \frac{1}{\mu~m_\mathrm{H}}\frac{M_\mathrm{d+g}}{(4/3)~\pi~R^3},
\end{equation}
where $\mu$ is the mean molecular mass per Hydrogen atom (equal to 2.3), $m_\mathrm{H}$ is the Hydrogen mass, and $R$ is the radius of the core (see sizes in Tables~\ref{t:SgrB2Nsources} and \ref{t:SgrB2Msources}). Finally, the column density, $N_\mathrm{H_2}$, is determined from
\begin{equation}
N_\mathrm{H_2} = \int_\mathrm{line~of~sight}{n_\mathrm{H_2}}~dl ,
\end{equation}
with $l$ corresponding to the size of the cloud.

The masses of the cores in both \SgrB(N) and \SgrB(M) range from a few \mo\ up to a few thousand \mo. We note that our sensitivity limit ($\sim8$~mJy~beam$^{-1}$) corresponds to a mass limit of about 1--3~\mo. The distribution of the masses is shown in panel (a) of Fig.~\ref{f:physical_histograms}. The mean and median masses (for a temperature of 50~K) are 495~\mo\ and 103~\mo\ for the whole sample of cores (including both regions). While the median mass remains around 100 for both regions separately, the mean mass is 750~\mo\ for the cores in \SgrB(N) and 315~\mo\ for the cores in \SgrB(M). These observations confirm a previous finding obtained with SMA observations \citep{Qin2011}: most of the mass in \SgrB(N) is concentrated in one single core (source AN01), that accounts for about 73\% (about 9000~\mo) of the total mass in the region. Differently, \SgrB(M) appears more fragmented with the two most massive cores (sources AM01 and AM02) accounting only for 50\% (about 3000~\mo) of the total mass. The brightest and more massive ALMA continuum sources are likely optically thick towards their centers (cf.\ spectral index maps in Fig.~\ref{f:SgrB2spindex}). This results in lower limits for the mass estimated listed in Table~\ref{t:properties}. At the same time, the assumed temperatures of 50~K and 100~K may be underestimated, since many hot molecular cores may have temperatures of about 200--300~K, and result in an upper limit for the determined dust (and gas) masses. In particular, sources AM01 and AM02 in \SgrB(M) and source AN01 in \SgrB(N) have brightness temperatures above 150~K (cf.\ Figs.~\ref{f:SgrB2Nsources} and \ref{f:SgrB2Msources}). This suggests that the physical temperature is at least of 150~K, which would result in a 30\% lower mass than the one derived with 100~K. A detailed analysis of the chemical content, including the determination of more accurate temperatures (and therefore masses) will be presented in a forthcoming paper.

The monolithic structure and large mass of \SgrB(N)-AN01 is remarkable, but it is not completely structureless. A number of filamentary structures are detected in our maps, which may suggest that it is possible that a very dense cluster of cores overlapping along the line-of-sight can produce the monolithic appearance of AN01. We calculate the Jeans mass of such an object to determine if it should fragment. For a density of $10^9$~cm$^{-3}$, a temperature of 150~K (corresponding to a thermal sound speed of 0.73~km~s$^{-1}$), and a linewidth of 10~km~s$^{-1}$ (from the observed spectra), we derive a thermal Jeans mass of about 1~\mo, and a non-thermal Jeans mass of about 2000~\mo, and Jeans lengths of 300~au and 4000~au, respectively. Therefore, considering non-thermal support, it is not inconceivable that a very dense fragment cluster is formed that might not be resolved. The non-thermal support may arise from the feedback of the O7.5 star that has been found inside the core, from the emission of the \hii\ region (see \citealt{Schmiedeke2016}).

The H$_2$ volume densities of the cores are usually in the range $10^7$--$10^8$~cm$^{-3}$, with the three most massive cores with densities above $10^9$~cm$^{-3}$. These high densities correspond to $10^7$~\mo~pc$^{-3}$, up to two orders of magnitude larger than the typical stellar densities found in super star clusters like Arches or Quintuplet ($\sim10^5$~\mo~pc$^{-3}$; \citealt{PortegiesZwart2010}), suggesting that \SgrB\ may evolve into a super star cluster. The H$_2$ column densities are above $10^{25}$~cm$^{-2}$ at the scales of $\sim5000$~au, for the brightest sources. In panels (b) and (c) of Figure~\ref{f:physical_histograms} we show the distribution of the H$_2$ volume densities and H$_2$ column densities for all the cores detected in \SgrB\ in the ALMA observations.

\begin{figure}[t!]
\begin{center}
\begin{tabular}[b]{c}
        \includegraphics[width=0.9\columnwidth]{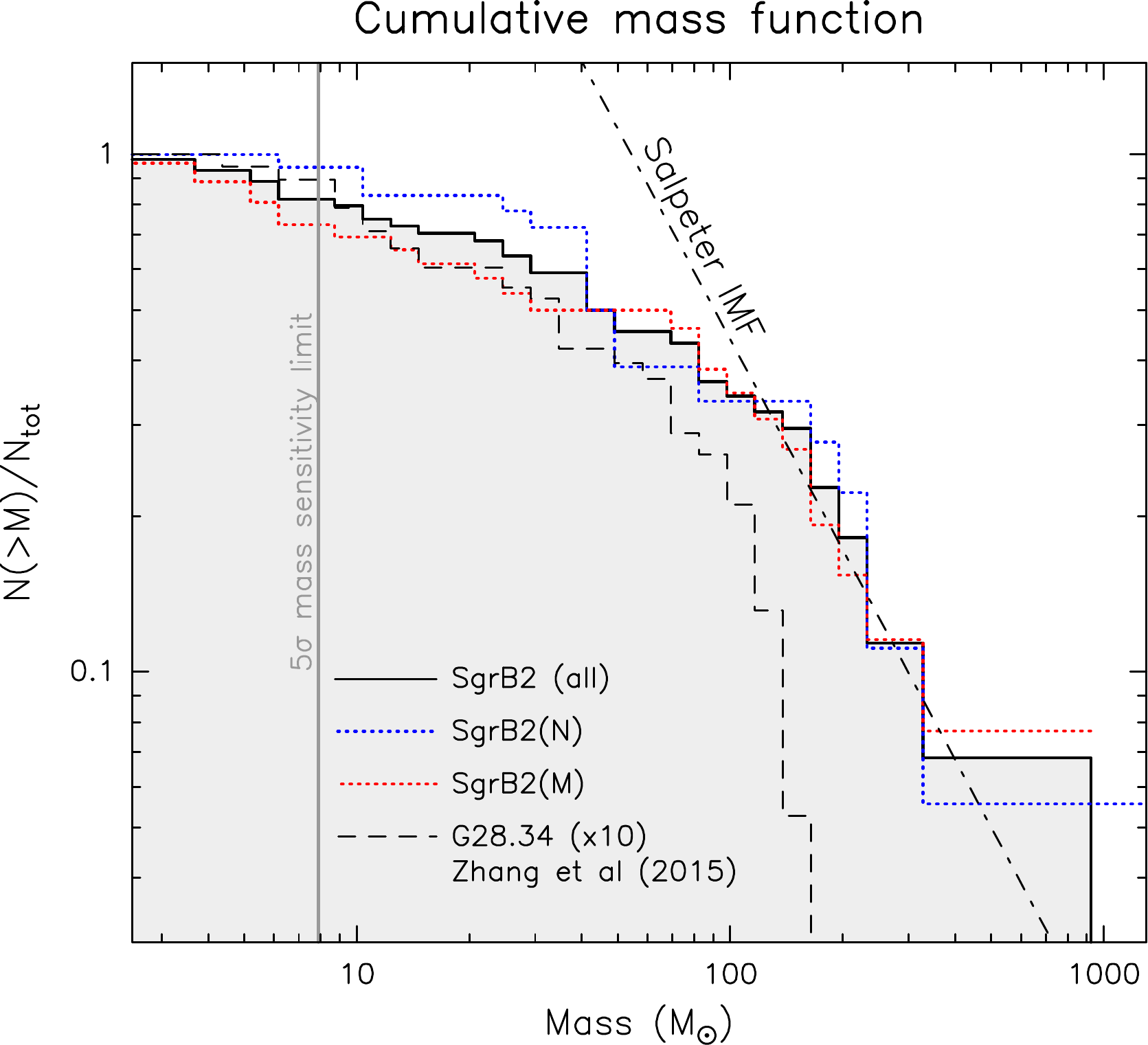} \\
\end{tabular}
\caption{Cumulative mass function of the sources identified towards \SgrB. The blue and red dotted lines correspond to the sources identified in \SgrB(N) and \SgrB(M), respectively. The dashed line shows the cumulative mass function of the fragments identified in the massive infrared dark cloud G28.34 by \citet{Zhang2015}, scaled by a factor of 10 in mass. The dot-dashed line marks the slope of the Salpeter initial mass funciton. The vertical gray line indicates the 5$\sigma$ mass sensitivity limit of 7.9~\mo\ for a temperature of 100~K.}
\label{f:PDF}
\end{center}
\end{figure}

In Figure~\ref{f:PDF} we present the normalized cumulative mass function $N(>M)/N_\mathrm{tot}$, where $N$ is the number of sources in the mass range $>M$ and $N_\mathrm{tot}$ is the total number of sources (see Tables~\ref{t:SgrB2Nsources} and \ref{t:SgrB2Msources}). The black solid line histogram corresponds to the whole sample containing \SgrB(N) and \SgrB(M), the blue and red dotted line histograms correspond to the individual populations of cores in \SgrB(N) and \SgrB(M), respectively. The dashed line denotes the slope of the Salpeter IMF. As seen in Fig.~\ref{f:PDF}, if the cumulative mass functions follows the shape of the IMF, there seems to be a deficit of lower-mass cores. It is worth noting that dynamic range effects in the vicinity of the brightest sources may hinder the detection of low-mass, faint cores. A similar result is found in the high-mass star-forming region G28.34 \citep{Zhang2015}. The core mass function of this region is shown in Fig.~\ref{f:PDF} with a dashed line. The \SgrB\ star-forming complex contains even more high-mass dense cores compared to the G28.34 star-forming region. As argued by \citet{Zhang2015}, the lack of a population of low-mass cores can be understood if (i) low-mass stars form close to the high-mass stars and remain unresolved with the currently achieved angular resolutions, (ii) the population of low-mass stars form far away from the central, most massive stars and move inwards during the collapse and transport of material from the outskirts to the center, or (iii) low-mass stars form later compared to high-mass stars. The different scenarios can be tested with new observations. Higher angular resolution observations with ALMA can probe the surroundings of the most massive cores and allow us to search for fainter fragments associated with low-mass dense cores. A larger map covering the whole \SgrB\ complex, with high enough angular resolution (Ginsburg et al.\ in prep), will permit us to search for low-mass cores distributed in the envelope of \SgrB\ originally formed far from the central high-mass star forming sites \SgrB(N) and \SgrB(M).

Finally, we have determined the physical parameters of the ionized gas associated with the ALMA continuum sources, assuming they are optically thin \hii\ regions. We have used the expressions given in the Appendix~C of \citet{Schmiedeke2016} to derive the electron density ($n_\mathrm{e}$), the ionized gas mass ($M_\mathrm{i}$) and the number of ionizing photons per second ($\dot{N}_\mathrm{i}$). In panels (d), (e) and (f) of Fig.~\ref{f:physical_histograms} we show the distribution of these three parameters for all the sources in both \SgrB(N) and \SgrB(M). We compare them with the distribution of the same parameters derived for all the \hii\ regions detected in both regions \citep[see Appendix~B in][and references therein]{Schmiedeke2016}. No clear differences are found between both distributions, suggesting that the physical and chemical properties (see below) derived for our sub-sample of \hii\ regions may be extrapolated to the whole sample of \hii\ regions in the \SgrB\ complex. However, this should be confirmed by conducting a detailed study of all the objects and not only those located in the vicinities of the densest parts of \SgrB(N) and \SgrB(M). 

\begin{figure*}[t!]
\begin{center}
\begin{tabular}[b]{c c}
        \includegraphics[width=0.82\columnwidth]{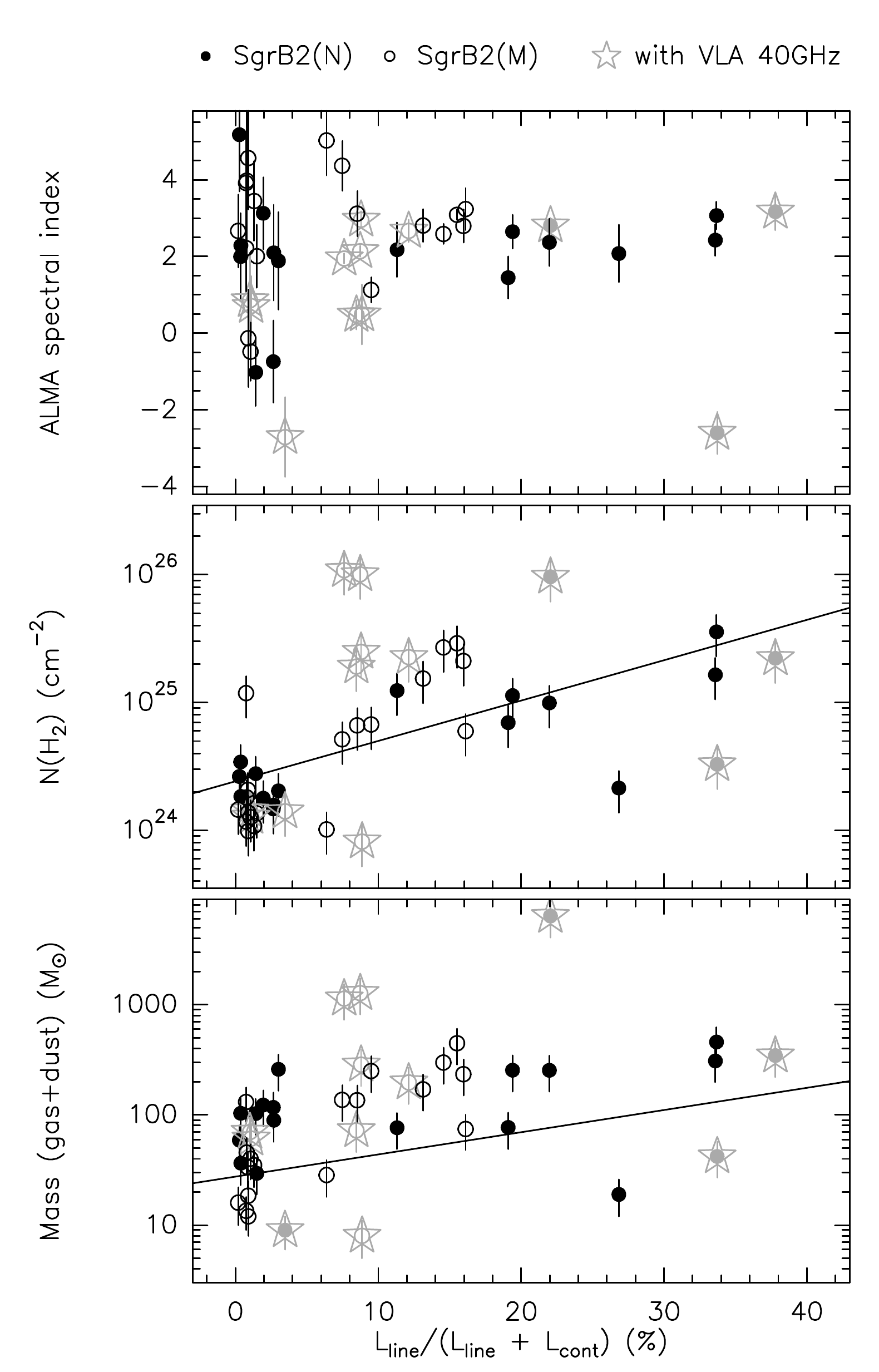} &
        \includegraphics[width=0.82\columnwidth]{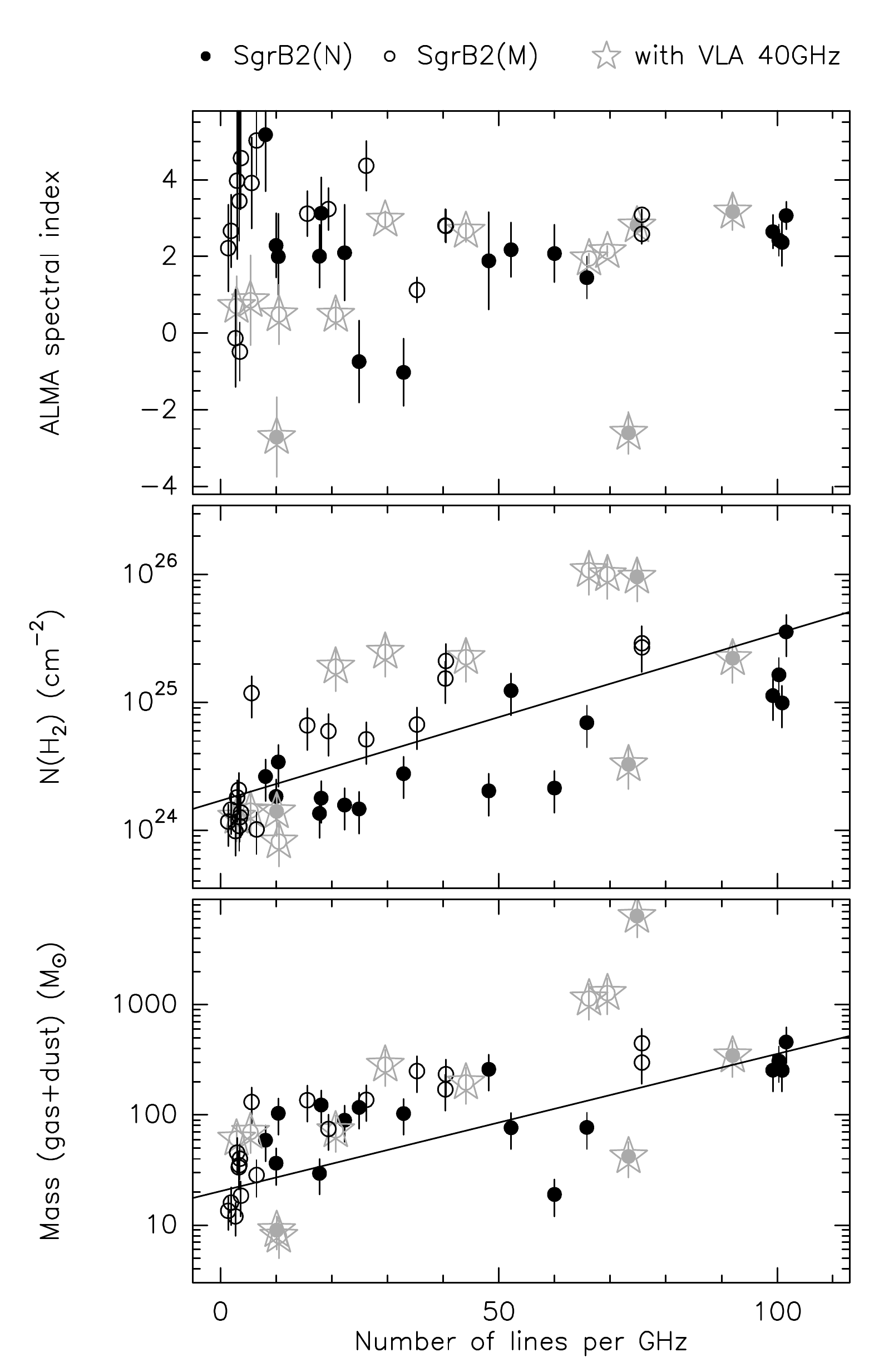} \\
\end{tabular}
\caption{Scatter plots showing the relation between the properties derived from the dust emission: spectral index determine in the range 211--275~GHz (top panels), the H$_2$ column density (middle panels) and the dust and gas mass (bottom panels); with respect to the percentage of line luminosity over the total luminosity (left column) and the number of lines per GHz (right column). Filled circles correspond to the sources identified in \SgrB(N), while open circles show the sources found in \SgrB(M). Gray symbols and stars mark those ALMA continuum sources associated with VLA 40~GHz continuum emission. The solid lines in four panels correspond to linear fits to the black symbols in each panel (see more details in Sect.~\ref{s:chemistry}).}
\label{f:Llines}
\end{center}
\end{figure*}

\begin{table}[h!]
\caption{\label{t:summary} Summary of the nature and chemical content}
\centering
\begin{tabular}{l c c c}
\hline\hline

&1.3~mm cont.
&chemical
&associated
\\
\multicolumn{1}{c}{ID}
&nature
&content
&with...
\\
\hline
\multicolumn{4}{l}{\phn \SgrB(N)} \\
\hline
AN01 &mixed   &rich   &SMA1; K2 \\
AN02 &dust    &rich   &SMA2 \\
AN03 &mixed   &rich   &SMA2 \\
AN04 &dust    &rich   &\\
AN05 &dust    &rich   &\\
AN06 &dust    &rich   &\\
AN07 &dust    &rich ? &\\
AN08 &dust    &rich   &\\
AN09 &dust    &\ldots &\\
AN10 &ionized &rich   &K1\\
AN11 &dust    &\ldots &\\
AN12 &dust    &rich ? &\\
AN13 &dust    &rich ? &\\
AN14 &dust    &\ldots &\\
AN15 &dust    &\ldots &\\
AN16 &mixed   &rich ? &K4 \\
AN17 &dust    &rich   &\\
AN18 &dust    &rich ? &\\
AN19 &dust    &\ldots &\\
AN20 &mixed   &\ldots &\\
\hline
\multicolumn{4}{l}{\phn \SgrB(M)} \\
\hline
AM01 &mixed   &rich ? &SMA1; F3 \\
AM02 &mixed   &rich ? &SMA2; F1 \\
AM03 &dust    &rich   &\\
AM04 &mixed   &rich ? &SMA6 \\
AM05 &dust    &rich ? &SMA7 \\
AM06 &mixed   &rich   &SMA11; F10.30 \\
AM07 &dust    &rich ? &\\
AM08 &mixed   &rich ? &SMA8 \\
AM09 &dust    &rich ? &SMA9 \\
AM10 &dust    &rich ? &SMA10 \\
AM11 &dust    &\ldots &\\
AM12 &mixed   &\ldots &SMA12; F10.318 \\
AM13 &dust    &rich ? &SMA10 \\
AM14 &dust    &rich ? &\\
AM15 &ionized &\ldots &B \\
AM16 &ionized &\ldots &I \\
AM17 &mixed   &\ldots &A1 \\
AM18 &dust    &\ldots &\\
AM19 &dust    &\ldots &\\
AM20 &dust    &\ldots &\\
AM21 &mixed   &\ldots &A1 \\
AM22 &dust    &\ldots &\\
AM23 &mixed   &\ldots &A1 \\
AM24 &ionized &\ldots &\\
AM25 &dust    &\ldots &\\
AM26 &dust    &\ldots &\\
AM27 &dust    &\ldots &\\
\hline
\end{tabular}
\tablefoot{A source is `ionized' if the ALMA continuum emission is consistent within a factor of 3 with ionized gas emission. A smaller contribution is listed as `mixed'. A source is chemically `rich' if lines contribute with more than 15\% to the total luminosity, and the source has more than 20 lines per GHz. If only one of the two criteria is fulfilled it is listed as `rich ?'. See Sect.~\ref{s:chemistry} for details. References for the association with other sources: \citet{Qin2011, Schmiedeke2016}}
\end{table}

\begin{figure*}[h!]
\begin{center}
\begin{tabular}[b]{c}
        \includegraphics[width=0.82\textwidth]{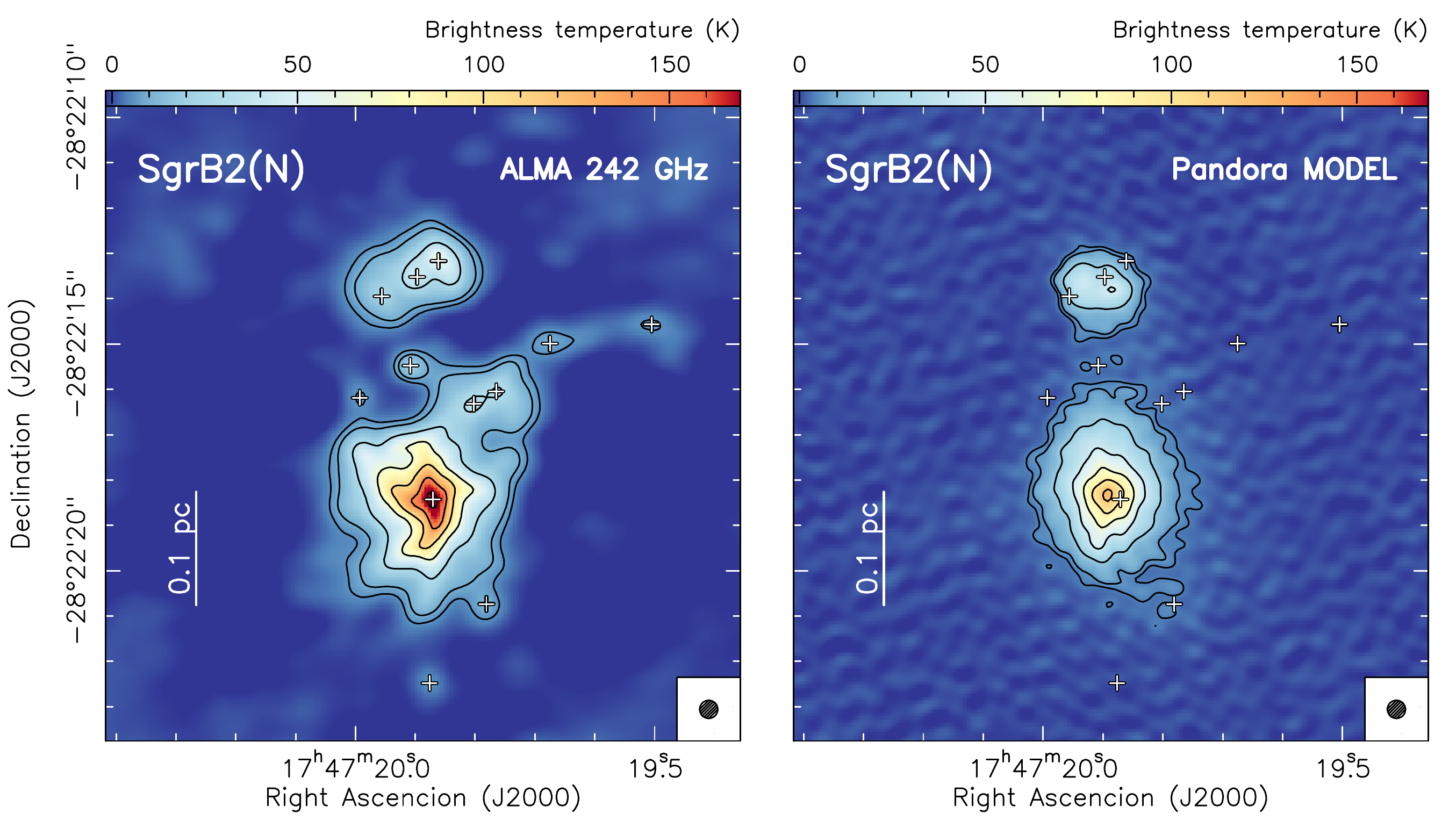} \\
        \includegraphics[width=0.82\textwidth]{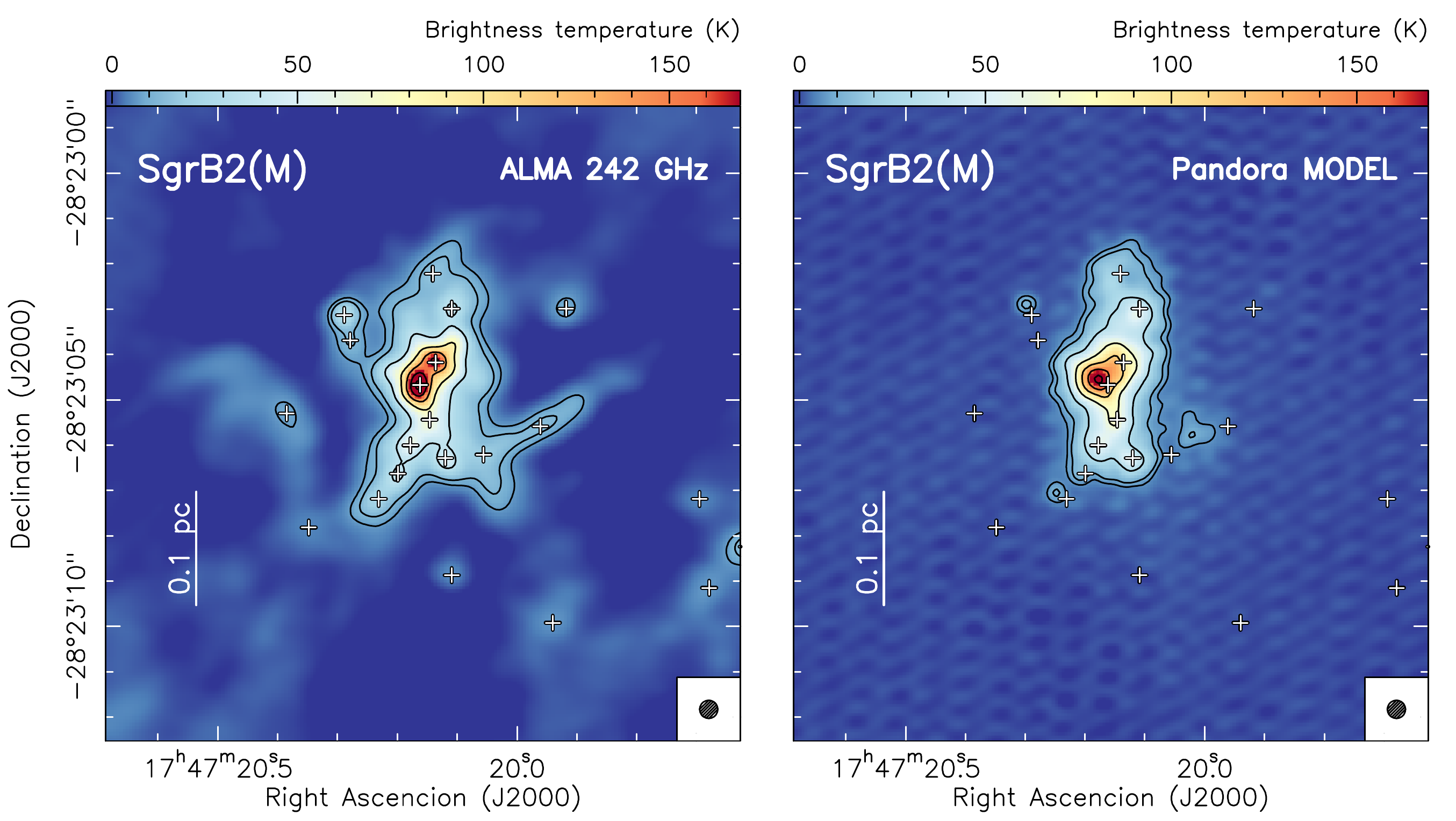} \\
\end{tabular}
\caption{Comparison of the ALMA continuum emission maps (left panels, see also Figs.~\ref{f:SgrB2Nsources} and \ref{f:SgrB2Msources}) to synthetic images at 242~GHz (right panels) of the 3D radiative transfer model presented in \citet{Schmiedeke2016}. The top panels show the emission towards \SgrB(N), while the bottom panels show the emission towards \SgrB(M). The synthetic images have been post-processed to the same uv-sampling of the actual ALMA observations, adding thermal noise corresponding to a precipitable water vapor of 0.7~mm.}
\label{f:SgrB2model}
\end{center}
\end{figure*}

\begin{figure*}[h!]
\begin{center}
\begin{tabular}[b]{c}
        \includegraphics[width=0.82\textwidth]{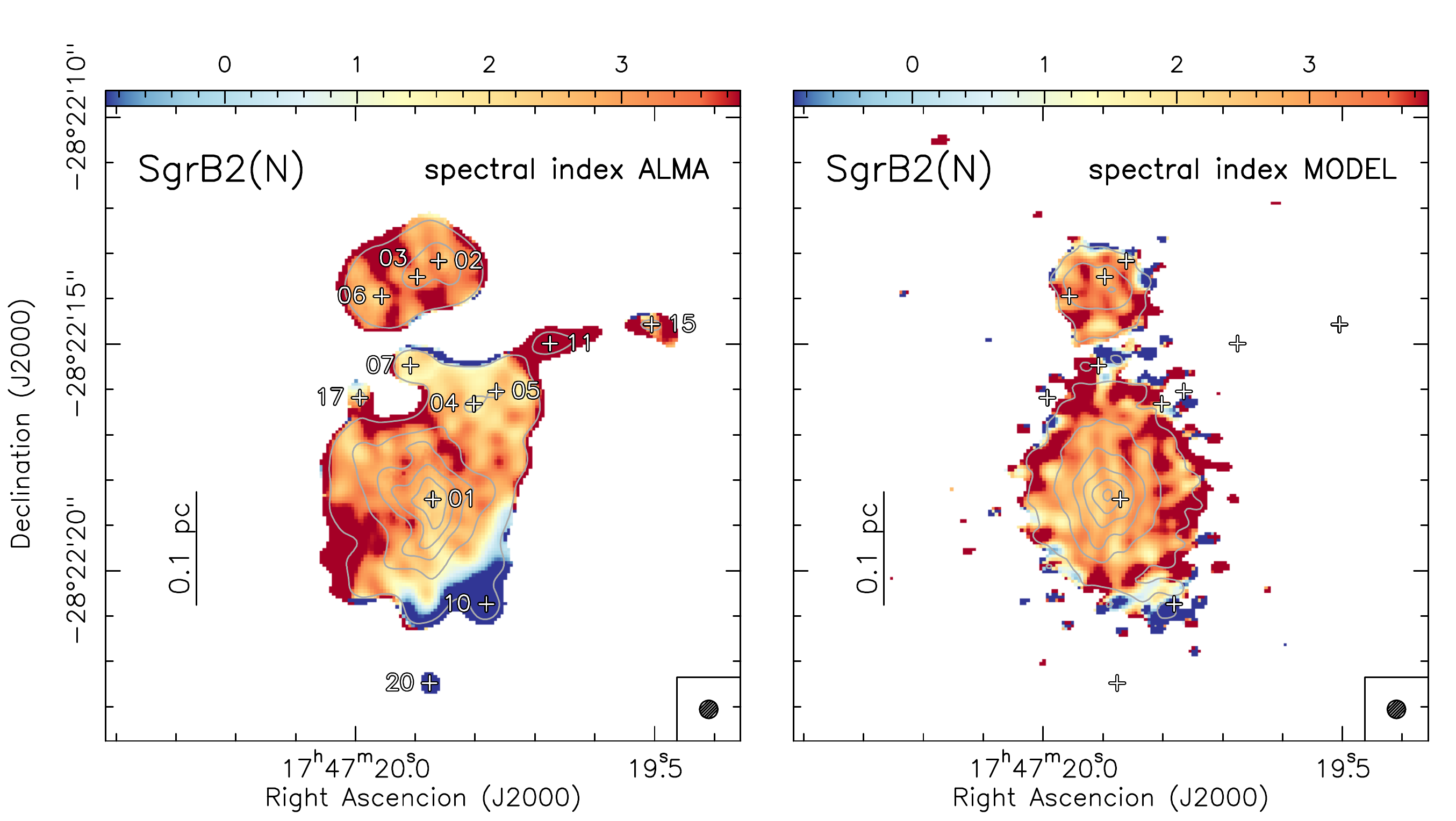} \\
        \includegraphics[width=0.82\textwidth]{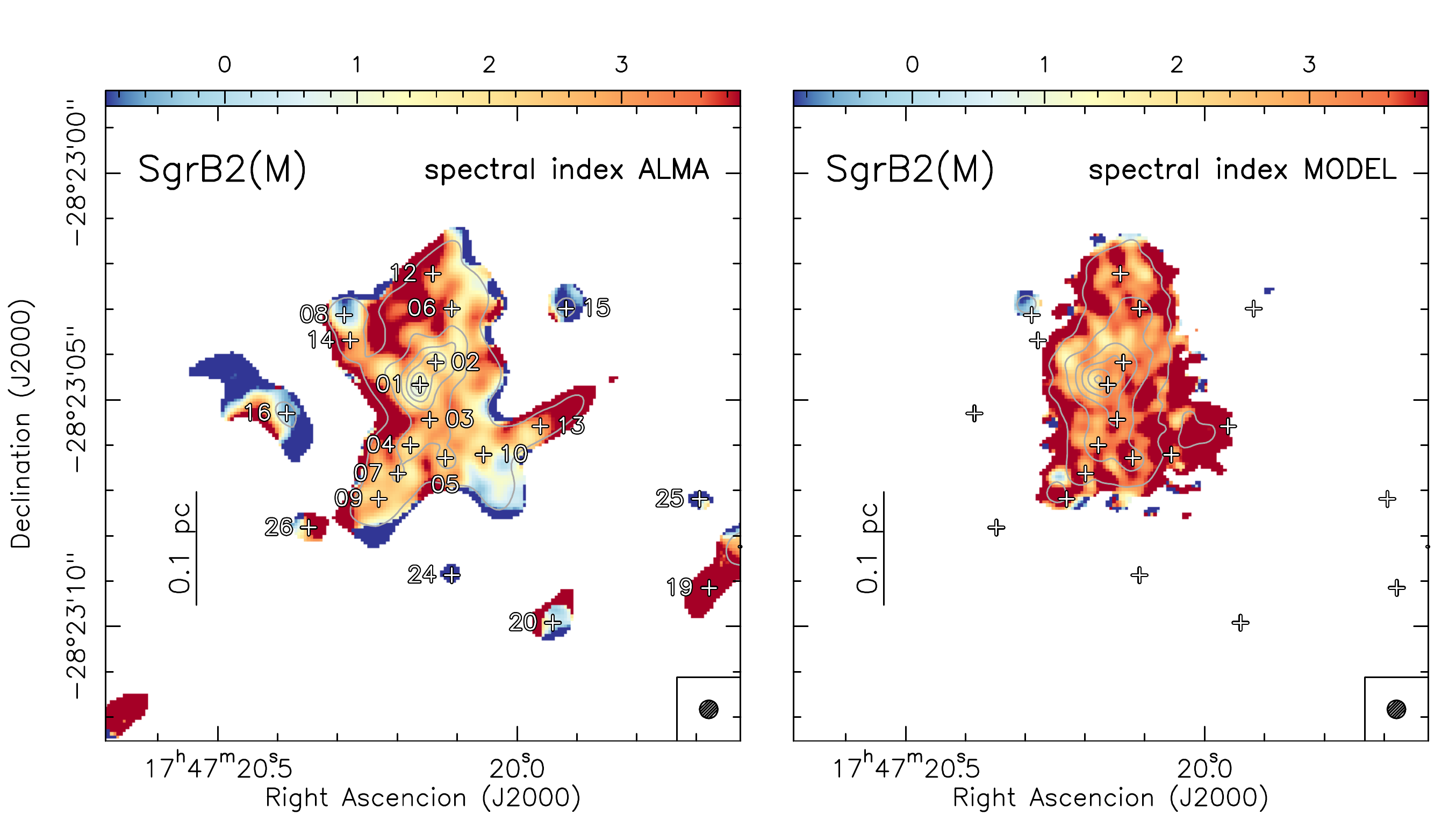} \\
\end{tabular}
\caption{Comparison of the spectral index map derived from the ALMA continuum measurements in the range 211-275~GHz (left panels, see also Fig.~\ref{f:SgrB2spindex}) to the spectral index map obtained from the 3D radiative transfer model (right panels). The top panels show the emission towards \SgrB(N), while the bottom panels show the emission towards \SgrB(M).}
\label{f:SgrB2spindexmodel}
\end{center}
\end{figure*}

\subsection{Chemical content in the continuum sources}\label{s:chemistry}

In the previous section we have studied the physical properties of the dense cores, considering both the dust and ionized gas content associated with each core. In this section we study the chemical content of each source. A detailed analysis of the chemical properties (e.g.\ molecular abundances, temperatures, kinematics) is beyond the scope of the present work and will be presented in a series of forthcoming papers (see also \citealt{Bonfand2017}). Here we present a statistical study of the chemical richness of each source detected in the ALMA observations. We follow two approaches: (1) number of emission line peaks, and (2) luminosity of the source contained in spectral lines compared to the luminosity of the continuum emission.

The last three columns of Table~\ref{t:properties} list the number of spectral line features (in emission) per GHz (or line density), the luminosity contained in the line spectral features, and the percentage of luminosity contained in the spectral lines with respect to the total luminosity (line plus continuum) of the core. The number of lines is automatically determined by searching for emission peaks above the 5$\sigma$ level across the whole spectrum, similar to the approach followed in ADMIT (ALMA Data Mining Toolkit; \citealt{Friedel2015}). In order to avoid fake peaks produced by noise fluctuations, we discard peaks if they are found to be closer than 5 channels (corresponding to 3~km~s$^{-1}$) with respect to another peak. The number in parenthesis listed in Table~\ref{t:properties} is a measure of the uniformity of the line density through the whole observed frequency range (i.e.\ from 211 to 275~GHz). This parameter is defined as the standard deviation of the line densities measured in the different spectral windows (see Fig.~\ref{f:freqcover}). A low value indicates that all spectral windows have a similar line density, while larger values reveal a significant variation of the line density across the spectral windows, likely pinpointing the presence of complex molecules like CH$_3$OH or C$_2$H$_5$OH dominating the spectrum, i.e.\ with a large number of spectral lines in some specific frequency ranges. The line luminosity is computed with the expression
\begin{equation}
L_\mathrm{line} = 4\,\pi\,D^2\,\sum_\mathrm{S_\nu>5\sigma}S_\nu\,\Delta\nu
\end{equation}
where $D$ is 8.34~kpc for \SgrB\ and the last term is the sum of the product of the intensity times the channel width for all those channels with an intensity above 5$\sigma$ above the continuum level, with $\sigma$ being 8~mJy~beam$^{-1}$.

The number of lines per GHz varies from a few to about 100. The sources with more line features are found in the clump located to the north of \SgrB(N) which contains sources AN02 and AN03 (cf.\ Fig.~\ref{f:SgrB2Nsources}). The main source AN01 in \SgrB(N), which corresponds to the well-known, chemically-rich LMH \citep[Large Molecule Heimat,][]{Snyder1994, Hollis2003, McGuire2013}, is slightly less chemically rich. The measured line luminosities span about four orders of magnitude ranging from $10^{-3}$~\lo\ to 10~\lo. This corresponds to a contribution of 2--50\% of the total luminosity in the 211--275~GHz frequency range. In panels (g), (h) and (i) of Fig.~\ref{f:physical_histograms} we show the distribution of these quantities for both \SgrB(N) and \SgrB(M). In general, the sources in \SgrB(N) seems to be more chemically rich than the dense cores in \SgrB(M). The mean (median) number of lines per GHz is 51 (50) for \SgrB(N), compared to 23 (13) for \SgrB(M). While the mean (median) percentage of line luminosity with respect to the total luminosity is 20\% (16\%) and 8\% (9\%) for \SgrB(N) and \SgrB(M), respectively.

In Fig.~\ref{f:Llines} we compare the dust and ionized gas properties to the chemical properties of the sources identified in \SgrB(N) (filled circles) and \SgrB(M) (open circles). Those sources associated with VLA 40~GHz emission, likely tracing ionized gas, are shown in gray color with a star symbol. In the top panels we compare the spectral index determined from the ALMA data (range 211 to 275~GHz) to the percentage of line luminosity and the number of lines. No clear correlation is found, although in general chemically-rich sources are found to have spectral indices in the range 2--4, suggesting they are preferentially associated with dust cores. There is a remarkable exception in the case of source AN10 in \SgrB(N). This source is one of the chemically richest sources but has a clearly negative spectral index. The source is found clearly associated with an \hii\ region (see Fig.~\ref{f:SgrB2VLASMA}; \citealt{Schmiedeke2016}, and references therein). The negative measured spectral index of $-2.6\pm0.6$, compared to the typical value of $-0.1$ for \hii\ regions, is likely due to the location of source AN10 within the extended envelope of the main source of \SgrB(N). As stated in Sect.~\ref{s:obs}, the sensitivity to angular scales of only $<5$\arcsec\ likely results in resolving out part of the emission of the envelope and, therefore, in negative lobes in the surroundings, affecting the fluxes measured towards some weak sources, as may happen in the case of source AN10. In summary, the chemical richness of an \hii\ region like AN10 suggests this source is still embedded in dust and gas that is heated by the UV radiation of the forming high-mass star. This is very similar to the situation found in W51~e2, where the source e2w is a chemically-rich source with the continuum emission likely dominated by free-free (i.e.\ with a flat spectral index), and is located within the dust envelope of source e2e (\citealt{Ginsburg2016}; Ginsburg et al.\ in prep.).

The middle and bottom panels of Fig.~\ref{f:Llines} compare the H$_2$ column density and dust (and gas) mass to the chemical richness. There are hints of a correlation where the most massive and dense cores are associated with the most chemically-rich sources. This trend is better seen if those cores associated with ionized gas, i.e.\ with the UV radiation altering the chemical properties, are excluded. The solid lines shown in the four panels correspond to linear fits to the data of those cores not associated with ionized gas. We derive the following empirical relations
\begin{equation}
    \log{N(\mathrm{H}_{2})} = (24.4\pm0.1)\,+\,\left(L_\mathrm{line}/L_\mathrm{line+cont}\right)\times(0.032\pm0.007) \nonumber
\end{equation}
\begin{equation}
    \log{N(\mathrm{H}_{2})} = (24.2\pm0.1)\,+\,\left(\mathrm{lines\,per\,GHz}\right)\times(0.013\pm0.002) \nonumber
\end{equation}
\begin{equation}
    \log{M_\mathrm{d+g}} = (1.4\pm0.1)\,+\,\left(L_\mathrm{line}/L_\mathrm{line+cont}\right)\times(20.1\pm8.3)\times10^{-3} \nonumber
\end{equation}
\begin{equation}
    \log{M_\mathrm{d+g}} = (1.3\pm0.1)\,+\,\left(\mathrm{lines\,per\,GHz}\right)\times(12.5\pm2.4)\times10^{-3} \nonumber
\end{equation}

These relations suggest that the poor chemical-richness of the less massive (and less dense) cores in both \SgrB(N) and \SgrB(M) can be due to a lack of sensitivity. A detailed study of the abundances (including upper limits for the fainter sources) of different molecules in all the cores, will allow us to determine if there are chemical differences between the cores in \SgrB, or if the different detection rates of spectral line features is sensitivity limited.

Finally, in Table~\ref{t:summary} we present a summary of the properties of the continuum sources detected in the ALMA images. The second column indicates the most probable origin of the continuum emission at 1.3~mm, from the analysis presented in Sects.~\ref{s:origin} and \ref{s:properties}. Complementary, in Fig.~\ref{f:findingchart} we present a finding chart of both \SgrB(N) and \SgrB(M) where we indicate the positions of already known \hii\ regions (yellow circles) and (sub)millimeter continuum sources (red crosses). We consider the emission to be dominated by ionized gas thermal emission, if the flux at 40~GHz (from VLA) is within a factor of 3 with respect to the flux at 242~GHz (from ALMA). If the source is associated with 40~GHz continuum, but its contribution is less than a factor of 3 with respect to the flux at 1.3~mm, we catalogue the source as a mixture of dust and ionized gas. All the other sources seem to be dust dominated. The chemical richness is evaluated from the number of lines per GHz and the fraction of line luminosity (see last columns in Table~\ref{t:properties}). Those sources with a fraction of the line luminosity above 15\% and more than 20 lines per GHz are considered to be chemically rich. If only one of the two criteria is fulfilled, the source is considered as potentially chemically rich. In the last column of Table~\ref{t:summary} we list the association of the ALMA continuum sources with other sources detected at different frequencies (mainly centimeter and millimeter). It is worth noting that even the smaller sources found in \SgrB(N) and \SgrB(M) showing a rich chemistry would be considered quite respectable if they were located far from the main, central hot molecular cores. For example, sources like AM09 and AM10 in \SgrB(M) or AN05 and AN06 in \SgrB(N) have masses of a few hundred \mo, similar to the masses and chemical richness of well-known hot molecular cores like Orion~KL, G31.41$+$0.31 or G29.96$-$0.02 (e.g.\ \citealt{Wyrowski1999, Schilke2001, Cesaroni2011}). Therefore, both \SgrB(N) and \SgrB(M) are harbouring rich clusters of hot molecular cores.

\subsection{Comparison with 3D-radiative transfer models}\label{s:pandora}

In this section we compare the ALMA continuum maps with predictions from a 3D radiative transfer model. \citet{Schmiedeke2016} used VLA 40~GHz data \citep{Rolffs2011} and SMA 345~GHz data \citep{Qin2011}, together with \textit{Herschel}-HIFI \citep{Bergin2010}, \textit{Herschel}-HiGAL \citep{Molinari2010} and APEX-ATLASGAL \citep{Schuller2009} observations of \SgrB\ to generate a 3D model of the structure of the whole region, from 100~au to 45~pc scales. The model contains the dust cores previously detected in the SMA images and the \hii\ regions reported by \citet{Gaume1995}, \citet{dePree1998, dePree2014} and \citet{Rolffs2011}. The model reproduces the intensity and structure of \SgrB(N) and \SgrB(M) as observed with the SMA and the VLA, and it is also able to create synthetic maps for any frequency. In Fig.~\ref{f:SgrB2model} we compare the continuum map obtained with ALMA, to the model prediction at a frequency of 242~GHz. The synthetic images produced with the model are post-processed to the same uv-sampling of the ALMA observations, and we have added thermal noise corresponding to 0.7~mm of precipitable water vapor. The general structure of the model agrees with the structure seen in the ALMA observations. However, there are some differences that deserve to be discussed.

First, the model does not include the new sources detected in the ALMA continuum images. This results in some structures missing in the model image, like e.g.\ the filamentary extension to the west of \SgrB(N). Similarly, the northern component of \SgrB(N) was catalogued as a single source in the SMA observations, while the ALMA images indicate a possible fragmentation in three different dense cores. This results in a disagreement in the extension and orientation of the emission between the observations and the model. The structure of the model of \SgrB(M) is more similar to the actual observations, although the new sources are not included in the model, like e.g.\ the sources to the south-east, which are located at the edge of the primary beam of the SMA observations. Second, the model considers the dense cores to have a plummer-like density structure. This implies that more complex structures like the filamentary structures seen towards the center of \SgrB(N) are not reproduced in the model. Finally, the intensity scale in the model image is off with respect to the observations by a factor of 1.5, with the observations being stronger than the model prediction. This can be related to uncertainties in the flux calibration of the SMA image used to create the model, that can be of up to 20--30\%. The uncertainty in the flux calibration of the ALMA observations is also about 20\%. Moreover, the dust mass opacity coefficient and the dust emissivity index were not well constrained in the creation of the 3D model, since only one high-angular resolution image at (sub)millimeter wavelengths was available (i.e.\ the SMA sub-arcsecond observations of \citealt{Qin2011}). This can result in the flux offset seen at the frequencies observed with ALMA.

In Fig.~\ref{f:SgrB2spindexmodel} we compare the spectral index maps derived from the ALMA observations (see Sect.~\ref{s:origin}) with spectral index maps built from the output of the 3D model. The synthetic spectral index maps have been created following the same approach as for the ALMA observations. A number of continuum images were produced from the model at the central frequencies of the ALMA spectral windows, and post-processed with the uv-sampling of the ALMA observations. Afterwards, we compute linear fits to the flux on a pixel-to-pixel basis to create the spectral index map. The synthetic spectral index maps have the same general properties as the observed ones, with the brightest sources having spectral indices around $+2$ likely tracing optically thick dust emission, and the outskirts being more optically thin. The model is also able to reproduce the flat spectral indices observed towards the sources AN10 in \SgrB(N) and AM08 in \SgrB(M), likely associated with ionized gas (see Sect.~\ref{s:properties} and Table~\ref{t:summary}).

In summary, most of the discrepancies found between the model and the observations are caused by a lack of information when the 3D structure model was created, i.e.\ lack of observations at other frequencies to better constrain the structure, number of sources, and dependence of flux with frequency. These issues will be solved in the near future thanks to a number of on-going observational projects aimed at studying \SgrB\ with an angular resolution of $\sim0\farcs5$ at frequencies of 5~GHz and 10~GHz (using the VLA) and at 100~GHz, 150~GHz and 180~GHz (using ALMA). These observations together with the presented ALMA band 6 observations will permit us to better characterize the physical properties of the different sources found in \SgrB, and to improve the current 3D model of the region.

\section{Summary}\label{s:summary}

We have observed the \SgrB(M) and \SgrB(N) high-mass star-forming regions with ALMA in the frequency range 211--275~GHz, i.e.\ covering the entire band~6 of ALMA in the spectral scan mode. The observations were conducted in one of the most extended configurations available in cycle~2, resulting in a synthesized beam that varies from $0\farcs39$ to $0\farcs65$ across the whole band. Our main results can be summarized as follows:

\begin{itemize}

\item We have applied a new continuum determination method to the ALMA \SgrB\ data in order to produce continuum images of both \SgrB(N) and \SgrB(M). The final images have an angular resolution of $0\farcs4$ (or 3400~au) and a rms noise level of about 8~mJy~beam$^{-1}$. We produce continuum images at different frequencies, which permits us to characterize the spectral index and study the nature of the continuum sources.

\item We have identified 20 sources in \SgrB(N) and 27 sources in \SgrB(M). The number of detected sources in the central 10\arcsec\ of each source is increased with respect to previous SMA observations at 345~GHz (12 against 2 sources in \SgrB(N), and 18 against 12 sources in \SgrB(M)). This suggests that the ALMA observations at 1.3~mm are sensitive not only to the dust cores detected with the SMA but to a different population of objects (e.g.\ fainter dust condensations and sources with ionized gas emission). As found in previous SMA observations, \SgrB(M) is highly fragmented, while \SgrB(N) consist of one major source surrounded by a few fainter objects. The ALMA maps reveal filamentary-like structures associated with the main core (AN01) in \SgrB(N), and converging towards the center, suggestive of accretion channels transferring mass from the outskirts to the center.

\item Spectral indices are derived for each source using the continuum images produced across all the ALMA band~6. We find that the sources with higher continuum intensities show spectral indices in the range 2--4, typical of dust continuum emission. Fainter sources have spectral indices that vary from negative or flat (typical of \hii\ regions) to positive spectral indices (characteristic of optically thick \hii\ regions or dust cores). The presence of ionized gas emission at the frequencies 211--275~GHz is confirmed when comparing with VLA 40~GHz continuum images. Spectral index maps show that the dense, dust-dominated cores are optically thick towards the center and optically thin in the outskirts.

\item We have derived physical properties of the dust and ionized gas for the sources identified in the ALMA images. The gas and dust mass of the sources range from a few to a few 1000~\mo, the H$_2$ volume density ranges from $10^7$ to $10^9$~cm$^{-3}$ and the H$_2$ column densities range from $10^{24}$ to $10^{26}$~cm$^{-2}$. While \SgrB(M) has most of the mass distributed among different sources, in \SgrB(N) most of the mass (about 73\%) is contained in one single core (AN01), corresponding to about 9000~\mo\ in a 0.05~pc size structure. The cumulative mass function (or core mass functon) suggests a lack of low-mass dense cores, similar to what has been found in other regions forming high-mass stars. The high densities found in \SgrB(N) and \SgrB(M) of about $10^5$--$10^7$~\mo~pc$^{-3}$ are one to two orders of magnitude larger than the stellar densities found in super star clusters, suggesting \SgrB\ has the potential to form a super star cluster.

\item We have statistically characterized the chemical content of the sources by studying the number of lines per GHz, and the percentage of luminosity contained in the lines with respect to the total luminosity (line plus continuum). In general, \SgrB(N) is chemically richer than \SgrB(M). The chemically richest sources have about 100 lines per GHz, and the fraction of luminosity contained in spectral lines is about 35\% for the most rich sources and in the range 10--20\% for the others. We find a correlation between the chemical richness and the mass (and density of the cores) that may suggest that less massive objects appear as less chemically rich because of sensitivity limitations. A more accurate analysis of the chemical content will be presented in forthcoming papers. \SgrB(N) and \SgrB(M) are harbouring clusters of massive and chemically rich hot molecular cores, similar to other well-known hot cores in the Galactic disk, like Orion~KL or G31.41$+$0.31.

\item Finally, we have compared the continuum images as well as spectral index maps obtained from the ALMA observations with predictions from a 3D radiative transfer model that reproduces the structure of \SgrB\ from 45~pc scales down to 100~au scales. The general structure of the model prediction agrees well with the structure seen in the ALMA images. However, there are some discrepancies caused by a lack of information when the 3D model was created, i.e.\ lack of new sources discovered in the ALMA observations, and lack of (sub-arcsecond) observations at different frequencies to better constrain the dependence of flux on frequency. The dataset presented here, together with ongoing (sub-arcsecond) observational projects in the frequency range from 5~GHz to 200~GHz will help to better constrain the 3D structure of \SgrB\ and derive more accurate physical parameters for the sources around the \SgrB(N) and \SgrB(M) star-forming regions.

\end{itemize}

\begin{acknowledgements}
This work was supported by Deutsche Forschungsgemeinschaft through grant SFB\,956 (subproject A6). S.-L.~Q.\ is supported by NSFC under grant No.\ 11373026, and Top Talents Program of Yunnan Province (2015HA030). This paper makes use of the following ALMA data: ADS/JAO.ALMA\#2013.1.00332.S. ALMA is a partnership of ESO (representing its member states), NSF (USA) and NINS (Japan), together with NRC (Canada) and NSC and ASIAA (Taiwan), in cooperation with the Republic of Chile. The Joint ALMA Observatory is operated by ESO, AUI/NRAO and NAOJ.
\end{acknowledgements}

%
\begin{appendix}

\section{Extra figures\label{a:extrafigures}}

\begin{figure*}[h!]
\begin{center}
\begin{tabular}[b]{c}
        \includegraphics[width=0.85\textwidth]{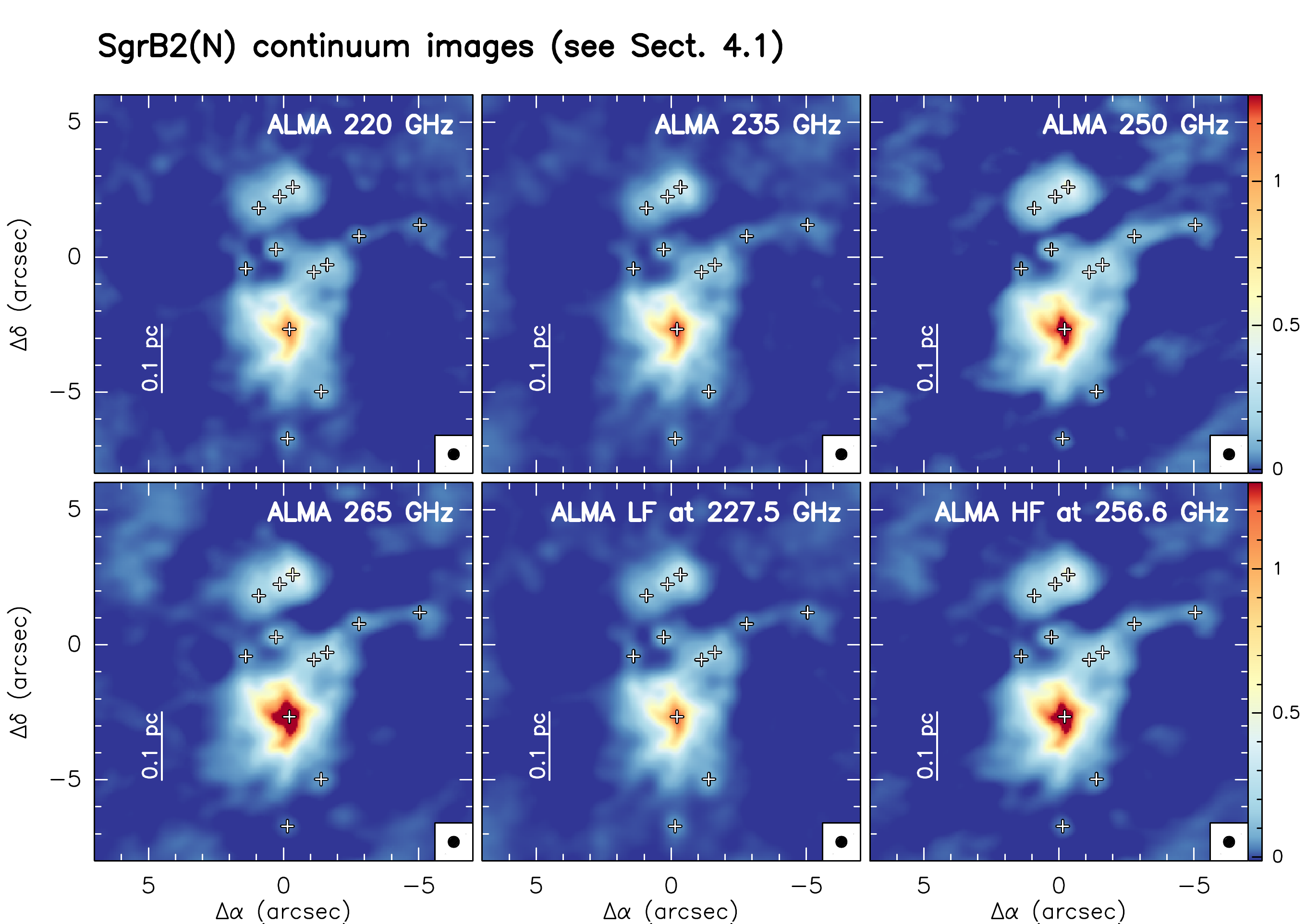} \\
        \includegraphics[width=0.85\textwidth]{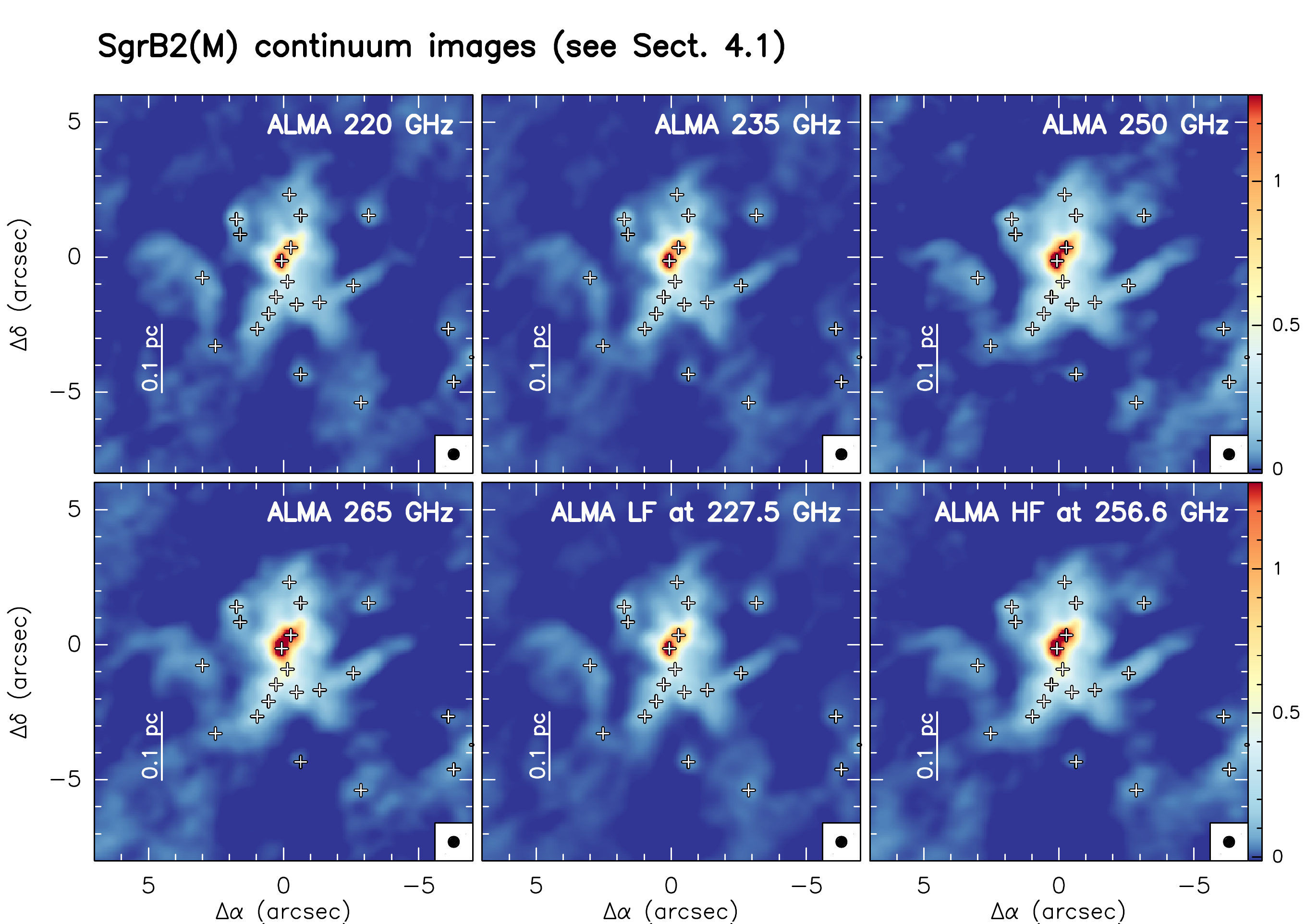} \\
\end{tabular}
\caption{Continuum emission maps for \SgrB(N) (top panels) and \SgrB(M) (bottom panels) obtained after averaging different continuum emission maps. The central frequencies are indicated in each panel. The color bar is units of Jy~beam$^{-1}$, with a round synthesized beam of 0\farcs4. See more details in Sect.~\ref{s:sources} and Appendix~\ref{a:extrafigures}.}
\label{f:SgrB2ALLcontinuum}
\end{center}
\end{figure*}

\begin{figure*}[h!]
\begin{center}
\begin{tabular}[b]{c}
        \includegraphics[width=0.85\textwidth]{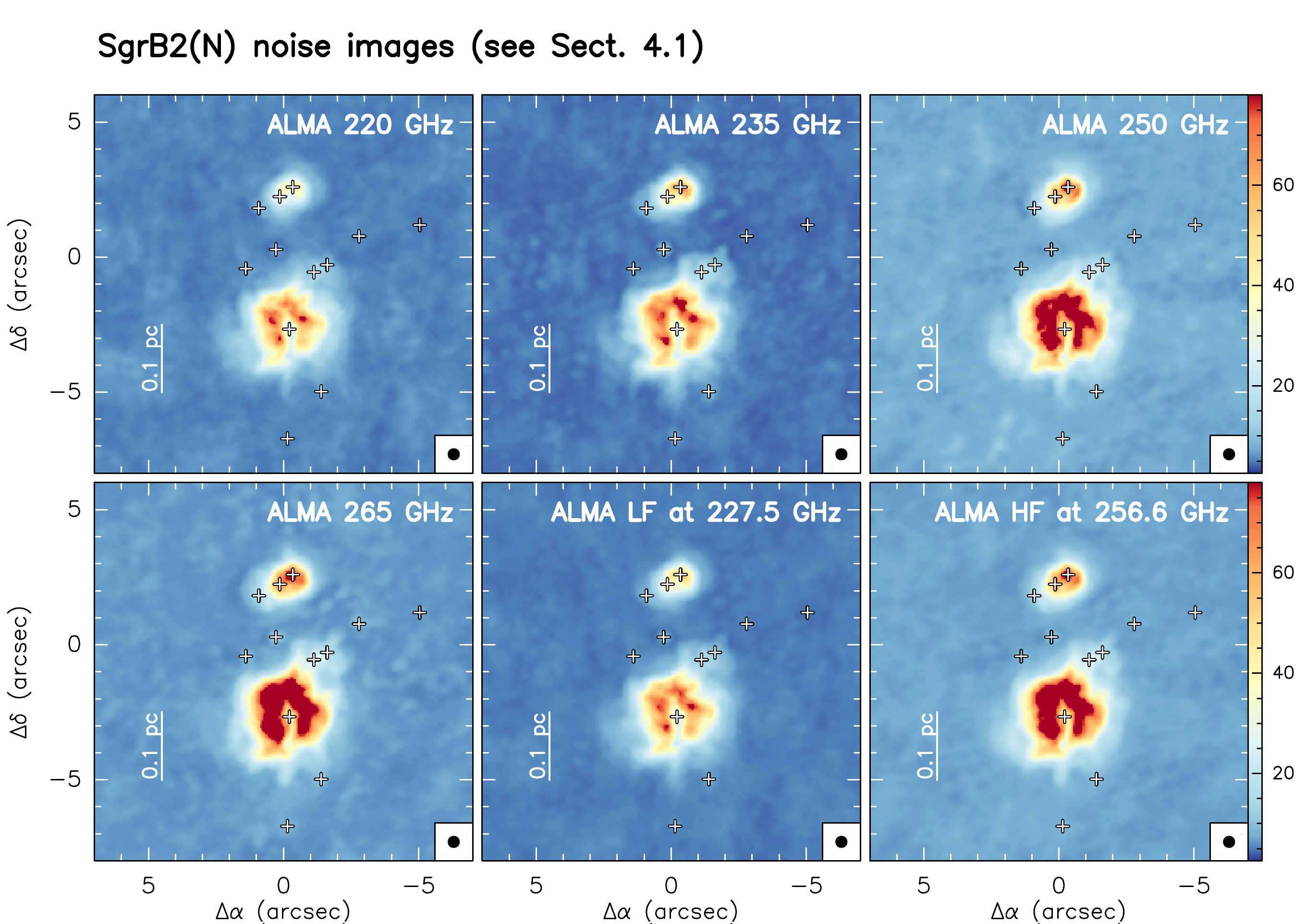} \\
        \includegraphics[width=0.85\textwidth]{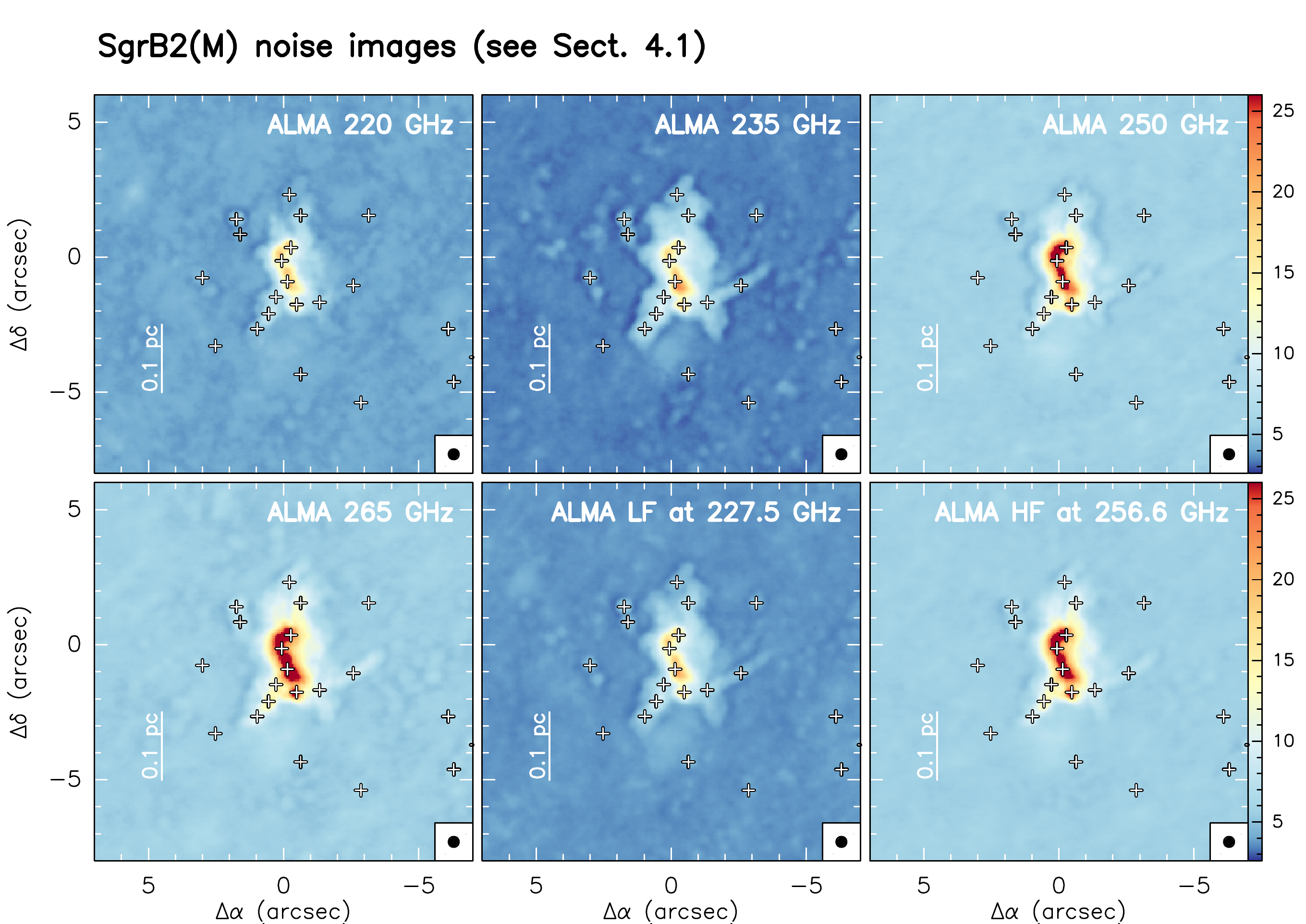} \\
\end{tabular}
\caption{Noise maps for \SgrB(N) (top panels) and \SgrB(M) (bottom panels) obtained from the determination of the continuum level, and after averaging the different noise maps produced for each spectral window. The central frequencies are indicated in each panel. The color bar is units of mJy~beam$^{-1}$, with a round synthesized beam of 0\farcs4. See more details in Sect.~\ref{s:sources} and Appendix~\ref{a:extrafigures}.}
\label{f:SgrB2ALLcontinuumnoise}
\end{center}
\end{figure*}

In Fig.~\ref{f:SgrB2ALLcontinuum} we show the continuum maps created to determine the sources in both \SgrB(N) and \SgrB(M) as described in Sect.~\ref{s:sources}. These maps have been used to search for common structures considered to be real emission, and structures appearing only in few maps considered to be artifacts produced during the cleaning and continuum determination processes. The first four panels correspond to the continuum images produced after averaging the continuum maps of the spectral windows contained in each of the frequency blocks shown in Fig.~\ref{f:freqcover}. The central frequencies are 220~GHz, 235~GHz, 250~GHz and 265~GHz. The last two panels show the averaged continuum images for the low-frequency tuning centered at 227.5~GHz and the high-frequency tuning centered at 256.6~GHz. Figure~\ref{f:SgrB2ALLcontinuumnoise} show the noise level emission, or error in the continuum determination, of the continuum emission for the maps shown in Fig.~\ref{f:SgrB2ALLcontinuum}. The error is obtained from the cSCM method in STATCONT, as described in S\'anchez-Monge et al.\ (submitted, see also Sect.~\ref{s:cont1mm}). The central pixels, associated with complex spectra have larger uncertainties than those pixels associated with just noise or faint emission.

\begin{figure*}[h!]
\begin{center}
\begin{tabular}[b]{c}
        \includegraphics[width=0.85\textwidth]{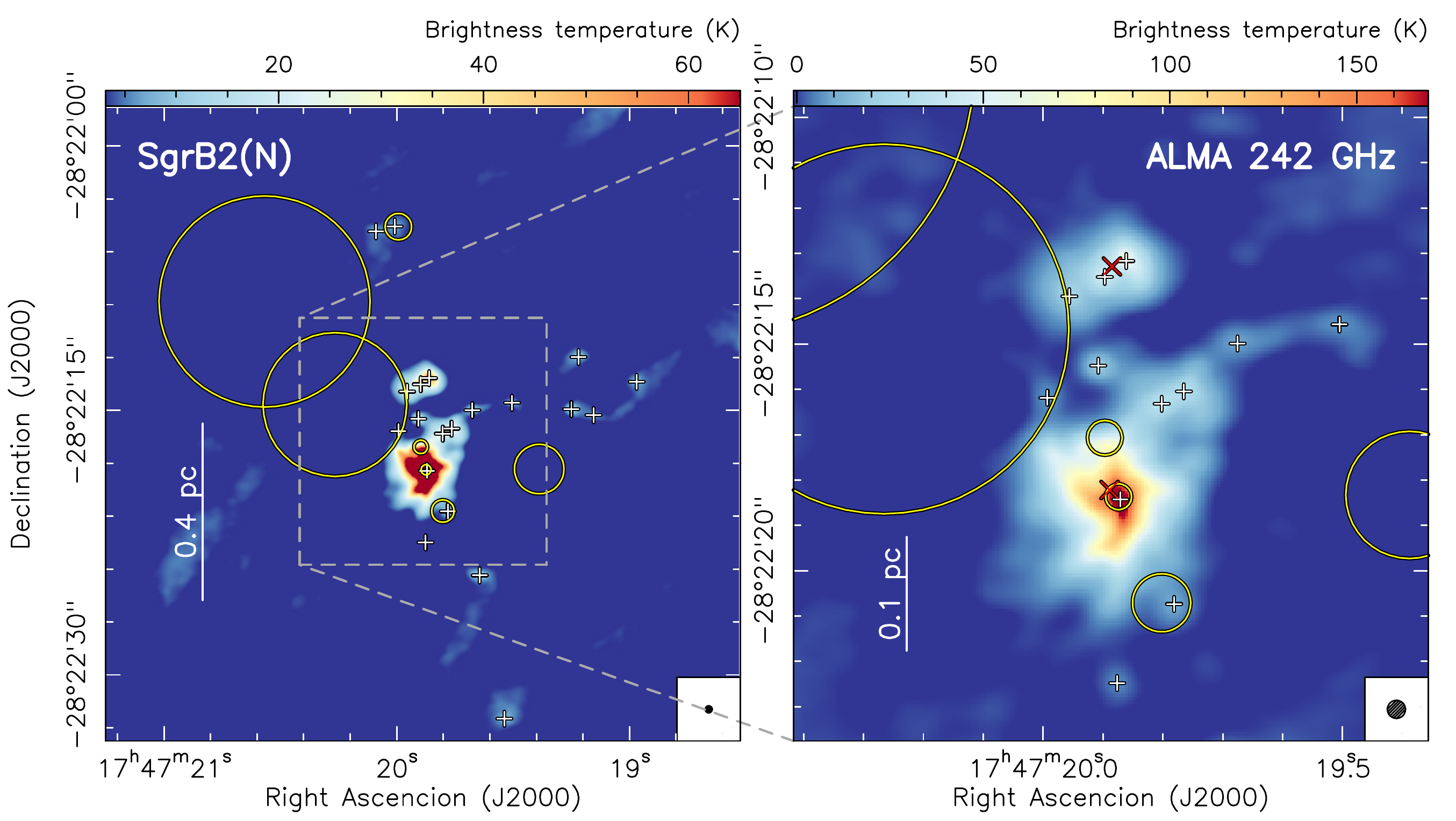} \\
        \includegraphics[width=0.85\textwidth]{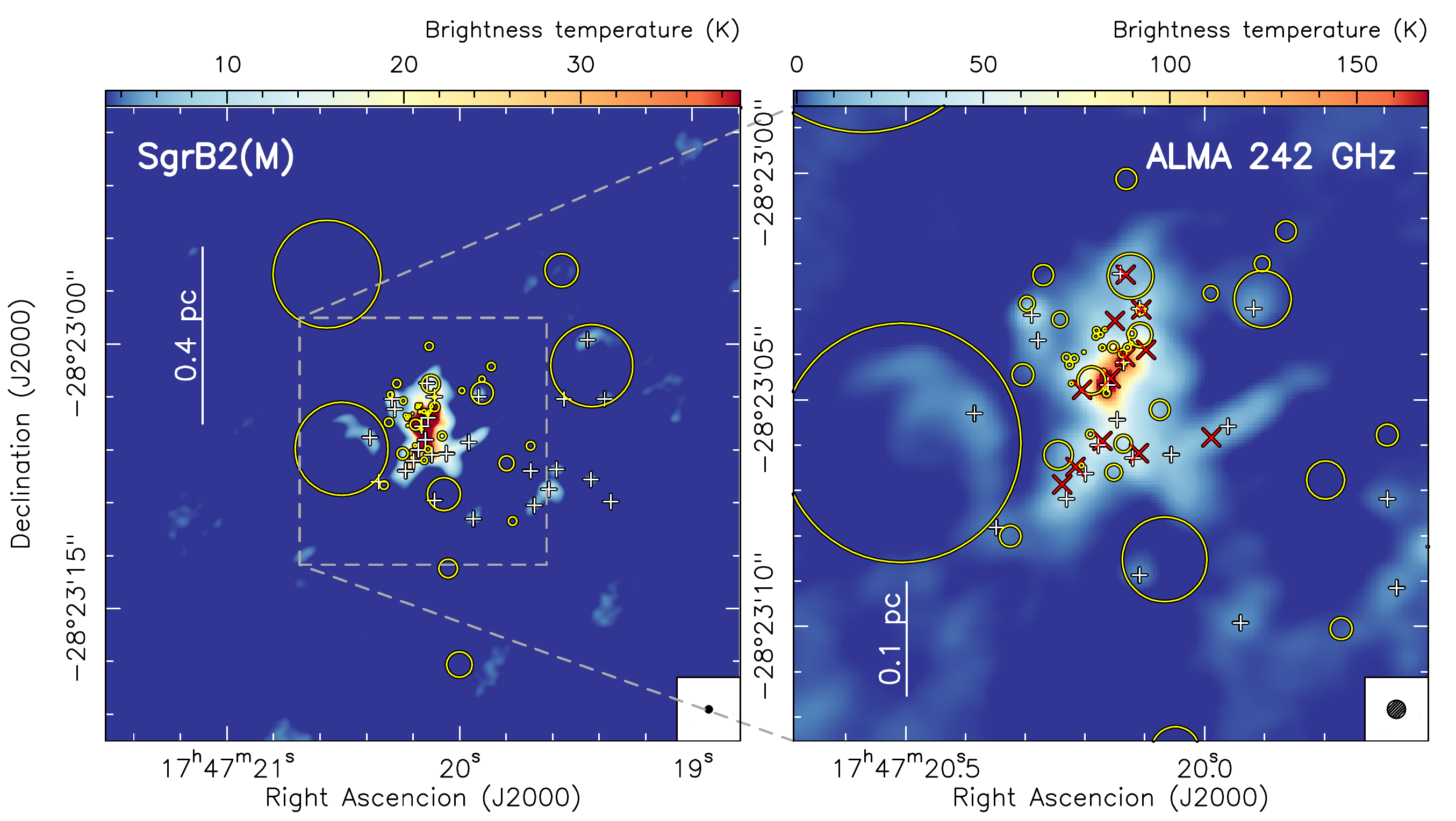} \\
\end{tabular}
\caption{Finding charts of the different known continuum sources in \SgrB(N) (top panels) and \SgrB(M) (bottom panels). The yellow circles indicate well-known \hii\ regions (see \citealt{Schmiedeke2016} and references therein), while the red crosses show the position of the sub-millimeter sources identified by \citet{Qin2011}. The association of the ALMA sources with the \hii\ regions and sub-millimeter sources is summarized in Table~\ref{t:summary}.}
\label{f:findingchart}
\end{center}
\end{figure*}

In Fig.~\ref{f:findingchart} we show the ALMA continumm emission maps of \SgrB(N) and \SgrB(M), and overlay the size and position of the known \hii\ regions (see \citealt{Schmiedeke2016} and references therein) and the position of the sub-millimeter continuum sources identified by \citet{Qin2011} with the SMA at 345~GHz.

\section{Continuum flux and chemical content of \SgrB\ sources\label{a:fluxes}}

Tables~\ref{t:SgrB2Nfluxes} and \ref{t:SgrB2Mfluxes} list the fluxes derived for each source identified in both \SgrB(N) and \SgrB(M), for all the continuum maps created across the band~6 of ALMA (from 211~GHz to 275~GHz). The fluxes of each source are listed in different columns, while each row contains the values for a given continuum image or frequency. The frequencies listed in the first column indicate the central frequency used during the determination of the continuum level emission. The first block of rows contains the fluxes for the individual spectral windows (see Fig.~\ref{f:freqcover}). The last blocks of rows contain the fluxes for the extra continuum images created to identify the ALMA 1.3~mm continuum sources (see Sect.~\ref{s:sources} for more details).

Figures~\ref{f:SgrB2NSEDs} and \ref{f:SgrB2MSEDs} show a summary of the mm-SED and chemical content for each identified source in \SgrB2(N) and \SgrB(M), respectively. The left panel shows the mm-SED with the measured fluxes and 1$\sigma$ errors shown in grey. The linear fit and the value of the spectral index ($\alpha$) are shown in black. The right panel shows a portion of the spectral line survey for each source. The spectra have been obtained after averaging the emission inside the 3$\sigma$ polygon defined to measure the continuum fluxes. The brightness temperature scale has been fixed to better show differences between sources.

\begin{sidewaystable*}
\centering
\caption{\SgrB(N) fluxes and errors for all the continuum maps. Each column corresponds to an ALMA 1.3~mm continuum source (see Fig.~\ref{f:SgrB2Nsources}). For details on the creation of the continuum maps and the polygons used to derive the fluxes, see Section~\ref{s:sources}.}
\label{t:SgrB2Nfluxes}
\begin{tabular}{l c c c c c c c c c c c c c c}
\hline\hline

\multicolumn{1}{c}{Frequency}
&
\\

\multicolumn{1}{c}{(MHz)}
&AN01
&AN02
&AN03
&AN04
&AN05
&AN06
&AN07
&AN08
&AN09
&AN10
\\
\hline
212000.00  &$15.56\pm1.49$ &$0.97\pm0.15$ &$0.69\pm0.12$ &$0.78\pm0.09$ &$0.70\pm0.06$ &$0.48\pm0.07$ &$0.19\pm0.02$ &$0.25\pm0.03$ &$0.11\pm0.04$ &$0.30\pm0.02$ \\
213875.00  &$16.62\pm1.48$ &$1.09\pm0.17$ &$0.78\pm0.13$ &$0.87\pm0.08$ &$0.78\pm0.05$ &$0.61\pm0.06$ &$0.21\pm0.02$ &$0.27\pm0.03$ &$0.22\pm0.04$ &$0.32\pm0.02$ \\
215750.00  &$18.93\pm2.18$ &$1.40\pm0.23$ &$0.99\pm0.15$ &$1.00\pm0.12$ &$0.84\pm0.07$ &$0.75\pm0.08$ &$0.22\pm0.03$ &$0.36\pm0.04$ &$0.18\pm0.06$ &$0.37\pm0.03$ \\
217625.00  &$17.99\pm1.59$ &$1.20\pm0.16$ &$0.83\pm0.13$ &$0.97\pm0.09$ &$0.84\pm0.06$ &$0.65\pm0.08$ &$0.23\pm0.02$ &$0.25\pm0.04$ &$0.31\pm0.05$ &$0.33\pm0.02$ \\
219500.00  &$22.30\pm3.15$ &$1.56\pm0.21$ &$1.34\pm0.17$ &$1.11\pm0.12$ &$0.78\pm0.07$ &$1.18\pm0.08$ &$0.33\pm0.03$ &$0.24\pm0.06$ &$0.50\pm0.06$ &$0.31\pm0.03$ \\
221375.00  &$21.59\pm2.60$ &$1.65\pm0.27$ &$1.44\pm0.20$ &$1.13\pm0.13$ &$0.76\pm0.07$ &$1.21\pm0.09$ &$0.31\pm0.03$ &$0.32\pm0.05$ &$0.53\pm0.06$ &$0.32\pm0.03$ \\
227000.00  &$22.48\pm2.65$ &$1.90\pm0.27$ &$1.36\pm0.20$ &$1.18\pm0.13$ &$0.86\pm0.09$ &$1.06\pm0.10$ &$0.40\pm0.03$ &$0.16\pm0.06$ &$0.53\pm0.09$ &$0.30\pm0.03$ \\
228000.00  &$20.08\pm2.55$ &$1.18\pm0.20$ &$0.79\pm0.16$ &$0.82\pm0.09$ &$0.78\pm0.06$ &$0.40\pm0.08$ &$0.16\pm0.03$ &$0.23\pm0.03$ &$0.22\pm0.04$ &$0.30\pm0.03$ \\
228875.00  &$22.06\pm2.24$ &$1.83\pm0.20$ &$1.42\pm0.15$ &$1.20\pm0.11$ &$0.83\pm0.07$ &$1.09\pm0.09$ &$0.40\pm0.03$ &$0.18\pm0.06$ &$0.57\pm0.07$ &$0.25\pm0.03$ \\
229875.00  &$20.18\pm1.69$ &$1.12\pm0.16$ &$0.77\pm0.13$ &$0.84\pm0.10$ &$0.75\pm0.07$ &$0.47\pm0.07$ &$0.17\pm0.02$ &$0.29\pm0.04$ &$0.23\pm0.05$ &$0.34\pm0.03$ \\
231750.00  &$21.49\pm2.18$ &$1.37\pm0.23$ &$0.86\pm0.15$ &$0.99\pm0.12$ &$0.81\pm0.07$ &$0.50\pm0.08$ &$0.20\pm0.03$ &$0.27\pm0.04$ &$0.15\pm0.06$ &$0.28\pm0.03$ \\
233625.00  &$23.37\pm3.18$ &$1.58\pm0.42$ &$0.88\pm0.27$ &$1.09\pm0.17$ &$0.85\pm0.09$ &$0.42\pm0.10$ &$0.18\pm0.03$ &$0.29\pm0.05$ &$0.03\pm0.06$ &$0.25\pm0.03$ \\
235500.00  &$20.99\pm2.02$ &$1.50\pm0.27$ &$1.07\pm0.20$ &$0.95\pm0.14$ &$0.73\pm0.07$ &$0.87\pm0.09$ &$0.26\pm0.02$ &$0.22\pm0.04$ &$0.32\pm0.05$ &$0.16\pm0.03$ \\
237375.00  &$22.24\pm3.06$ &$1.41\pm0.23$ &$1.07\pm0.20$ &$0.95\pm0.12$ &$0.80\pm0.07$ &$0.87\pm0.09$ &$0.17\pm0.03$ &$0.49\pm0.05$ &$0.33\pm0.07$ &$0.21\pm0.03$ \\
241077.50  &$28.24\pm3.80$ &$1.96\pm0.41$ &$1.60\pm0.35$ &$1.36\pm0.17$ &$1.14\pm0.09$ &$1.10\pm0.13$ &$0.26\pm0.04$ &$0.30\pm0.07$ &$0.46\pm0.08$ &$0.17\pm0.04$ \\
242952.50  &$27.32\pm2.90$ &$1.89\pm0.36$ &$1.58\pm0.27$ &$1.40\pm0.18$ &$1.14\pm0.10$ &$1.10\pm0.13$ &$0.24\pm0.04$ &$0.33\pm0.07$ &$0.59\pm0.08$ &$0.16\pm0.04$ \\
243000.00  &$24.04\pm2.40$ &$1.90\pm0.32$ &$1.37\pm0.20$ &$1.15\pm0.14$ &$0.84\pm0.08$ &$0.99\pm0.09$ &$0.31\pm0.03$ &$0.38\pm0.05$ &$0.66\pm0.06$ &$0.20\pm0.03$ \\
244827.50  &$26.64\pm2.60$ &$2.05\pm0.30$ &$1.68\pm0.24$ &$1.45\pm0.15$ &$1.20\pm0.08$ &$1.35\pm0.11$ &$0.41\pm0.03$ &$0.34\pm0.05$ &$0.41\pm0.07$ &$0.13\pm0.04$ \\
244875.00  &$23.71\pm2.43$ &$1.90\pm0.29$ &$1.44\pm0.22$ &$1.13\pm0.15$ &$0.87\pm0.08$ &$1.10\pm0.10$ &$0.36\pm0.03$ &$0.33\pm0.05$ &$0.51\pm0.05$ &$0.15\pm0.02$ \\
246702.50  &$28.01\pm3.34$ &$2.05\pm0.27$ &$1.73\pm0.25$ &$1.50\pm0.14$ &$1.23\pm0.09$ &$1.32\pm0.12$ &$0.41\pm0.03$ &$0.36\pm0.05$ &$0.50\pm0.08$ &$0.09\pm0.04$ \\
248577.50  &$27.89\pm3.08$ &$2.09\pm0.29$ &$1.71\pm0.23$ &$1.74\pm0.16$ &$1.54\pm0.11$ &$1.23\pm0.13$ &$0.51\pm0.04$ &$0.28\pm0.08$ &$0.93\pm0.09$ &$0.03\pm0.04$ \\
250452.50  &$30.90\pm3.83$ &$2.49\pm0.51$ &$1.86\pm0.33$ &$1.84\pm0.21$ &$1.61\pm0.13$ &$1.23\pm0.15$ &$0.51\pm0.04$ &$0.27\pm0.08$ &$0.80\pm0.11$ &$0.04\pm0.05$ \\
256077.50  &$30.57\pm4.36$ &$2.41\pm0.33$ &$1.75\pm0.27$ &$1.55\pm0.19$ &$1.43\pm0.12$ &$1.22\pm0.14$ &$0.34\pm0.04$ &$0.36\pm0.07$ &$0.29\pm0.09$ &$0.30\pm0.04$ \\
257077.50  &$34.24\pm4.86$ &$1.85\pm0.35$ &$1.65\pm0.31$ &$1.32\pm0.18$ &$1.02\pm0.10$ &$1.24\pm0.13$ &$0.24\pm0.04$ &$0.43\pm0.07$ &$0.74\pm0.08$ &$0.19\pm0.05$ \\
257952.50  &$33.86\pm4.52$ &$2.61\pm0.42$ &$1.91\pm0.33$ &$1.69\pm0.22$ &$1.52\pm0.13$ &$1.24\pm0.16$ &$0.32\pm0.04$ &$0.34\pm0.08$ &$0.34\pm0.09$ &$0.31\pm0.05$ \\
258952.50  &$31.32\pm3.63$ &$1.93\pm0.40$ &$1.60\pm0.34$ &$1.26\pm0.19$ &$0.98\pm0.10$ &$1.14\pm0.13$ &$0.28\pm0.03$ &$0.39\pm0.06$ &$0.50\pm0.06$ &$0.19\pm0.03$ \\
260827.50  &$33.15\pm4.56$ &$2.30\pm0.53$ &$1.80\pm0.41$ &$1.42\pm0.24$ &$1.04\pm0.11$ &$1.21\pm0.16$ &$0.33\pm0.03$ &$0.33\pm0.05$ &$0.36\pm0.05$ &$0.15\pm0.03$ \\
262702.50  &$30.48\pm2.84$ &$2.14\pm0.39$ &$1.83\pm0.34$ &$1.53\pm0.22$ &$1.14\pm0.11$ &$1.38\pm0.14$ &$0.34\pm0.04$ &$0.44\pm0.06$ &$0.51\pm0.07$ &$0.17\pm0.04$ \\
264577.50  &$30.27\pm3.03$ &$2.09\pm0.33$ &$1.76\pm0.29$ &$1.21\pm0.17$ &$1.00\pm0.11$ &$1.35\pm0.14$ &$0.44\pm0.04$ &$0.27\pm0.05$ &$0.34\pm0.07$ &$0.11\pm0.04$ \\
266452.50  &$35.33\pm4.12$ &$2.34\pm0.28$ &$1.94\pm0.25$ &$1.30\pm0.17$ &$1.10\pm0.12$ &$1.61\pm0.13$ &$0.52\pm0.04$ &$0.47\pm0.06$ &$0.60\pm0.10$ &$0.05\pm0.04$ \\
272077.50  &$30.09\pm2.69$ &$2.45\pm0.39$ &$1.67\pm0.27$ &$1.42\pm0.21$ &$1.37\pm0.13$ &$1.11\pm0.13$ &$0.28\pm0.03$ &$0.22\pm0.06$ &$0.28\pm0.06$ &$0.22\pm0.04$ \\
273952.50  &$31.72\pm3.67$ &$2.53\pm0.45$ &$1.74\pm0.34$ &$1.41\pm0.25$ &$1.34\pm0.16$ &$1.07\pm0.16$ &$0.26\pm0.04$ &$0.19\pm0.07$ &$0.25\pm0.08$ &$0.20\pm0.04$ \\
\hline
\multicolumn{8}{l}{\phn Average maps centered at 220~GHz, 235~GHz, 250~GHz and 265~GHz} \\
\hline
219500.00  &$19.69\pm2.17$ &$1.45\pm0.21$ &$1.11\pm0.16$ &$1.03\pm0.11$ &$0.80\pm0.07$ &$0.88\pm0.08$ &$0.29\pm0.03$ &$0.26\pm0.05$ &$0.37\pm0.06$ &$0.31\pm0.03$ \\
235500.00  &$22.01\pm2.48$ &$1.50\pm0.27$ &$1.03\pm0.20$ &$0.99\pm0.13$ &$0.80\pm0.07$ &$0.70\pm0.09$ &$0.23\pm0.03$ &$0.31\pm0.04$ &$0.31\pm0.05$ &$0.24\pm0.03$ \\
248577.50  &$29.18\pm3.55$ &$2.19\pm0.36$ &$1.73\pm0.28$ &$1.56\pm0.18$ &$1.35\pm0.11$ &$1.22\pm0.13$ &$0.37\pm0.04$ &$0.32\pm0.07$ &$0.54\pm0.09$ &$0.14\pm0.04$ \\
264577.50  &$32.08\pm3.67$ &$2.20\pm0.39$ &$1.75\pm0.32$ &$1.36\pm0.20$ &$1.12\pm0.12$ &$1.26\pm0.14$ &$0.34\pm0.04$ &$0.34\pm0.06$ &$0.45\pm0.07$ &$0.16\pm0.04$ \\
\hline
\multicolumn{8}{l}{\phn Average maps for the LF (227.5~GHz) and HF (256.0~GHz) spectral windows} \\
\hline
227500.00  &$20.85\pm2.33$ &$1.47\pm0.24$ &$1.07\pm0.18$ &$1.01\pm0.12$ &$0.80\pm0.07$ &$0.79\pm0.09$ &$0.26\pm0.03$ &$0.28\pm0.05$ &$0.34\pm0.05$ &$0.27\pm0.03$ \\
256577.50  &$30.63\pm3.61$ &$2.20\pm0.38$ &$1.74\pm0.30$ &$1.46\pm0.19$ &$1.24\pm0.11$ &$1.24\pm0.14$ &$0.35\pm0.04$ &$0.33\pm0.06$ &$0.49\pm0.08$ &$0.15\pm0.04$ \\
\hline
\multicolumn{8}{l}{\phn Average of all the continuum maps} \\
\hline
242038.75  &$25.74\pm2.97$ &$1.84\pm0.31$ &$1.40\pm0.24$ &$1.24\pm0.15$ &$1.02\pm0.09$ &$1.02\pm0.11$ &$0.31\pm0.03$ &$0.31\pm0.05$ &$0.42\pm0.07$ &$0.21\pm0.03$ \\
\hline
\end{tabular}
\end{sidewaystable*}
\begin{sidewaystable*}
\ContinuedFloat
\centering
\caption{Continued.}
\begin{tabular}{l c c c c c c c c c c c c c c}
\hline\hline

\multicolumn{1}{c}{Frequency}
&
\\

\multicolumn{1}{c}{(MHz)}
&AN11
&AN12
&AN13
&AN14
&AN15
&AN16
&AN17
&AN18
&AN19
&AN20
\\
\hline
212000.00  &$0.15\pm0.02$ &$0.09\pm0.05$ &$0.71\pm0.13$ &$0.45\pm0.05$ &$0.15\pm0.02$ &$0.27\pm0.06$ &$0.04\pm0.01$ &$0.08\pm0.04$ &$0.11\pm0.01$ &$0.08\pm0.01$ \\
213875.00  &$0.14\pm0.02$ &$0.15\pm0.05$ &$0.89\pm0.12$ &$0.56\pm0.06$ &$0.09\pm0.02$ &$0.38\pm0.06$ &$0.04\pm0.01$ &$0.17\pm0.04$ &$0.10\pm0.02$ &$0.09\pm0.01$ \\
215750.00  &$0.04\pm0.04$ &$0.46\pm0.07$ &$1.03\pm0.19$ &$0.67\pm0.09$ &$0.00\pm0.03$ &$0.72\pm0.09$ &$0.08\pm0.02$ &$0.23\pm0.06$ &$0.17\pm0.02$ &$0.11\pm0.01$ \\
217625.00  &$0.10\pm0.03$ &$0.43\pm0.06$ &$1.36\pm0.17$ &$0.59\pm0.08$ &$0.05\pm0.02$ &$0.64\pm0.07$ &$0.09\pm0.01$ &$0.29\pm0.05$ &$0.14\pm0.02$ &$0.08\pm0.01$ \\
219500.00  &$0.02\pm0.04$ &$0.70\pm0.09$ &$0.63\pm0.19$ &$0.18\pm0.09$ &$0.09\pm0.03$ &$0.53\pm0.09$ &$0.10\pm0.01$ &$0.35\pm0.07$ &$0.16\pm0.02$ &$0.06\pm0.01$ \\
221375.00  &$0.06\pm0.04$ &$0.91\pm0.09$ &$0.82\pm0.18$ &$0.20\pm0.09$ &$0.11\pm0.03$ &$0.67\pm0.09$ &$0.06\pm0.02$ &$0.34\pm0.06$ &$0.17\pm0.02$ &$0.06\pm0.01$ \\
227000.00  &$0.06\pm0.04$ &$0.62\pm0.08$ &$1.75\pm0.27$ &$0.15\pm0.10$ &$0.13\pm0.03$ &$0.65\pm0.10$ &$0.08\pm0.02$ &$0.45\pm0.07$ &$0.03\pm0.02$ &$0.00\pm0.01$ \\
228000.00  &$0.19\pm0.02$ &$0.04\pm0.05$ &$0.64\pm0.13$ &$0.32\pm0.06$ &$0.10\pm0.02$ &$0.33\pm0.05$ &$0.02\pm0.01$ &$0.16\pm0.04$ &$0.10\pm0.01$ &$0.10\pm0.01$ \\
228875.00  &$0.05\pm0.04$ &$0.59\pm0.07$ &$1.41\pm0.21$ &$0.24\pm0.10$ &$0.09\pm0.03$ &$0.55\pm0.09$ &$0.07\pm0.02$ &$0.44\pm0.06$ &$0.06\pm0.02$ &$0.01\pm0.01$ \\
229875.00  &$0.14\pm0.02$ &$0.04\pm0.05$ &$0.61\pm0.13$ &$0.37\pm0.06$ &$0.12\pm0.02$ &$0.33\pm0.06$ &$0.03\pm0.01$ &$0.09\pm0.04$ &$0.13\pm0.02$ &$0.11\pm0.01$ \\
231750.00  &$0.12\pm0.04$ &$0.09\pm0.07$ &$0.81\pm0.19$ &$0.47\pm0.09$ &$0.07\pm0.03$ &$0.54\pm0.09$ &$0.06\pm0.02$ &$0.26\pm0.06$ &$0.09\pm0.02$ &$0.08\pm0.01$ \\
233625.00  &$0.18\pm0.03$ &$0.04\pm0.08$ &$0.80\pm0.16$ &$0.52\pm0.08$ &$0.18\pm0.02$ &$0.50\pm0.08$ &$0.05\pm0.01$ &$0.28\pm0.06$ &$0.12\pm0.02$ &$0.09\pm0.01$ \\
235500.00  &$0.09\pm0.02$ &$0.44\pm0.05$ &$0.44\pm0.12$ &$0.15\pm0.05$ &$0.09\pm0.02$ &$0.26\pm0.06$ &$0.09\pm0.01$ &$0.13\pm0.05$ &$0.05\pm0.02$ &$0.04\pm0.01$ \\
237375.00  &$0.10\pm0.04$ &$0.54\pm0.08$ &$0.59\pm0.20$ &$0.06\pm0.09$ &$0.14\pm0.03$ &$0.45\pm0.10$ &$0.06\pm0.01$ &$0.16\pm0.07$ &$0.15\pm0.02$ &$0.02\pm0.01$ \\
241077.50  &$0.33\pm0.04$ &$0.51\pm0.08$ &$2.20\pm0.23$ &$0.84\pm0.12$ &$0.21\pm0.03$ &$0.98\pm0.11$ &$0.05\pm0.02$ &$0.81\pm0.09$ &$0.16\pm0.03$ &$0.06\pm0.01$ \\
242952.50  &$0.26\pm0.04$ &$0.58\pm0.09$ &$2.35\pm0.25$ &$0.88\pm0.12$ &$0.18\pm0.03$ &$1.00\pm0.12$ &$0.06\pm0.02$ &$0.80\pm0.09$ &$0.20\pm0.03$ &$0.07\pm0.01$ \\
243000.00  &$0.21\pm0.03$ &$0.33\pm0.07$ &$1.47\pm0.18$ &$0.20\pm0.09$ &$0.20\pm0.03$ &$0.26\pm0.09$ &$0.07\pm0.01$ &$0.28\pm0.07$ &$0.16\pm0.02$ &$0.03\pm0.01$ \\
244827.50  &$0.40\pm0.03$ &$0.54\pm0.07$ &$1.82\pm0.19$ &$1.03\pm0.09$ &$0.26\pm0.03$ &$0.35\pm0.08$ &$0.13\pm0.01$ &$0.41\pm0.07$ &$0.22\pm0.02$ &$0.05\pm0.01$ \\
244875.00  &$0.22\pm0.03$ &$0.41\pm0.06$ &$1.23\pm0.15$ &$0.12\pm0.08$ &$0.15\pm0.02$ &$0.32\pm0.07$ &$0.08\pm0.01$ &$0.26\pm0.05$ &$0.10\pm0.02$ &$0.02\pm0.01$ \\
246702.50  &$0.37\pm0.04$ &$0.59\pm0.08$ &$1.85\pm0.22$ &$1.13\pm0.10$ &$0.24\pm0.03$ &$0.42\pm0.09$ &$0.10\pm0.02$ &$0.47\pm0.08$ &$0.19\pm0.02$ &$0.03\pm0.01$ \\
248577.50  &$0.63\pm0.05$ &$0.36\pm0.09$ &$1.12\pm0.28$ &$0.06\pm0.13$ &$0.25\pm0.04$ &$0.72\pm0.11$ &$0.03\pm0.02$ &$0.61\pm0.09$ &$0.07\pm0.03$ &$0.02\pm0.01$ \\
250452.50  &$0.71\pm0.05$ &$0.27\pm0.10$ &$1.41\pm0.30$ &$0.15\pm0.12$ &$0.24\pm0.04$ &$0.66\pm0.11$ &$0.05\pm0.02$ &$0.61\pm0.09$ &$0.06\pm0.03$ &$0.02\pm0.02$ \\
256077.50  &$0.51\pm0.04$ &$0.42\pm0.10$ &$0.63\pm0.28$ &$0.38\pm0.12$ &$0.14\pm0.04$ &$0.18\pm0.13$ &$0.12\pm0.03$ &$0.37\pm0.10$ &$0.03\pm0.03$ &$0.06\pm0.02$ \\
257077.50  &$0.19\pm0.05$ &$0.50\pm0.08$ &$1.74\pm0.26$ &$0.95\pm0.10$ &$0.12\pm0.03$ &$0.60\pm0.10$ &$0.06\pm0.02$ &$0.41\pm0.08$ &$0.13\pm0.03$ &$0.05\pm0.01$ \\
257952.50  &$0.55\pm0.05$ &$0.44\pm0.10$ &$0.64\pm0.29$ &$0.38\pm0.12$ &$0.16\pm0.04$ &$0.22\pm0.14$ &$0.10\pm0.03$ &$0.42\pm0.11$ &$0.01\pm0.03$ &$0.05\pm0.02$ \\
258952.50  &$0.22\pm0.04$ &$0.41\pm0.07$ &$1.29\pm0.19$ &$0.69\pm0.08$ &$0.13\pm0.03$ &$0.44\pm0.08$ &$0.09\pm0.01$ &$0.26\pm0.06$ &$0.11\pm0.02$ &$0.06\pm0.01$ \\
260827.50  &$0.27\pm0.03$ &$0.36\pm0.06$ &$0.78\pm0.15$ &$0.80\pm0.10$ &$0.18\pm0.02$ &$0.39\pm0.06$ &$0.10\pm0.02$ &$0.27\pm0.04$ &$0.11\pm0.02$ &$0.03\pm0.01$ \\
262702.50  &$0.27\pm0.04$ &$0.46\pm0.07$ &$1.10\pm0.20$ &$1.05\pm0.10$ &$0.21\pm0.03$ &$0.47\pm0.09$ &$0.11\pm0.02$ &$0.38\pm0.07$ &$0.15\pm0.02$ &$0.02\pm0.01$ \\
264577.50  &$0.13\pm0.05$ &$0.50\pm0.07$ &$0.57\pm0.20$ &$0.87\pm0.09$ &$0.12\pm0.03$ &$0.32\pm0.09$ &$0.12\pm0.02$ &$0.46\pm0.06$ &$0.22\pm0.02$ &$0.02\pm0.01$ \\
266452.50  &$0.17\pm0.06$ &$0.76\pm0.10$ &$0.55\pm0.24$ &$1.13\pm0.11$ &$0.24\pm0.04$ &$0.42\pm0.12$ &$0.14\pm0.02$ &$0.67\pm0.08$ &$0.28\pm0.03$ &$0.04\pm0.02$ \\
272077.50  &$0.34\pm0.04$ &$0.37\pm0.08$ &$0.03\pm0.19$ &$0.18\pm0.09$ &$0.14\pm0.03$ &$0.21\pm0.08$ &$0.09\pm0.02$ &$0.29\pm0.06$ &$0.02\pm0.02$ &$0.02\pm0.01$ \\
273952.50  &$0.31\pm0.05$ &$0.36\pm0.09$ &$0.03\pm0.23$ &$0.16\pm0.11$ &$0.12\pm0.03$ &$0.17\pm0.10$ &$0.06\pm0.02$ &$0.28\pm0.07$ &$0.01\pm0.03$ &$0.00\pm0.01$ \\\hline
\hline
\multicolumn{8}{l}{\phn Average maps centered at 220~GHz, 235~GHz, 250~GHz and 265~GHz} \\
\hline
219500.00  &$0.08\pm0.03$ &$0.49\pm0.07$ &$1.08\pm0.18$ &$0.38\pm0.08$ &$0.09\pm0.02$ &$0.55\pm0.08$ &$0.07\pm0.01$ &$0.29\pm0.06$ &$0.12\pm0.02$ &$0.06\pm0.01$ \\
235500.00  &$0.16\pm0.03$ &$0.23\pm0.06$ &$0.82\pm0.15$ &$0.26\pm0.07$ &$0.13\pm0.02$ &$0.37\pm0.07$ &$0.06\pm0.01$ &$0.20\pm0.05$ &$0.11\pm0.02$ &$0.06\pm0.01$ \\
248577.50  &$0.47\pm0.04$ &$0.46\pm0.09$ &$1.50\pm0.25$ &$0.61\pm0.11$ &$0.21\pm0.04$ &$0.57\pm0.11$ &$0.08\pm0.02$ &$0.56\pm0.09$ &$0.12\pm0.03$ &$0.04\pm0.01$ \\
264577.50  &$0.24\pm0.04$ &$0.46\pm0.08$ &$0.75\pm0.21$ &$0.73\pm0.10$ &$0.16\pm0.03$ &$0.38\pm0.09$ &$0.10\pm0.02$ &$0.38\pm0.07$ &$0.12\pm0.02$ &$0.02\pm0.01$ \\
\hline
\multicolumn{8}{l}{\phn Average maps for the LF (227.5~GHz) and HF (256.0~GHz) spectral windows} \\
\hline
227500.00  &$0.12\pm0.03$ &$0.36\pm0.07$ &$0.95\pm0.17$ &$0.32\pm0.08$ &$0.11\pm0.02$ &$0.46\pm0.08$ &$0.06\pm0.01$ &$0.25\pm0.06$ &$0.12\pm0.02$ &$0.06\pm0.01$ \\
256577.50  &$0.35\pm0.04$ &$0.46\pm0.08$ &$1.13\pm0.23$ &$0.67\pm0.11$ &$0.18\pm0.03$ &$0.47\pm0.10$ &$0.09\pm0.02$ &$0.47\pm0.08$ &$0.12\pm0.03$ &$0.03\pm0.01$ \\
\hline
\multicolumn{8}{l}{\phn Average of all the continuum maps} \\
\hline
242038.75  &$0.23\pm0.04$ &$0.41\pm0.07$ &$1.04\pm0.20$ &$0.49\pm0.09$ &$0.15\pm0.03$ &$0.47\pm0.09$ &$0.08\pm0.02$ &$0.36\pm0.07$ &$0.12\pm0.02$ &$0.04\pm0.01$ \\
\hline
\end{tabular}
\end{sidewaystable*}

\begin{sidewaystable*}
\centering
\caption{\SgrB(M) fluxes and errors for all the continuum maps. Each column corresponds to an ALMA 1.3~mm continuum source (see Fig.~\ref{f:SgrB2Msources}). For details on the creation of the continuum maps and the polygons used to derive the fluxes, see Section~\ref{s:sources}.}
\label{t:SgrB2Mfluxes}
\begin{tabular}{l c c c c c c c c c c c c c c}
\hline\hline

\multicolumn{1}{c}{Frequency}
&
\\

\multicolumn{1}{c}{(MHz)}
&AM01
&AM02
&AM03
&AM04
&AM05
&AM06
&AM07
&AM08
&AM09
&AM10
\\
\hline
212000.00  &$ 4.23\pm0.09$ &$4.01\pm0.10$ &$1.12\pm0.06$ &$0.75\pm0.03$ &$0.78\pm0.06$ &$0.56\pm0.04$ &$0.54\pm0.03$ &$0.31\pm0.02$ &$0.45\pm0.04$ &$0.79\pm0.05$ \\
213875.00  &$ 4.43\pm0.07$ &$4.34\pm0.08$ &$1.22\pm0.04$ &$0.82\pm0.03$ &$0.91\pm0.04$ &$0.66\pm0.03$ &$0.62\pm0.03$ &$0.33\pm0.01$ &$0.51\pm0.04$ &$0.86\pm0.05$ \\
215750.00  &$ 4.56\pm0.08$ &$4.59\pm0.10$ &$1.24\pm0.05$ &$0.88\pm0.03$ &$0.91\pm0.05$ &$0.77\pm0.04$ &$0.60\pm0.03$ &$0.43\pm0.02$ &$0.57\pm0.04$ &$1.02\pm0.05$ \\
217625.00  &$ 4.61\pm0.08$ &$4.71\pm0.09$ &$1.27\pm0.06$ &$0.92\pm0.03$ &$0.95\pm0.05$ &$0.82\pm0.04$ &$0.64\pm0.03$ &$0.43\pm0.02$ &$0.60\pm0.04$ &$0.99\pm0.05$ \\
219500.00  &$ 4.98\pm0.16$ &$4.96\pm0.15$ &$1.45\pm0.09$ &$1.00\pm0.04$ &$1.02\pm0.06$ &$0.80\pm0.04$ &$0.69\pm0.04$ &$0.50\pm0.02$ &$0.60\pm0.05$ &$1.05\pm0.06$ \\
221375.00  &$ 4.95\pm0.11$ &$4.88\pm0.11$ &$1.46\pm0.07$ &$1.02\pm0.03$ &$1.00\pm0.06$ &$0.79\pm0.04$ &$0.71\pm0.03$ &$0.47\pm0.02$ &$0.58\pm0.05$ &$0.95\pm0.06$ \\
227000.00  &$ 4.94\pm0.12$ &$5.25\pm0.13$ &$1.44\pm0.08$ &$0.93\pm0.04$ &$1.00\pm0.07$ &$0.78\pm0.05$ &$0.63\pm0.04$ &$0.40\pm0.02$ &$0.54\pm0.04$ &$0.95\pm0.06$ \\
228000.00  &$ 4.68\pm0.10$ &$4.58\pm0.11$ &$1.39\pm0.07$ &$0.89\pm0.03$ &$0.92\pm0.06$ &$0.63\pm0.04$ &$0.62\pm0.03$ &$0.29\pm0.01$ &$0.51\pm0.04$ &$0.81\pm0.05$ \\
228875.00  &$ 4.84\pm0.10$ &$5.18\pm0.12$ &$1.43\pm0.06$ &$0.90\pm0.03$ &$0.97\pm0.05$ &$0.76\pm0.04$ &$0.62\pm0.03$ &$0.39\pm0.02$ &$0.57\pm0.04$ &$0.89\pm0.05$ \\
229875.00  &$ 5.06\pm0.09$ &$5.07\pm0.10$ &$1.53\pm0.06$ &$1.03\pm0.03$ &$1.12\pm0.06$ &$0.75\pm0.04$ &$0.76\pm0.03$ &$0.32\pm0.02$ &$0.62\pm0.04$ &$1.03\pm0.06$ \\
231750.00  &$ 4.75\pm0.08$ &$4.84\pm0.10$ &$1.39\pm0.05$ &$0.95\pm0.03$ &$0.87\pm0.05$ &$0.70\pm0.04$ &$0.59\pm0.03$ &$0.33\pm0.02$ &$0.43\pm0.04$ &$0.76\pm0.05$ \\
233625.00  &$ 4.95\pm0.09$ &$5.17\pm0.11$ &$1.50\pm0.08$ &$1.00\pm0.04$ &$0.96\pm0.07$ &$0.81\pm0.04$ &$0.64\pm0.04$ &$0.34\pm0.02$ &$0.54\pm0.04$ &$0.85\pm0.06$ \\
235500.00  &$ 4.94\pm0.08$ &$4.96\pm0.10$ &$1.48\pm0.08$ &$0.99\pm0.04$ &$0.92\pm0.06$ &$0.70\pm0.04$ &$0.61\pm0.04$ &$0.39\pm0.02$ &$0.49\pm0.04$ &$0.78\pm0.05$ \\
237375.00  &$ 5.11\pm0.15$ &$5.11\pm0.15$ &$1.54\pm0.10$ &$1.03\pm0.04$ &$0.92\pm0.07$ &$0.72\pm0.04$ &$0.63\pm0.04$ &$0.38\pm0.02$ &$0.50\pm0.05$ &$0.78\pm0.06$ \\
241077.50  &$ 6.07\pm0.13$ &$5.53\pm0.14$ &$1.89\pm0.09$ &$1.32\pm0.05$ &$1.26\pm0.08$ &$0.94\pm0.05$ &$0.93\pm0.04$ &$0.40\pm0.02$ &$0.89\pm0.06$ &$1.04\pm0.07$ \\
242952.50  &$ 5.96\pm0.10$ &$5.46\pm0.12$ &$1.85\pm0.09$ &$1.28\pm0.05$ &$1.22\pm0.07$ &$0.88\pm0.04$ &$0.87\pm0.04$ &$0.39\pm0.02$ &$0.85\pm0.05$ &$1.06\pm0.06$ \\
243000.00  &$ 5.23\pm0.09$ &$5.78\pm0.12$ &$1.67\pm0.08$ &$1.04\pm0.04$ &$1.08\pm0.07$ &$0.89\pm0.05$ &$0.68\pm0.04$ &$0.35\pm0.02$ &$0.58\pm0.05$ &$0.90\pm0.06$ \\
244827.50  &$ 6.06\pm0.11$ &$5.95\pm0.12$ &$1.95\pm0.08$ &$1.33\pm0.04$ &$1.36\pm0.08$ &$1.09\pm0.05$ &$0.94\pm0.04$ &$0.47\pm0.02$ &$0.86\pm0.06$ &$1.12\pm0.07$ \\
244875.00  &$ 5.28\pm0.10$ &$5.76\pm0.11$ &$1.67\pm0.08$ &$1.04\pm0.04$ &$1.06\pm0.07$ &$0.87\pm0.04$ &$0.67\pm0.04$ &$0.38\pm0.02$ &$0.59\pm0.05$ &$0.91\pm0.06$ \\
246702.50  &$ 5.94\pm0.15$ &$5.80\pm0.15$ &$1.88\pm0.09$ &$1.25\pm0.04$ &$1.27\pm0.08$ &$0.99\pm0.05$ &$0.85\pm0.04$ &$0.43\pm0.02$ &$0.78\pm0.05$ &$1.04\pm0.07$ \\
248577.50  &$ 6.03\pm0.16$ &$5.96\pm0.16$ &$1.94\pm0.09$ &$1.33\pm0.05$ &$1.41\pm0.07$ &$1.18\pm0.05$ &$0.82\pm0.05$ &$0.38\pm0.03$ &$0.46\pm0.07$ &$1.02\pm0.08$ \\
250452.50  &$ 6.20\pm0.15$ &$6.13\pm0.16$ &$2.08\pm0.10$ &$1.39\pm0.06$ &$1.55\pm0.09$ &$1.23\pm0.06$ &$0.88\pm0.05$ &$0.39\pm0.02$ &$0.49\pm0.08$ &$0.94\pm0.09$ \\
256077.50  &$ 6.26\pm0.27$ &$7.46\pm0.29$ &$2.24\pm0.13$ &$1.52\pm0.06$ &$1.58\pm0.10$ &$1.51\pm0.08$ &$1.05\pm0.06$ &$0.51\pm0.03$ &$0.93\pm0.07$ &$1.38\pm0.09$ \\
257077.50  &$ 6.66\pm0.24$ &$6.16\pm0.22$ &$2.15\pm0.13$ &$1.42\pm0.06$ &$1.33\pm0.09$ &$0.93\pm0.05$ &$0.92\pm0.05$ &$0.38\pm0.02$ &$0.83\pm0.06$ &$1.08\pm0.07$ \\
257952.50  &$ 6.57\pm0.21$ &$7.80\pm0.25$ &$2.39\pm0.12$ &$1.62\pm0.06$ &$1.69\pm0.10$ &$1.59\pm0.07$ &$1.12\pm0.06$ &$0.54\pm0.03$ &$0.96\pm0.07$ &$1.45\pm0.10$ \\
258952.50  &$ 6.73\pm0.16$ &$6.24\pm0.15$ &$2.21\pm0.11$ &$1.48\pm0.06$ &$1.38\pm0.10$ &$0.96\pm0.06$ &$0.96\pm0.05$ &$0.39\pm0.02$ &$0.85\pm0.06$ &$1.12\pm0.08$ \\
260827.50  &$ 6.49\pm0.14$ &$6.47\pm0.16$ &$2.16\pm0.12$ &$1.38\pm0.06$ &$1.37\pm0.10$ &$0.98\pm0.06$ &$0.89\pm0.06$ &$0.40\pm0.02$ &$0.74\pm0.06$ &$0.99\pm0.09$ \\
262702.50  &$ 6.76\pm0.13$ &$6.79\pm0.15$ &$2.33\pm0.12$ &$1.51\pm0.06$ &$1.58\pm0.11$ &$1.16\pm0.06$ &$1.03\pm0.06$ &$0.44\pm0.02$ &$0.91\pm0.07$ &$1.25\pm0.09$ \\
264577.50  &$ 6.56\pm0.15$ &$6.29\pm0.16$ &$2.25\pm0.10$ &$1.61\pm0.05$ &$1.35\pm0.08$ &$0.86\pm0.06$ &$1.06\pm0.05$ &$0.42\pm0.02$ &$1.01\pm0.07$ &$0.94\pm0.07$ \\
266452.50  &$ 6.88\pm0.22$ &$6.66\pm0.23$ &$2.36\pm0.11$ &$1.71\pm0.06$ &$1.48\pm0.08$ &$0.90\pm0.07$ &$1.19\pm0.06$ &$0.45\pm0.03$ &$1.10\pm0.08$ &$1.03\pm0.09$ \\
272077.50  &$ 6.45\pm0.17$ &$7.87\pm0.21$ &$2.38\pm0.12$ &$1.52\pm0.06$ &$1.52\pm0.10$ &$1.34\pm0.09$ &$0.93\pm0.07$ &$0.42\pm0.03$ &$0.77\pm0.07$ &$1.16\pm0.10$ \\
273952.50  &$ 6.64\pm0.26$ &$8.11\pm0.30$ &$2.48\pm0.15$ &$1.52\pm0.08$ &$1.55\pm0.12$ &$1.42\pm0.11$ &$0.93\pm0.08$ &$0.39\pm0.04$ &$0.73\pm0.09$ &$1.14\pm0.13$ \\
\hline
\multicolumn{8}{l}{\phn Average maps centered at 220~GHz, 235~GHz, 250~GHz and 265~GHz} \\
\hline
219500.00  &$ 4.69\pm0.10$ &$4.74\pm0.11$ &$1.33\pm0.07$ &$0.90\pm0.03$ &$0.94\pm0.05$ &$0.74\pm0.04$ &$0.63\pm0.03$ &$0.41\pm0.02$ &$0.55\pm0.04$ &$0.94\pm0.06$ \\
235500.00  &$ 5.00\pm0.10$ &$5.16\pm0.11$ &$1.52\pm0.08$ &$1.00\pm0.04$ &$0.98\pm0.06$ &$0.76\pm0.04$ &$0.65\pm0.04$ &$0.35\pm0.02$ &$0.53\pm0.04$ &$0.85\pm0.06$ \\
248577.50  &$ 6.14\pm0.16$ &$6.26\pm0.17$ &$2.03\pm0.10$ &$1.38\pm0.05$ &$1.42\pm0.08$ &$1.18\pm0.06$ &$0.93\pm0.05$ &$0.44\pm0.02$ &$0.78\pm0.06$ &$1.13\pm0.08$ \\
264577.50  &$ 6.65\pm0.18$ &$6.82\pm0.20$ &$2.29\pm0.12$ &$1.52\pm0.06$ &$1.44\pm0.10$ &$1.07\pm0.07$ &$0.99\pm0.06$ &$0.41\pm0.03$ &$0.87\pm0.07$ &$1.09\pm0.09$ \\
\hline
\multicolumn{8}{l}{\phn Average maps for the LF (227.5~GHz) and HF (256.0~GHz) spectral windows} \\
\hline
227500.00  &$ 4.85\pm0.10$ &$4.95\pm0.11$ &$1.43\pm0.07$ &$0.95\pm0.03$ &$0.96\pm0.06$ &$0.75\pm0.04$ &$0.64\pm0.03$ &$0.38\pm0.02$ &$0.54\pm0.04$ &$0.89\pm0.06$ \\
256577.50  &$ 6.39\pm0.17$ &$6.54\pm0.19$ &$2.16\pm0.11$ &$1.45\pm0.06$ &$1.43\pm0.09$ &$1.12\pm0.06$ &$0.96\pm0.05$ &$0.42\pm0.03$ &$0.82\pm0.07$ &$1.11\pm0.08$ \\
\hline
\multicolumn{8}{l}{\phn Average of all the continuum maps} \\
\hline
242038.75  &$ 5.62\pm0.14$ &$5.75\pm0.15$ &$1.79\pm0.09$ &$1.20\pm0.05$ &$1.20\pm0.08$ &$0.94\pm0.05$ &$0.80\pm0.04$ &$0.40\pm0.02$ &$0.68\pm0.05$ &$1.00\pm0.07$ \\
\hline
\end{tabular}
\end{sidewaystable*}
\begin{sidewaystable*}
\ContinuedFloat
\centering
\caption{Continued.}
\begin{tabular}{l c c c c c c c c c c c c c c}
\hline\hline

\multicolumn{1}{c}{Frequency}
&
\\

\multicolumn{1}{c}{(MHz)}
&AM11
&AM12
&AM13
&AM14
&AM15
&AM16
&AM17
&AM18
&AM19
&AM20
\\
\hline
212000.00  &$0.24\pm0.04$ &$0.26\pm0.03$ &$0.31\pm0.03$ &$0.16\pm0.02$ &$0.12\pm0.01$ &$0.45\pm0.05$ &$0.24\pm0.04$ &$0.07\pm0.01$ &$0.07\pm0.02$ &$0.13\pm0.02$ \\
213875.00  &$0.42\pm0.04$ &$0.34\pm0.03$ &$0.40\pm0.03$ &$0.15\pm0.02$ &$0.16\pm0.01$ &$0.51\pm0.05$ &$0.35\pm0.04$ &$0.10\pm0.02$ &$0.12\pm0.02$ &$0.21\pm0.02$ \\
215750.00  &$0.36\pm0.04$ &$0.38\pm0.03$ &$0.32\pm0.04$ &$0.18\pm0.02$ &$0.15\pm0.02$ &$0.41\pm0.05$ &$0.57\pm0.05$ &$0.10\pm0.02$ &$0.05\pm0.02$ &$0.22\pm0.02$ \\
217625.00  &$0.34\pm0.05$ &$0.43\pm0.03$ &$0.37\pm0.04$ &$0.22\pm0.02$ &$0.20\pm0.01$ &$0.42\pm0.05$ &$0.48\pm0.04$ &$0.12\pm0.02$ &$0.07\pm0.02$ &$0.28\pm0.02$ \\
219500.00  &$0.52\pm0.05$ &$0.55\pm0.04$ &$0.27\pm0.05$ &$0.29\pm0.02$ &$0.20\pm0.02$ &$0.71\pm0.06$ &$0.66\pm0.05$ &$0.12\pm0.02$ &$0.18\pm0.03$ &$0.20\pm0.02$ \\
221375.00  &$0.38\pm0.05$ &$0.58\pm0.03$ &$0.31\pm0.04$ &$0.31\pm0.02$ &$0.19\pm0.02$ &$0.60\pm0.06$ &$0.58\pm0.05$ &$0.04\pm0.02$ &$0.11\pm0.02$ &$0.19\pm0.02$ \\
227000.00  &$0.32\pm0.05$ &$0.49\pm0.03$ &$0.37\pm0.04$ &$0.30\pm0.02$ &$0.09\pm0.02$ &$0.63\pm0.05$ &$0.31\pm0.04$ &$0.07\pm0.02$ &$0.09\pm0.02$ &$0.01\pm0.02$ \\
228000.00  &$0.38\pm0.04$ &$0.26\pm0.03$ &$0.40\pm0.03$ &$0.17\pm0.02$ &$0.13\pm0.01$ &$0.23\pm0.04$ &$0.15\pm0.03$ &$0.10\pm0.01$ &$0.02\pm0.02$ &$0.15\pm0.01$ \\
228875.00  &$0.33\pm0.04$ &$0.49\pm0.03$ &$0.33\pm0.03$ &$0.30\pm0.02$ &$0.09\pm0.01$ &$0.61\pm0.04$ &$0.31\pm0.04$ &$0.11\pm0.01$ &$0.08\pm0.02$ &$0.03\pm0.02$ \\
229875.00  &$0.62\pm0.05$ &$0.33\pm0.03$ &$0.49\pm0.04$ &$0.19\pm0.02$ &$0.15\pm0.02$ &$0.35\pm0.05$ &$0.32\pm0.05$ &$0.12\pm0.02$ &$0.01\pm0.02$ &$0.21\pm0.02$ \\
231750.00  &$0.39\pm0.04$ &$0.31\pm0.03$ &$0.41\pm0.04$ &$0.17\pm0.02$ &$0.14\pm0.02$ &$0.23\pm0.05$ &$0.13\pm0.05$ &$0.10\pm0.02$ &$0.03\pm0.02$ &$0.12\pm0.02$ \\
233625.00  &$0.42\pm0.04$ &$0.34\pm0.03$ &$0.45\pm0.04$ &$0.18\pm0.02$ &$0.17\pm0.01$ &$0.24\pm0.04$ &$0.18\pm0.04$ &$0.10\pm0.01$ &$0.02\pm0.02$ &$0.10\pm0.02$ \\
235500.00  &$0.41\pm0.04$ &$0.41\pm0.03$ &$0.38\pm0.03$ &$0.23\pm0.02$ &$0.15\pm0.01$ &$0.46\pm0.04$ &$0.34\pm0.04$ &$0.09\pm0.01$ &$0.06\pm0.02$ &$0.13\pm0.02$ \\
237375.00  &$0.34\pm0.04$ &$0.39\pm0.03$ &$0.34\pm0.04$ &$0.24\pm0.02$ &$0.17\pm0.02$ &$0.42\pm0.05$ &$0.40\pm0.04$ &$0.07\pm0.01$ &$0.05\pm0.02$ &$0.13\pm0.02$ \\
241077.50  &$0.40\pm0.06$ &$0.56\pm0.04$ &$0.71\pm0.04$ &$0.32\pm0.03$ &$0.15\pm0.02$ &$0.54\pm0.07$ &$0.35\pm0.06$ &$0.03\pm0.02$ &$0.30\pm0.03$ &$0.14\pm0.03$ \\
242952.50  &$0.40\pm0.05$ &$0.54\pm0.04$ &$0.66\pm0.04$ &$0.33\pm0.02$ &$0.15\pm0.02$ &$0.58\pm0.05$ &$0.37\pm0.05$ &$0.00\pm0.02$ &$0.27\pm0.02$ &$0.13\pm0.02$ \\
243000.00  &$0.29\pm0.04$ &$0.49\pm0.04$ &$0.34\pm0.04$ &$0.30\pm0.02$ &$0.07\pm0.01$ &$0.47\pm0.05$ &$0.32\pm0.04$ &$0.04\pm0.02$ &$0.10\pm0.02$ &$0.09\pm0.02$ \\
244827.50  &$0.71\pm0.06$ &$0.64\pm0.04$ &$0.72\pm0.04$ &$0.35\pm0.02$ &$0.22\pm0.02$ &$0.53\pm0.07$ &$0.21\pm0.06$ &$0.16\pm0.02$ &$0.28\pm0.03$ &$0.21\pm0.02$ \\
244875.00  &$0.23\pm0.04$ &$0.52\pm0.04$ &$0.37\pm0.04$ &$0.31\pm0.02$ &$0.07\pm0.01$ &$0.46\pm0.05$ &$0.33\pm0.04$ &$0.02\pm0.02$ &$0.07\pm0.02$ &$0.07\pm0.02$ \\
246702.50  &$0.56\pm0.05$ &$0.61\pm0.04$ &$0.64\pm0.04$ &$0.34\pm0.02$ &$0.21\pm0.02$ &$0.54\pm0.05$ &$0.26\pm0.05$ &$0.10\pm0.02$ &$0.26\pm0.02$ &$0.15\pm0.02$ \\
248577.50  &$1.24\pm0.07$ &$0.76\pm0.05$ &$0.82\pm0.05$ &$0.31\pm0.04$ &$0.30\pm0.02$ &$0.36\pm0.08$ &$0.95\pm0.08$ &$0.53\pm0.02$ &$0.63\pm0.03$ &$0.24\pm0.03$ \\
250452.50  &$1.30\pm0.07$ &$0.78\pm0.06$ &$0.70\pm0.06$ &$0.22\pm0.03$ &$0.20\pm0.02$ &$0.19\pm0.08$ &$0.78\pm0.07$ &$0.53\pm0.03$ &$0.61\pm0.04$ &$0.27\pm0.03$ \\
256077.50  &$0.95\pm0.07$ &$0.83\pm0.06$ &$0.95\pm0.06$ &$0.47\pm0.03$ &$0.14\pm0.03$ &$0.66\pm0.09$ &$0.17\pm0.07$ &$0.26\pm0.03$ &$0.35\pm0.04$ &$0.22\pm0.03$ \\
257077.50  &$0.36\pm0.05$ &$0.51\pm0.04$ &$0.65\pm0.05$ &$0.31\pm0.02$ &$0.08\pm0.02$ &$0.39\pm0.06$ &$0.31\pm0.05$ &$0.10\pm0.02$ &$0.18\pm0.02$ &$0.17\pm0.02$ \\
257952.50  &$1.00\pm0.08$ &$0.87\pm0.05$ &$1.00\pm0.06$ &$0.50\pm0.03$ &$0.12\pm0.03$ &$0.61\pm0.09$ &$0.17\pm0.08$ &$0.28\pm0.03$ &$0.32\pm0.04$ &$0.22\pm0.03$ \\
258952.50  &$0.39\pm0.06$ &$0.49\pm0.04$ &$0.66\pm0.05$ &$0.31\pm0.03$ &$0.06\pm0.02$ &$0.35\pm0.07$ &$0.29\pm0.06$ &$0.12\pm0.02$ &$0.19\pm0.03$ &$0.19\pm0.03$ \\
260827.50  &$0.38\pm0.05$ &$0.49\pm0.05$ &$0.67\pm0.06$ &$0.31\pm0.03$ &$0.12\pm0.02$ &$0.38\pm0.06$ &$0.22\pm0.04$ &$0.05\pm0.02$ &$0.21\pm0.03$ &$0.14\pm0.02$ \\
262702.50  &$0.65\pm0.06$ &$0.65\pm0.05$ &$0.81\pm0.06$ &$0.40\pm0.03$ &$0.15\pm0.02$ &$0.41\pm0.08$ &$0.37\pm0.06$ &$0.11\pm0.02$ &$0.35\pm0.03$ &$0.26\pm0.03$ \\
264577.50  &$0.66\pm0.07$ &$0.77\pm0.05$ &$0.52\pm0.05$ &$0.43\pm0.03$ &$0.21\pm0.02$ &$0.88\pm0.07$ &$0.72\pm0.06$ &$0.14\pm0.03$ &$0.29\pm0.03$ &$0.12\pm0.03$ \\
266452.50  &$0.67\pm0.07$ &$0.87\pm0.06$ &$0.58\pm0.07$ &$0.47\pm0.04$ &$0.23\pm0.03$ &$0.93\pm0.10$ &$0.77\pm0.09$ &$0.16\pm0.03$ &$0.23\pm0.04$ &$0.09\pm0.04$ \\
272077.50  &$0.58\pm0.07$ &$0.70\pm0.06$ &$0.80\pm0.08$ &$0.37\pm0.04$ &$0.14\pm0.02$ &$0.34\pm0.07$ &$0.12\pm0.06$ &$0.18\pm0.02$ &$0.22\pm0.03$ &$0.16\pm0.03$ \\
273952.50  &$0.56\pm0.08$ &$0.73\pm0.08$ &$0.78\pm0.09$ &$0.36\pm0.04$ &$0.13\pm0.03$ &$0.29\pm0.09$ &$0.15\pm0.08$ &$0.18\pm0.03$ &$0.23\pm0.04$ &$0.14\pm0.03$ \\
\hline
\multicolumn{8}{l}{\phn Average maps centered at 220~GHz, 235~GHz, 250~GHz and 265~GHz} \\
\hline
219500.00  &$0.37\pm0.04$ &$0.44\pm0.03$ &$0.34\pm0.04$ &$0.24\pm0.02$ &$0.15\pm0.01$ &$0.54\pm0.05$ &$0.44\pm0.04$ &$0.09\pm0.02$ &$0.09\pm0.02$ &$0.16\pm0.02$ \\
235500.00  &$0.38\pm0.04$ &$0.38\pm0.03$ &$0.40\pm0.04$ &$0.22\pm0.02$ &$0.13\pm0.01$ &$0.36\pm0.04$ &$0.27\pm0.04$ &$0.08\pm0.01$ &$0.03\pm0.02$ &$0.12\pm0.02$ \\
248577.50  &$0.82\pm0.06$ &$0.70\pm0.05$ &$0.78\pm0.05$ &$0.36\pm0.03$ &$0.19\pm0.02$ &$0.50\pm0.07$ &$0.41\pm0.06$ &$0.24\pm0.02$ &$0.38\pm0.03$ &$0.20\pm0.03$ \\
264577.50  &$0.53\pm0.06$ &$0.65\pm0.05$ &$0.68\pm0.06$ &$0.37\pm0.03$ &$0.14\pm0.02$ &$0.49\pm0.07$ &$0.37\pm0.06$ &$0.13\pm0.02$ &$0.24\pm0.03$ &$0.16\pm0.03$ \\
\hline
\multicolumn{8}{l}{\phn Average maps for the LF (227.5~GHz) and HF (256.0~GHz) spectral windows} \\
\hline
227500.00  &$0.37\pm0.04$ &$0.41\pm0.03$ &$0.37\pm0.04$ &$0.23\pm0.02$ &$0.14\pm0.01$ &$0.45\pm0.05$ &$0.35\pm0.04$ &$0.08\pm0.02$ &$0.06\pm0.02$ &$0.14\pm0.02$ \\
256577.50  &$0.67\pm0.06$ &$0.68\pm0.05$ &$0.73\pm0.06$ &$0.36\pm0.03$ &$0.16\pm0.02$ &$0.50\pm0.07$ &$0.39\pm0.06$ &$0.18\pm0.02$ &$0.31\pm0.03$ &$0.18\pm0.03$ \\
\hline
\multicolumn{8}{l}{\phn Average of all the continuum maps} \\
\hline
242038.75  &$0.52\pm0.05$ &$0.54\pm0.04$ &$0.55\pm0.05$ &$0.30\pm0.02$ &$0.15\pm0.02$ &$0.47\pm0.06$ &$0.37\pm0.05$ &$0.13\pm0.02$ &$0.18\pm0.03$ &$0.16\pm0.02$ \\
\hline
\end{tabular}
\end{sidewaystable*}
\begin{sidewaystable*}
\ContinuedFloat
\centering
\caption{Continued.}
\begin{tabular}{l c c c c c c c c c c c c c c}
\hline\hline

\multicolumn{1}{c}{Frequency}
&
\\

\multicolumn{1}{c}{(MHz)}
&AM21
&AM22
&AM23
&AM24
&AM25
&AM26
&AM27
\\
\hline
212000.00  &$0.025\pm0.008$ &$0.039\pm0.010$ &$0.135\pm0.017$ &$0.050\pm0.005$ &$0.056\pm0.008$ &$0.085\pm0.016$ &$0.058\pm0.008$ \\
213875.00  &$0.022\pm0.008$ &$0.044\pm0.009$ &$0.198\pm0.019$ &$0.068\pm0.004$ &$0.074\pm0.008$ &$0.068\pm0.017$ &$0.070\pm0.009$ \\
215750.00  &$0.054\pm0.009$ &$0.051\pm0.011$ &$0.117\pm0.022$ &$0.077\pm0.005$ &$0.082\pm0.009$ &$0.026\pm0.019$ &$0.070\pm0.009$ \\
217625.00  &$0.036\pm0.009$ &$0.046\pm0.011$ &$0.170\pm0.020$ &$0.062\pm0.008$ &$0.075\pm0.015$ &$0.070\pm0.018$ &$0.071\pm0.009$ \\
219500.00  &$0.091\pm0.011$ &$0.063\pm0.012$ &$0.081\pm0.024$ &$0.059\pm0.006$ &$0.073\pm0.011$ &$0.028\pm0.022$ &$0.092\pm0.011$ \\
221375.00  &$0.086\pm0.010$ &$0.054\pm0.011$ &$0.108\pm0.021$ &$0.058\pm0.005$ &$0.032\pm0.010$ &$0.043\pm0.022$ &$0.101\pm0.010$ \\
227000.00  &$0.058\pm0.009$ &$0.116\pm0.013$ &$0.110\pm0.020$ &$0.025\pm0.006$ &$0.069\pm0.011$ &$0.100\pm0.021$ &$0.010\pm0.010$ \\
228000.00  &$0.054\pm0.006$ &$0.029\pm0.007$ &$0.097\pm0.013$ &$0.047\pm0.005$ &$0.023\pm0.007$ &$0.092\pm0.013$ &$0.035\pm0.006$ \\
228875.00  &$0.046\pm0.008$ &$0.094\pm0.009$ &$0.100\pm0.019$ &$0.032\pm0.004$ &$0.068\pm0.008$ &$0.128\pm0.015$ &$0.022\pm0.008$ \\
229875.00  &$0.064\pm0.011$ &$0.054\pm0.013$ &$0.107\pm0.021$ &$0.069\pm0.005$ &$0.060\pm0.009$ &$0.095\pm0.020$ &$0.061\pm0.010$ \\
231750.00  &$0.047\pm0.009$ &$0.028\pm0.011$ &$0.054\pm0.022$ &$0.045\pm0.005$ &$0.037\pm0.009$ &$0.086\pm0.019$ &$0.051\pm0.009$ \\
233625.00  &$0.062\pm0.008$ &$0.038\pm0.008$ &$0.075\pm0.015$ &$0.050\pm0.006$ &$0.043\pm0.008$ &$0.092\pm0.014$ &$0.057\pm0.008$ \\
235500.00  &$0.054\pm0.008$ &$0.021\pm0.009$ &$0.027\pm0.017$ &$0.027\pm0.005$ &$0.010\pm0.008$ &$0.097\pm0.015$ &$0.109\pm0.009$ \\
237375.00  &$0.056\pm0.009$ &$0.028\pm0.010$ &$0.018\pm0.019$ &$0.016\pm0.005$ &$0.041\pm0.008$ &$0.108\pm0.017$ &$0.119\pm0.011$ \\
241077.50  &$0.040\pm0.011$ &$0.174\pm0.014$ &$0.295\pm0.025$ &$0.018\pm0.007$ &$0.021\pm0.013$ &$0.042\pm0.022$ &$0.073\pm0.012$ \\
242952.50  &$0.049\pm0.010$ &$0.152\pm0.011$ &$0.256\pm0.022$ &$0.037\pm0.006$ &$0.014\pm0.012$ &$0.090\pm0.020$ &$0.048\pm0.011$ \\
243000.00  &$0.045\pm0.009$ &$0.044\pm0.010$ &$0.198\pm0.021$ &$0.032\pm0.005$ &$0.026\pm0.009$ &$0.155\pm0.018$ &$0.041\pm0.009$ \\
244827.50  &$0.042\pm0.011$ &$0.076\pm0.013$ &$0.143\pm0.026$ &$0.034\pm0.007$ &$0.066\pm0.012$ &$0.113\pm0.025$ &$0.036\pm0.013$ \\
244875.00  &$0.036\pm0.008$ &$0.052\pm0.009$ &$0.183\pm0.020$ &$0.033\pm0.005$ &$0.043\pm0.009$ &$0.158\pm0.016$ &$0.062\pm0.009$ \\
246702.50  &$0.042\pm0.011$ &$0.067\pm0.011$ &$0.127\pm0.022$ &$0.032\pm0.007$ &$0.070\pm0.010$ &$0.102\pm0.021$ &$0.011\pm0.011$ \\
248577.50  &$0.130\pm0.014$ &$0.067\pm0.015$ &$0.009\pm0.032$ &$0.021\pm0.008$ &$0.141\pm0.014$ &$0.109\pm0.029$ &$0.011\pm0.014$ \\
250452.50  &$0.096\pm0.013$ &$0.074\pm0.018$ &$0.017\pm0.032$ &$0.015\pm0.009$ &$0.146\pm0.015$ &$0.154\pm0.032$ &$0.040\pm0.015$ \\
256077.50  &$0.072\pm0.015$ &$0.077\pm0.017$ &$0.123\pm0.034$ &$0.074\pm0.010$ &$0.118\pm0.016$ &$0.201\pm0.031$ &$0.016\pm0.018$ \\
257077.50  &$0.092\pm0.009$ &$0.137\pm0.013$ &$0.265\pm0.026$ &$0.024\pm0.006$ &$0.009\pm0.011$ &$0.084\pm0.021$ &$0.044\pm0.011$ \\
257952.50  &$0.076\pm0.016$ &$0.077\pm0.018$ &$0.131\pm0.036$ &$0.073\pm0.010$ &$0.124\pm0.016$ &$0.204\pm0.032$ &$0.026\pm0.016$ \\
258952.50  &$0.107\pm0.011$ &$0.139\pm0.013$ &$0.251\pm0.027$ &$0.028\pm0.007$ &$0.018\pm0.013$ &$0.066\pm0.026$ &$0.039\pm0.012$ \\
260827.50  &$0.047\pm0.009$ &$0.036\pm0.010$ &$0.099\pm0.018$ &$0.020\pm0.005$ &$0.058\pm0.011$ &$0.089\pm0.017$ &$0.007\pm0.009$ \\
262702.50  &$0.060\pm0.012$ &$0.076\pm0.015$ &$0.197\pm0.030$ &$0.029\pm0.007$ &$0.104\pm0.014$ &$0.100\pm0.027$ &$0.003\pm0.014$ \\
264577.50  &$0.114\pm0.013$ &$0.173\pm0.015$ &$0.365\pm0.032$ &$0.037\pm0.007$ &$0.014\pm0.014$ &$0.300\pm0.027$ &$0.098\pm0.018$ \\
266452.50  &$0.140\pm0.017$ &$0.177\pm0.018$ &$0.379\pm0.041$ &$0.040\pm0.011$ &$0.000\pm0.017$ &$0.309\pm0.034$ &$0.105\pm0.019$ \\
272077.50  &$0.054\pm0.011$ &$0.016\pm0.014$ &$0.032\pm0.028$ &$0.035\pm0.006$ &$0.042\pm0.012$ &$0.151\pm0.022$ &$0.027\pm0.013$ \\
273952.50  &$0.057\pm0.014$ &$0.013\pm0.019$ &$0.021\pm0.037$ &$0.026\pm0.010$ &$0.032\pm0.016$ &$0.140\pm0.030$ &$0.016\pm0.017$ \\
\hline
\multicolumn{7}{l}{\phn Average maps centered at 220~GHz, 235~GHz, 250~GHz and 265~GHz} \\
\hline
219500.00  &$0.052\pm0.009$ &$0.063\pm0.011$ &$0.127\pm0.020$ &$0.054\pm0.006$ &$0.066\pm0.010$ &$0.069\pm0.019$ &$0.062\pm0.009$ \\
235500.00  &$0.052\pm0.008$ &$0.037\pm0.009$ &$0.095\pm0.018$ &$0.040\pm0.005$ &$0.033\pm0.008$ &$0.110\pm0.016$ &$0.067\pm0.009$ \\
248577.50  &$0.068\pm0.013$ &$0.096\pm0.015$ &$0.138\pm0.029$ &$0.038\pm0.008$ &$0.088\pm0.013$ &$0.127\pm0.027$ &$0.020\pm0.014$ \\
264577.50  &$0.084\pm0.012$ &$0.096\pm0.014$ &$0.201\pm0.030$ &$0.010\pm0.007$ &$0.031\pm0.013$ &$0.155\pm0.025$ &$0.042\pm0.014$ \\
\hline
\multicolumn{7}{l}{\phn Average maps for the LF (227.5~GHz) and HF (256.0~GHz) spectral windows} \\
\hline
227500.00  &$0.052\pm0.009$ &$0.050\pm0.010$ &$0.111\pm0.019$ &$0.047\pm0.005$ &$0.050\pm0.009$ &$0.089\pm0.017$ &$0.064\pm0.009$ \\
256577.50  &$0.076\pm0.012$ &$0.096\pm0.015$ &$0.169\pm0.029$ &$0.024\pm0.008$ &$0.059\pm0.013$ &$0.141\pm0.026$ &$0.031\pm0.014$ \\
\hline
\multicolumn{7}{l}{\phn Average of all the continuum maps} \\
\hline
242038.75  &$0.064\pm0.011$ &$0.073\pm0.012$ &$0.140\pm0.024$ &$0.036\pm0.007$ &$0.054\pm0.011$ &$0.115\pm0.022$ &$0.048\pm0.011$ \\
\hline
\end{tabular}
\end{sidewaystable*}

\clearpage
\begin{figure*}[ht]
\centering
\includegraphics[width=0.9\textwidth]{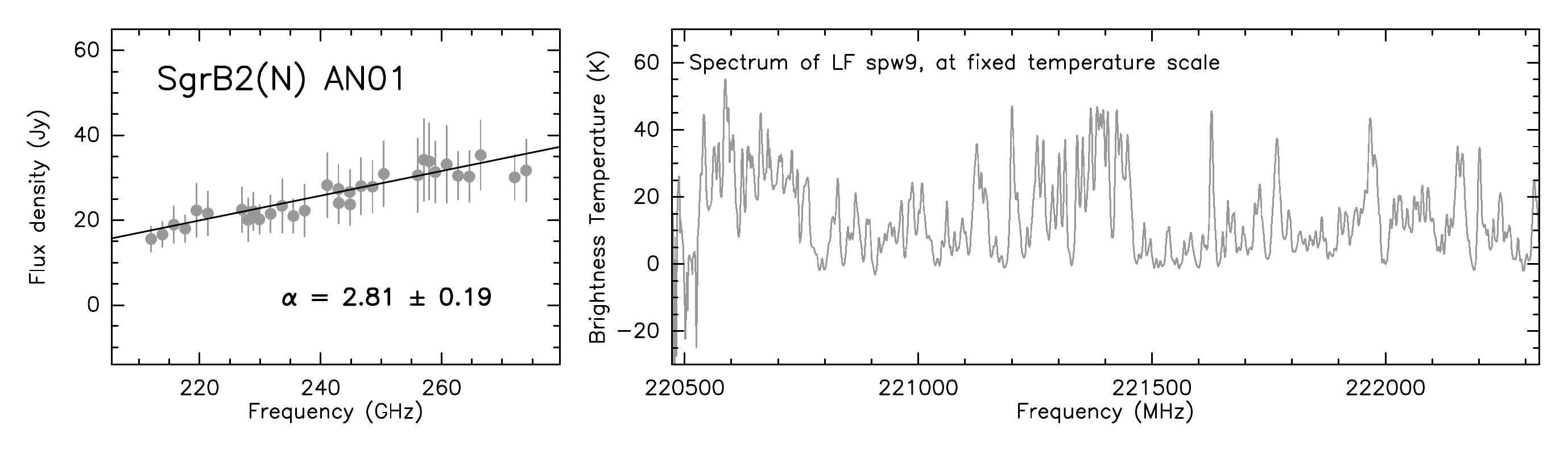} \\
\includegraphics[width=0.9\textwidth]{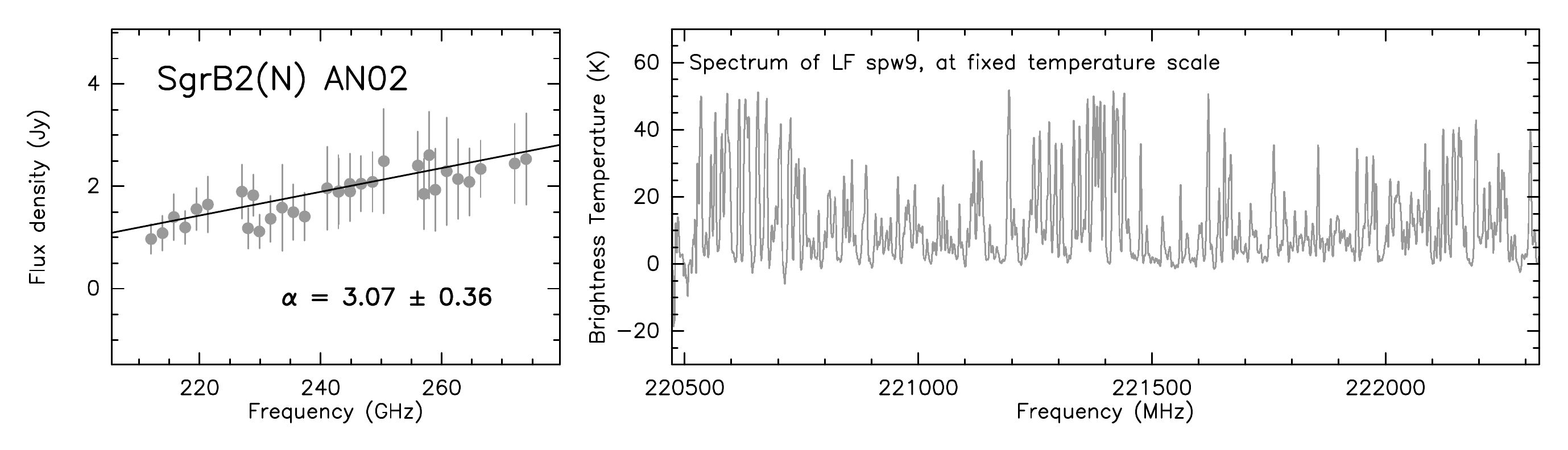} \\
\includegraphics[width=0.9\textwidth]{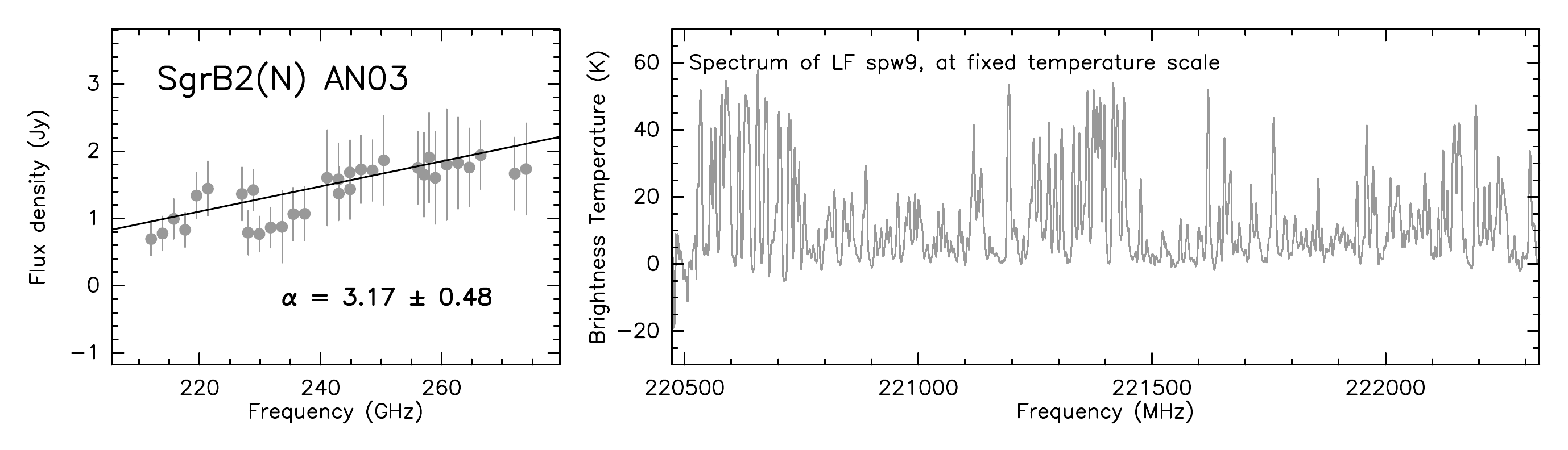} \\
\includegraphics[width=0.9\textwidth]{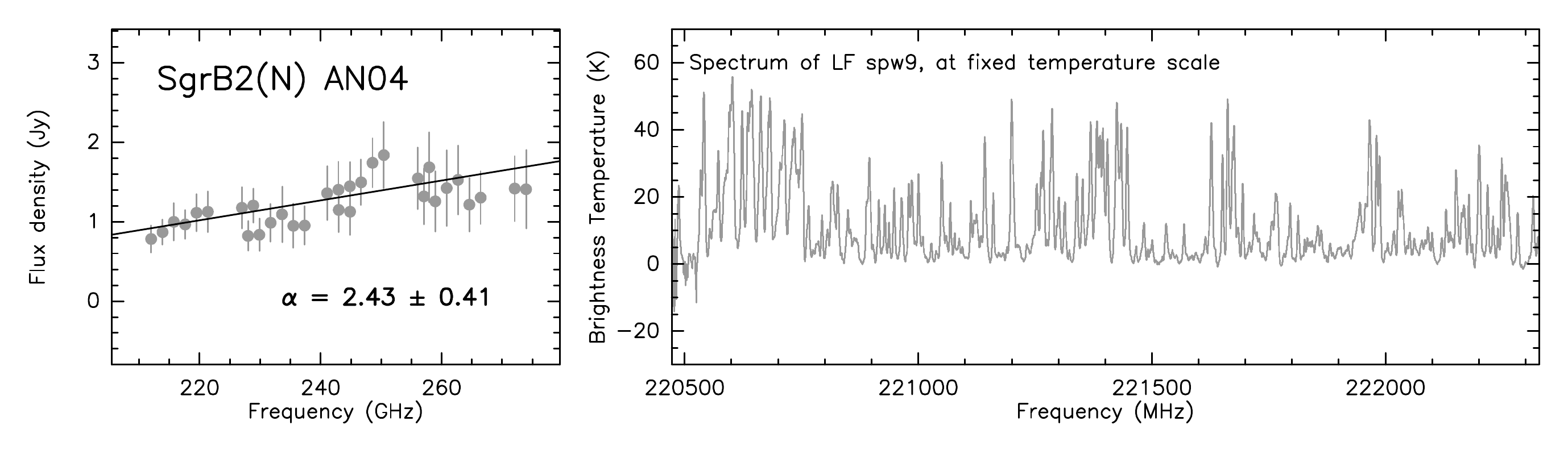} \\
\caption{\textit{Left}: \SgrB(N) spectral density distributions for the ALMA continuum sources shown in Fig.~\ref{f:SgrB2Nsources} and listed in Table~\ref{t:SgrB2Nsources}. Each gray dot corresponds to the integrated flux over the area of each source, when the flux is above the $3\sigma$ detection threshold. The fluxes at different frequencies are listed in Table~\ref{t:SgrB2Nfluxes}. The solid line is a linear fit to the data, with $\alpha$ corresponding to the spectral index ($S_\nu\propto\nu^\alpha$). \textit{Right}: Averaged spectrum over the 3$\sigma$-level polygon that defines the source, corresponding to the frequency range 211--213~GHz. Note that the intensity scale of the top panel is adjusted to better show the line features, while the scale in the bottom panel has been fixed for all the sources to show the relative brightness of the spectral lines between sources.}
\label{f:SgrB2NSEDs}
\end{figure*}
\begin{figure*}[ht]
\ContinuedFloat
\centering
\includegraphics[width=0.9\textwidth]{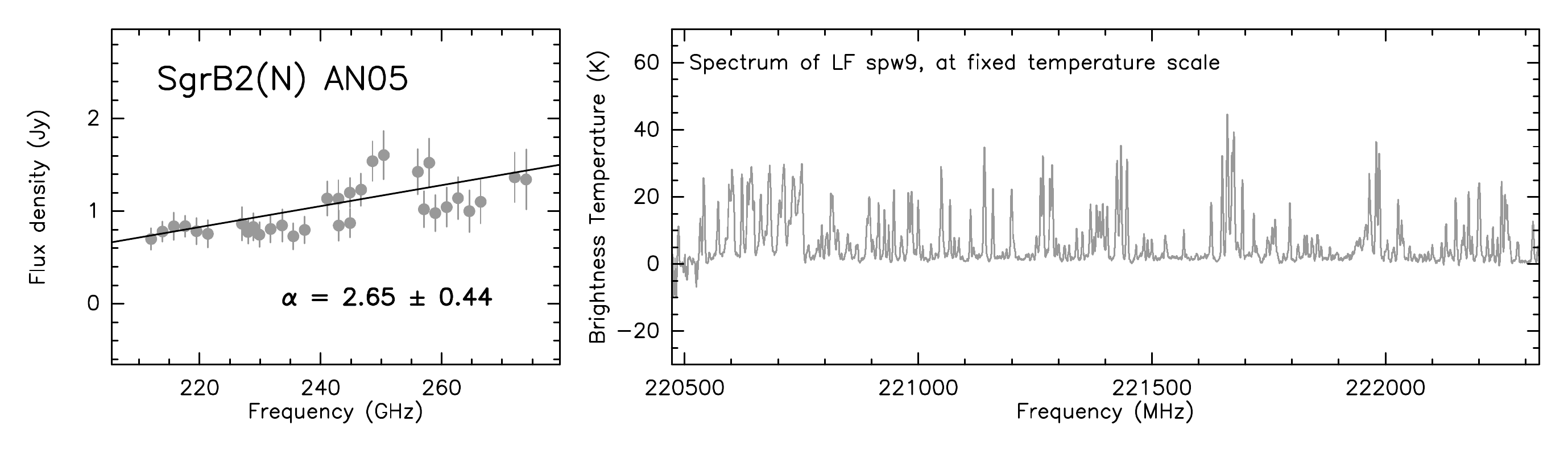} \\
\includegraphics[width=0.9\textwidth]{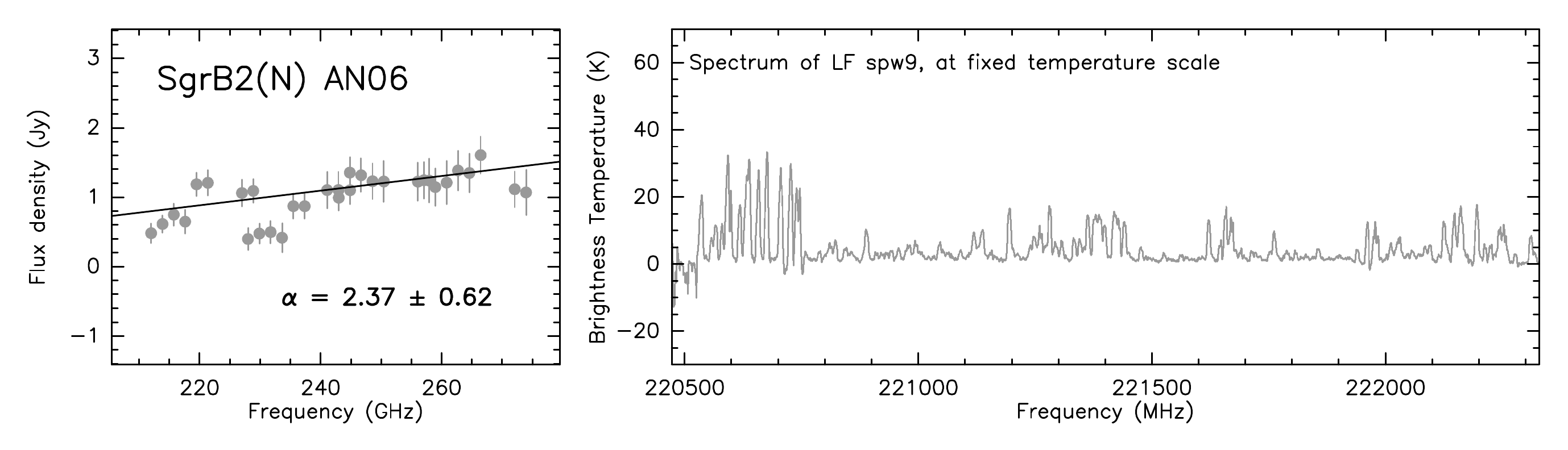} \\
\includegraphics[width=0.9\textwidth]{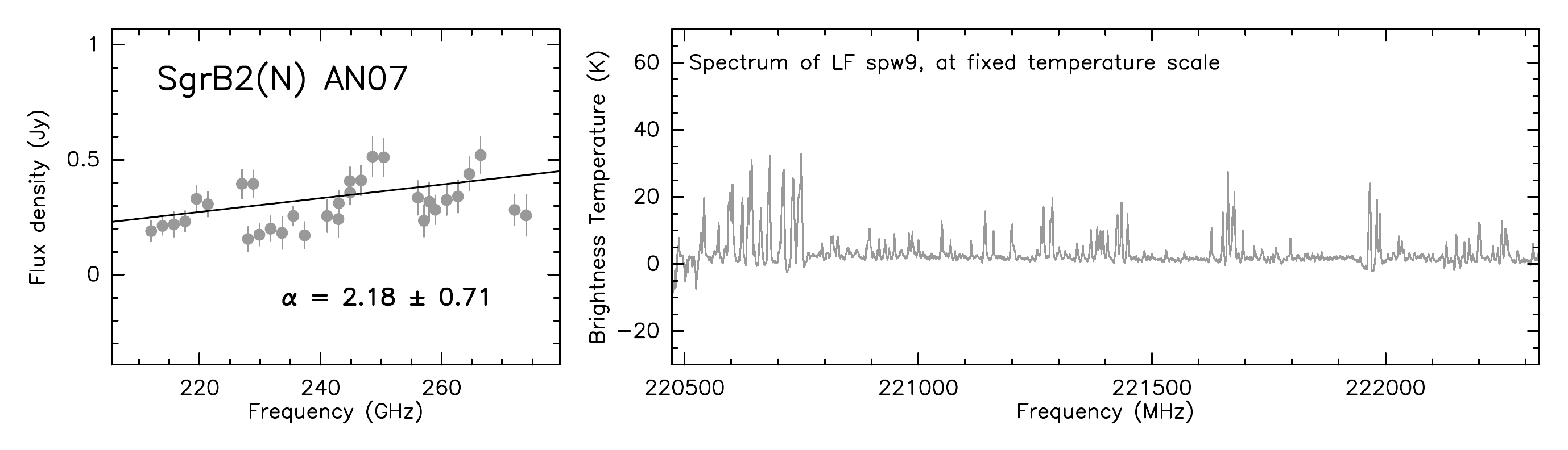} \\
\includegraphics[width=0.9\textwidth]{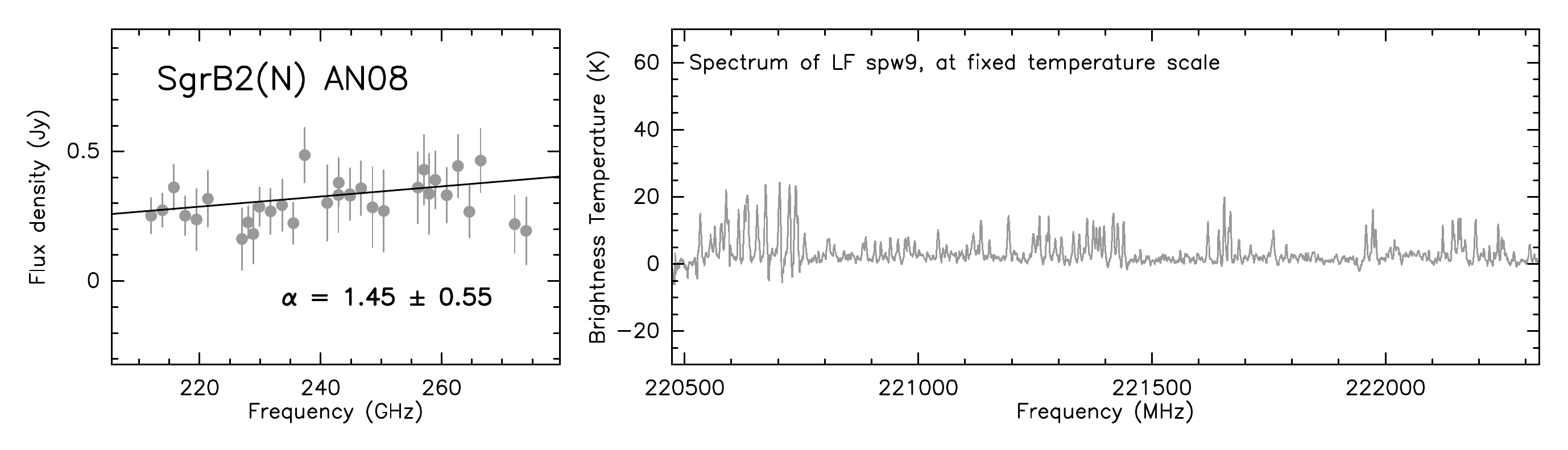} \\
\includegraphics[width=0.9\textwidth]{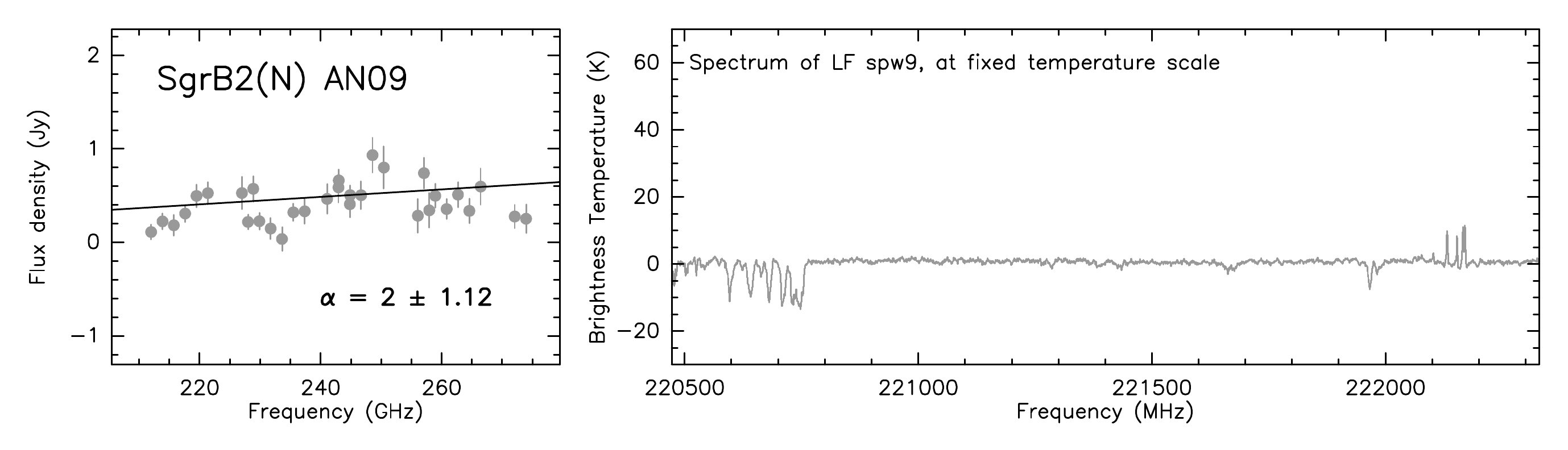} \\
\caption{\SgrB(N) mm-SEDs continued.}
\end{figure*}
\begin{figure*}[ht]
\ContinuedFloat
\centering
\includegraphics[width=0.9\textwidth]{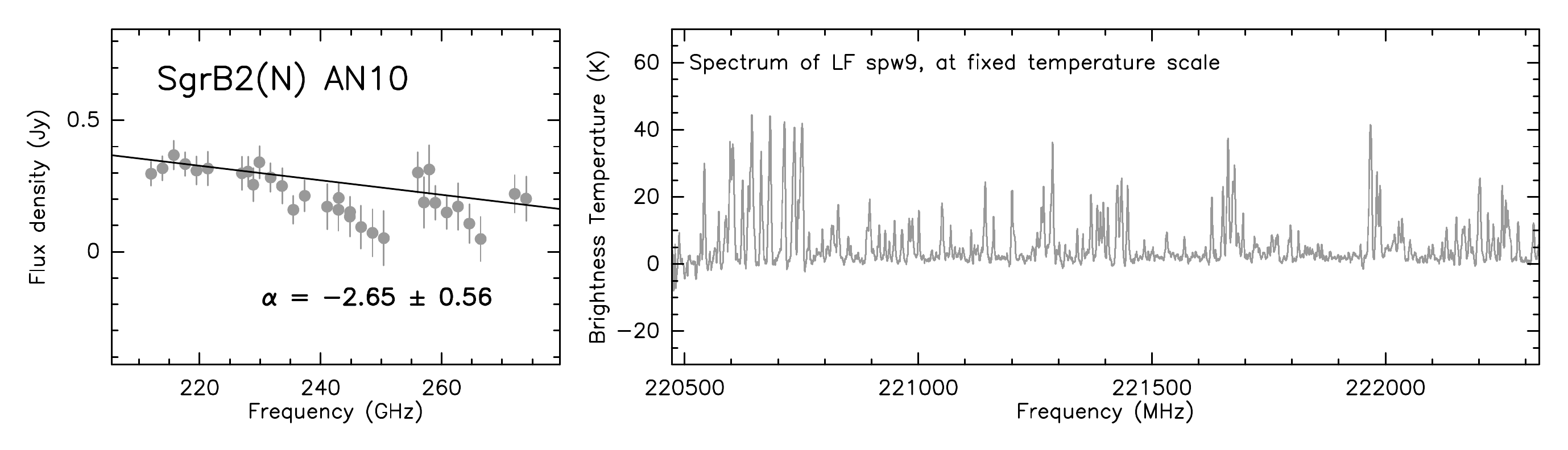} \\
\includegraphics[width=0.9\textwidth]{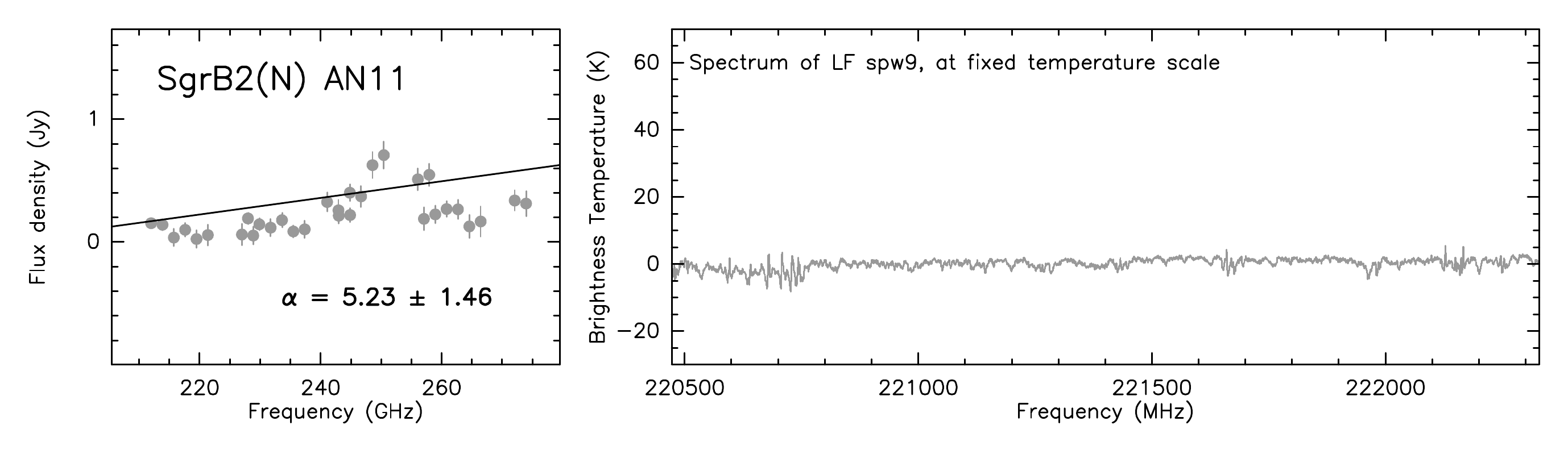} \\
\includegraphics[width=0.9\textwidth]{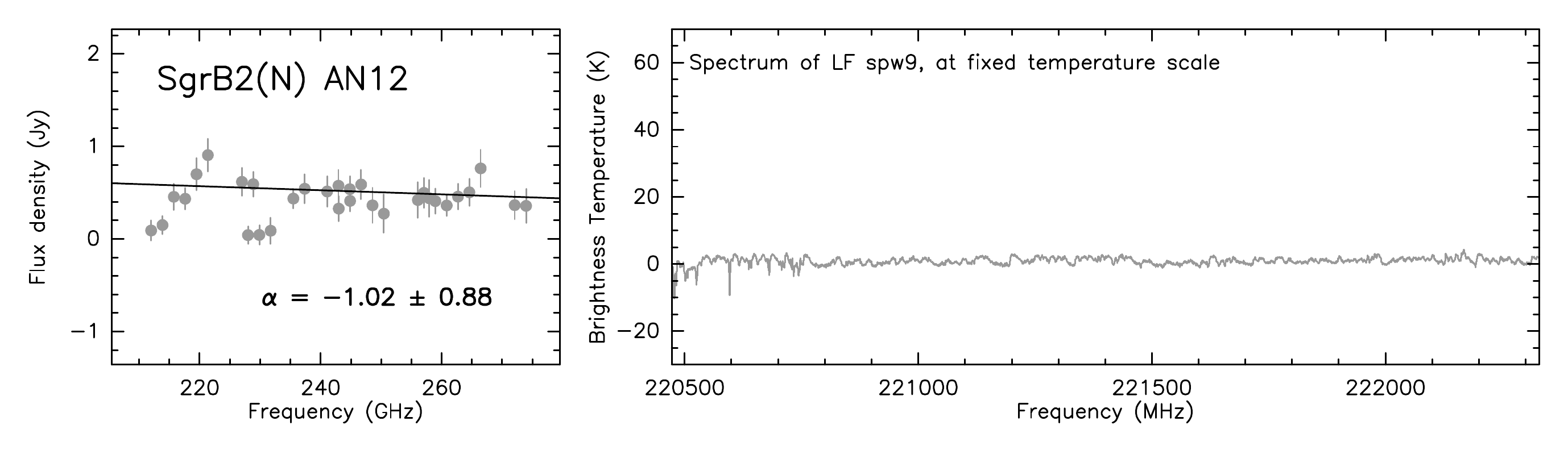} \\
\includegraphics[width=0.9\textwidth]{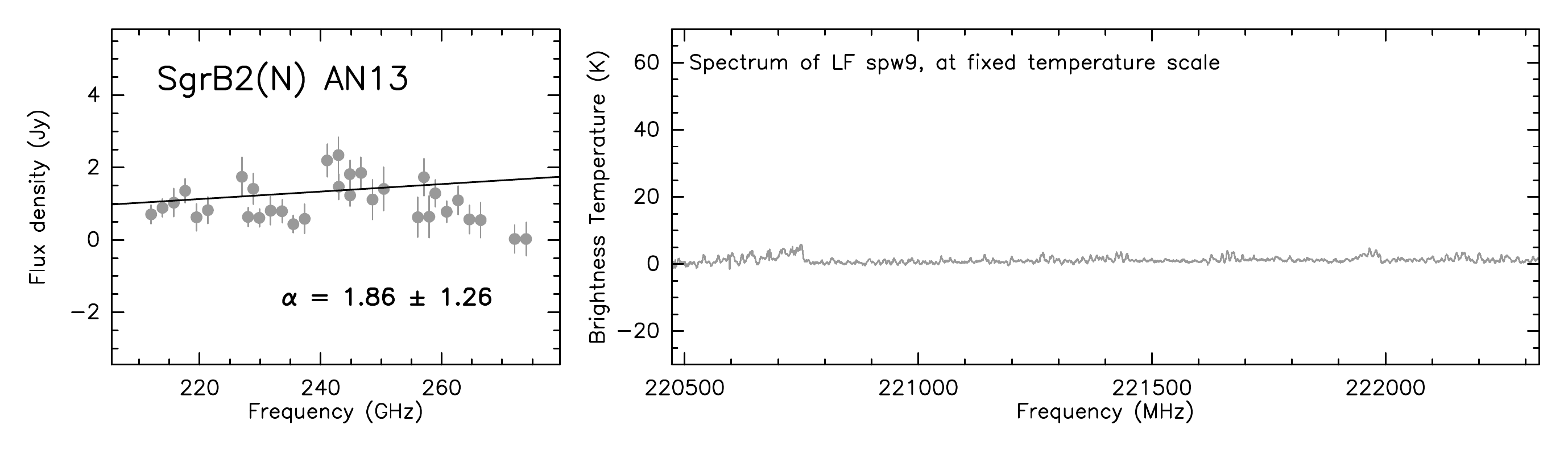} \\
\includegraphics[width=0.9\textwidth]{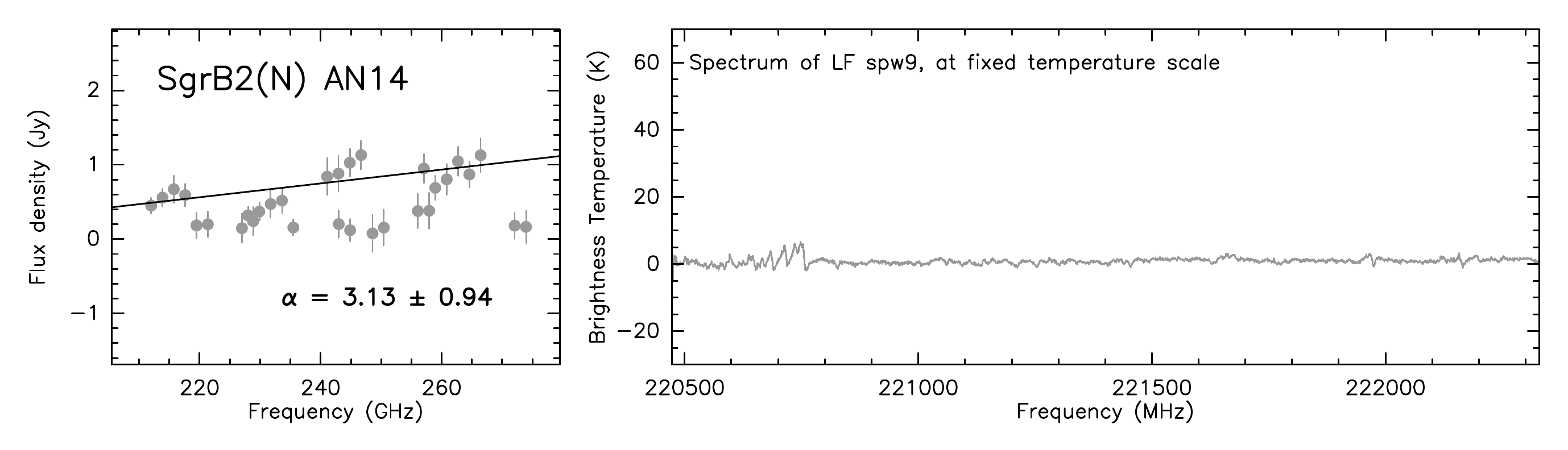} \\
\caption{\SgrB(N) mm-SEDs continued.}
\end{figure*}
\begin{figure*}[ht]
\ContinuedFloat
\centering
\includegraphics[width=0.9\textwidth]{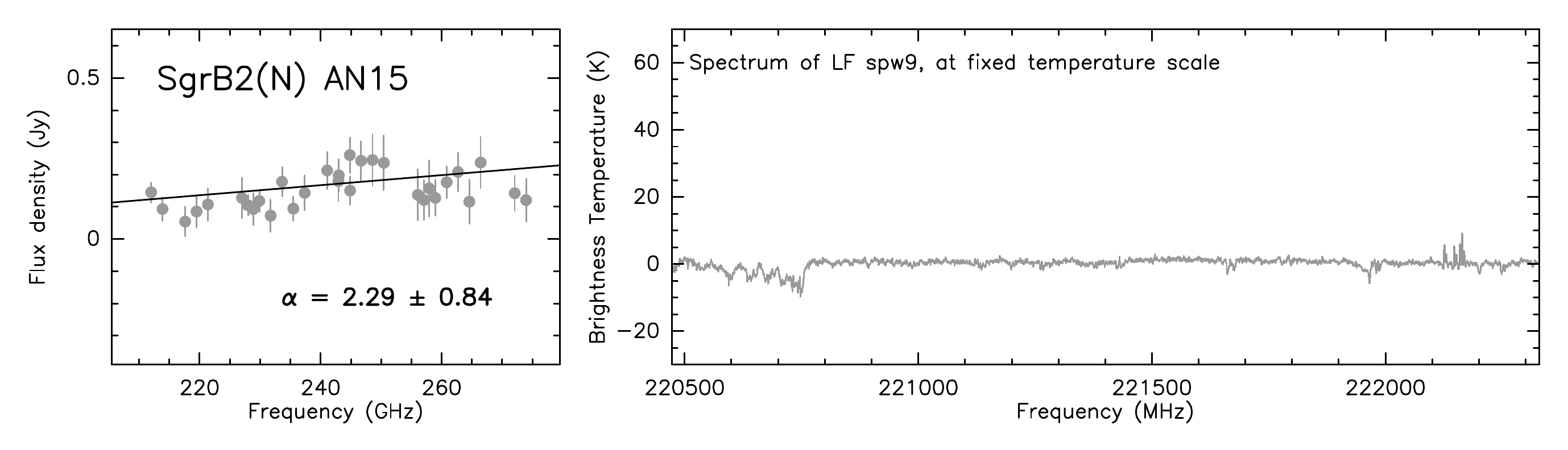} \\
\includegraphics[width=0.9\textwidth]{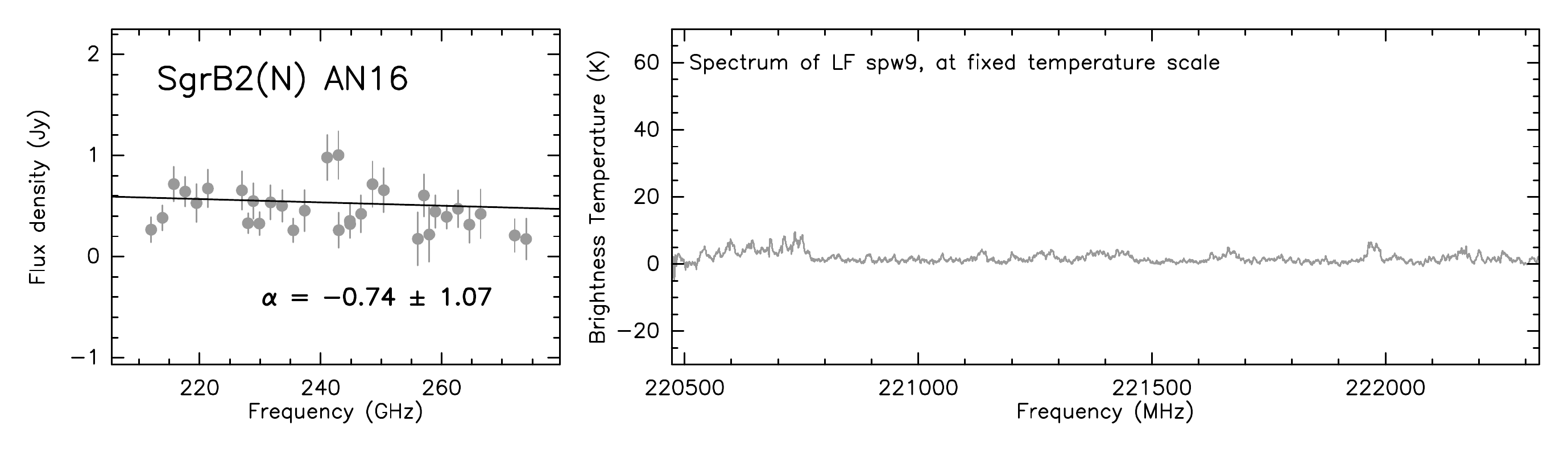} \\
\includegraphics[width=0.9\textwidth]{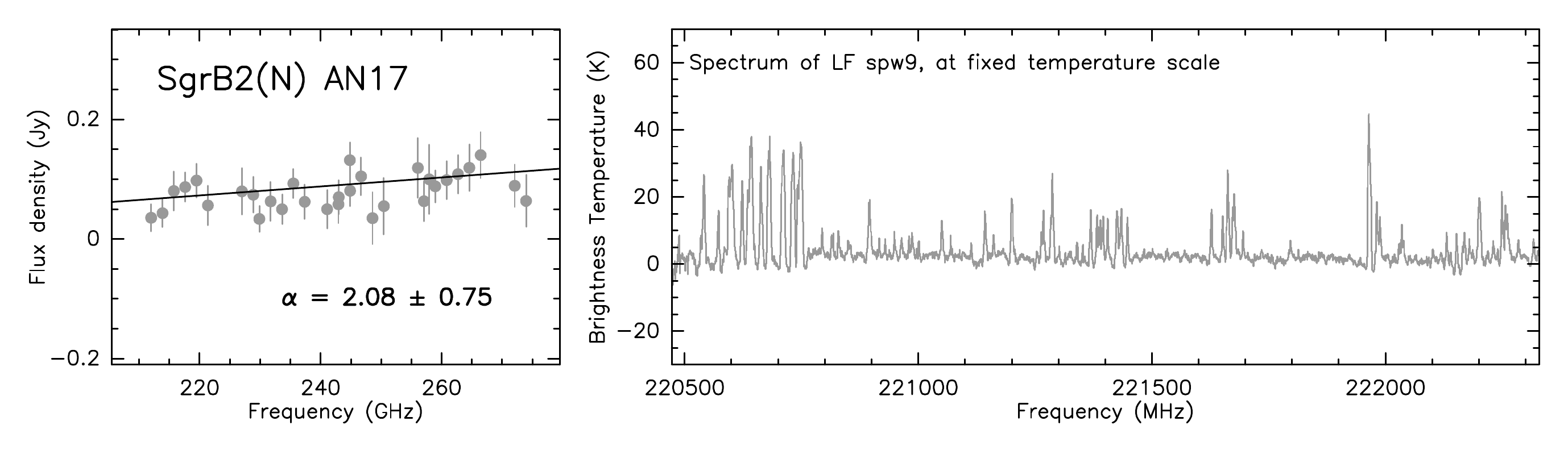} \\
\includegraphics[width=0.9\textwidth]{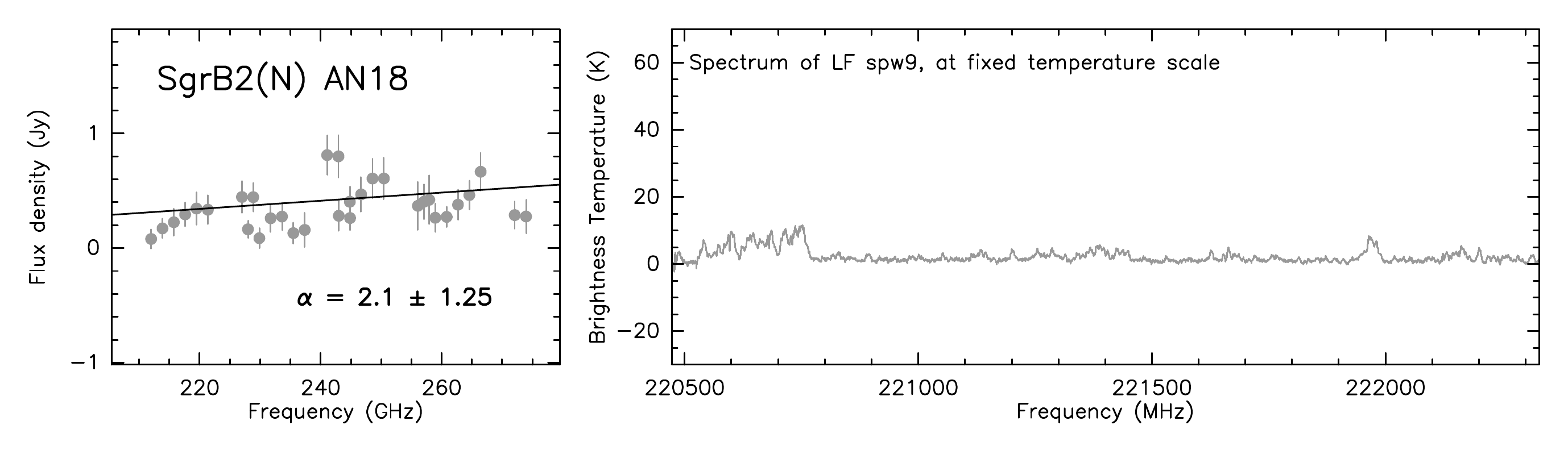} \\
\includegraphics[width=0.9\textwidth]{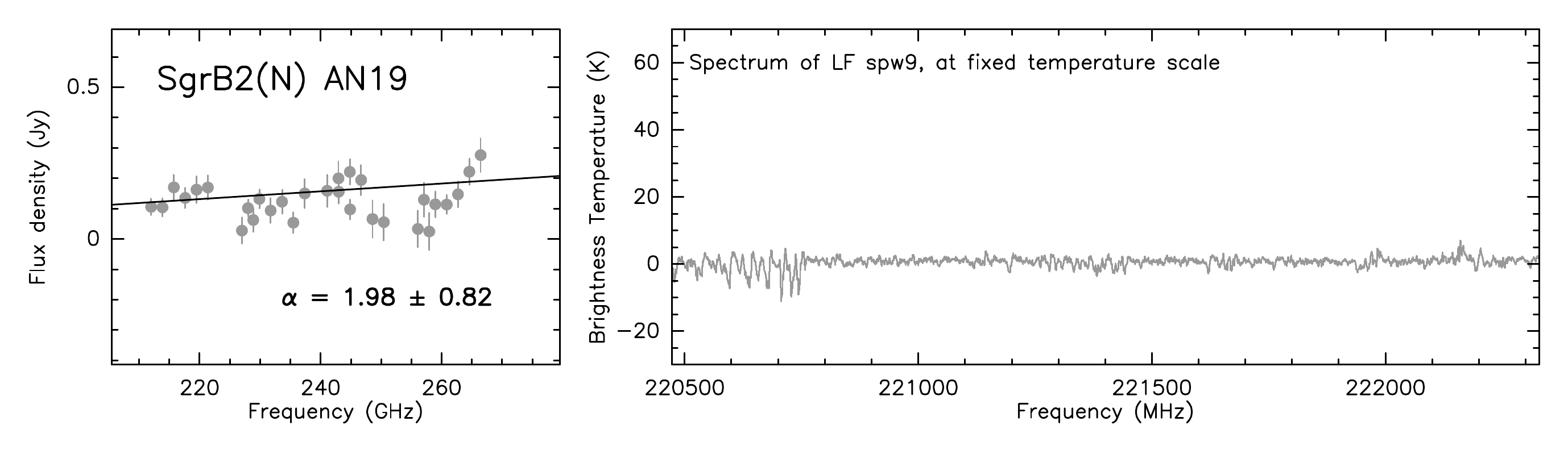} \\
\caption{\SgrB(N) mm-SEDs continued.}
\end{figure*}
\begin{figure*}[ht]
\ContinuedFloat
\centering
\includegraphics[width=0.9\textwidth]{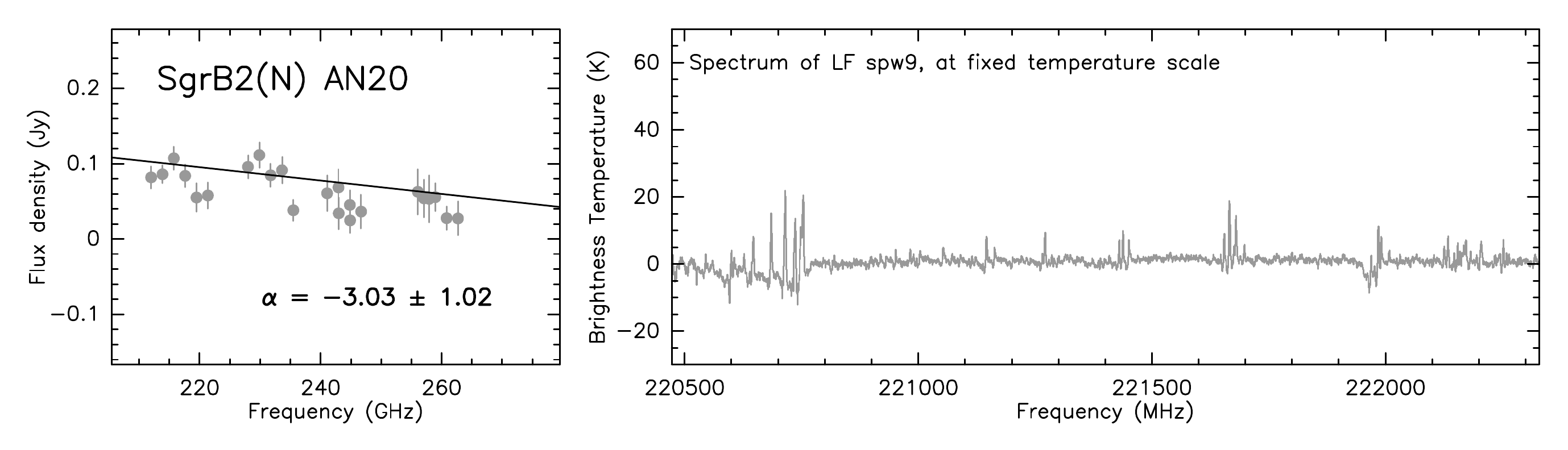} \\
\caption{\SgrB(N) mm-SEDs continued.}
\end{figure*}

\clearpage
\begin{figure*}[ht]
\centering
\includegraphics[width=0.9\textwidth]{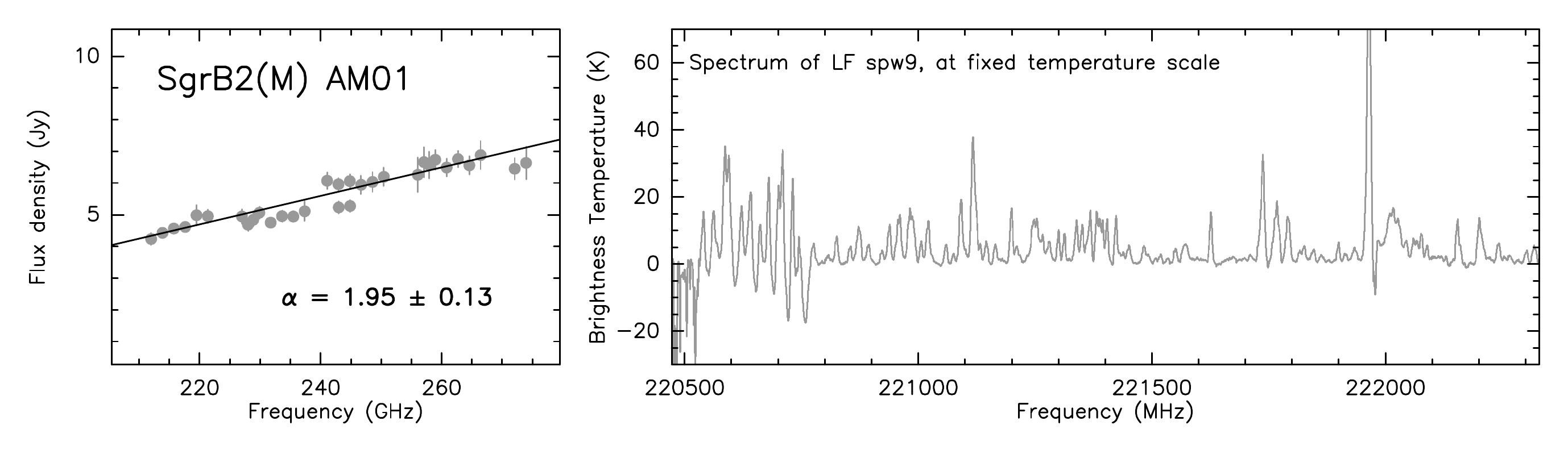} \\
\includegraphics[width=0.9\textwidth]{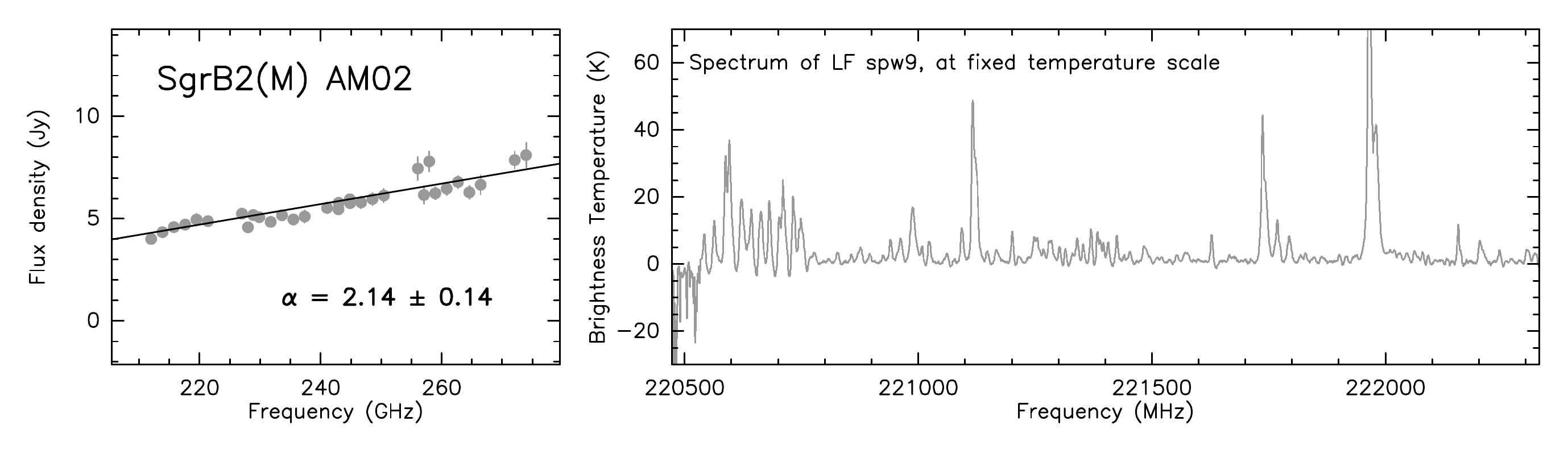} \\
\includegraphics[width=0.9\textwidth]{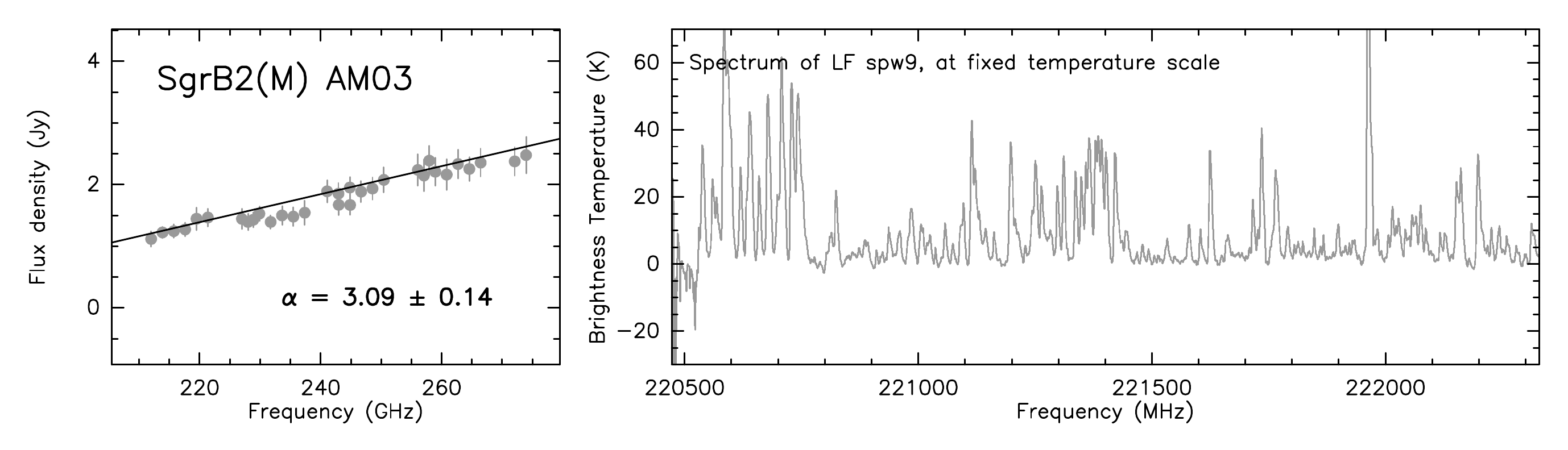} \\
\caption{\textit{Left}: \SgrB(M) spectral density distributions for the ALMA continuum sources shown in Fig.~\ref{f:SgrB2Msources} and listed in Table~\ref{t:SgrB2Nsources}. Each gray dot corresponds to the integrated flux over the area of each source, when the flux is above the $3\sigma$ detection threshold. The fluxes at different frequencies are listed in Table~\ref{t:SgrB2Mfluxes}. The solid line is a linear fit to the data, with $\alpha$ corresponding to the spectral index ($S_\nu\propto\nu^\alpha$). \textit{Right}: Averaged spectrum over the 3$\sigma$-level polygon that defines the source, corresponding to the frequency range 211--213~GHz. Note that the intensity scale of the top panel is adjusted to better show the line features, while the scale in the bottom panel has been fixed for all the sources to show the relative brightness of the spectral lines between sources.}
\label{f:SgrB2MSEDs}
\end{figure*}
\begin{figure*}[ht]
\ContinuedFloat
\centering
\includegraphics[width=0.9\textwidth]{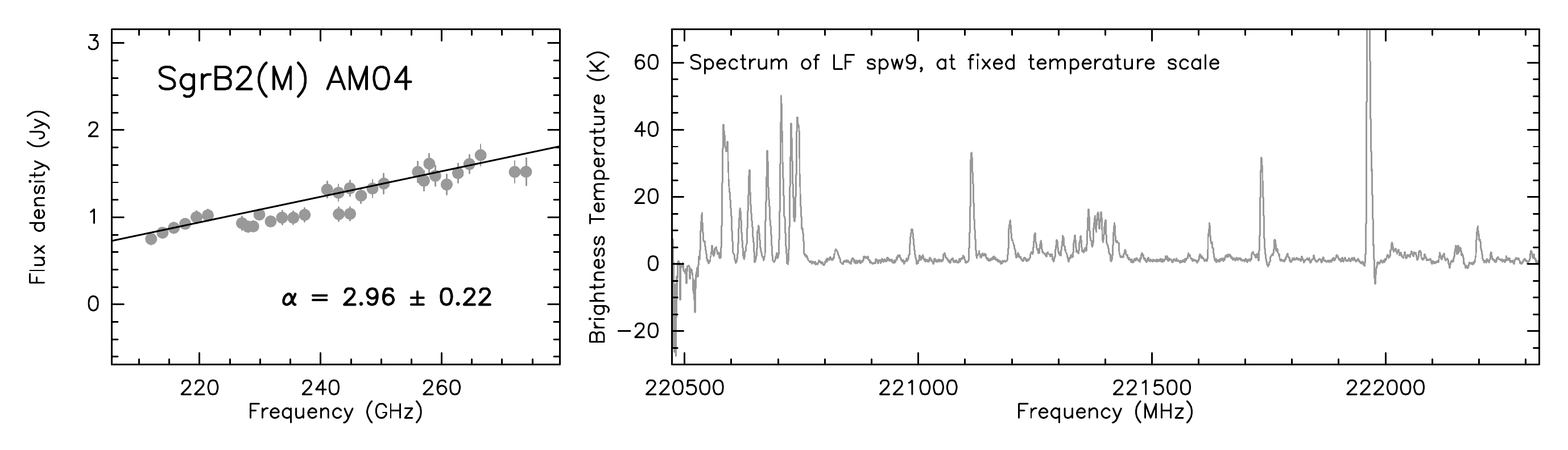} \\
\includegraphics[width=0.9\textwidth]{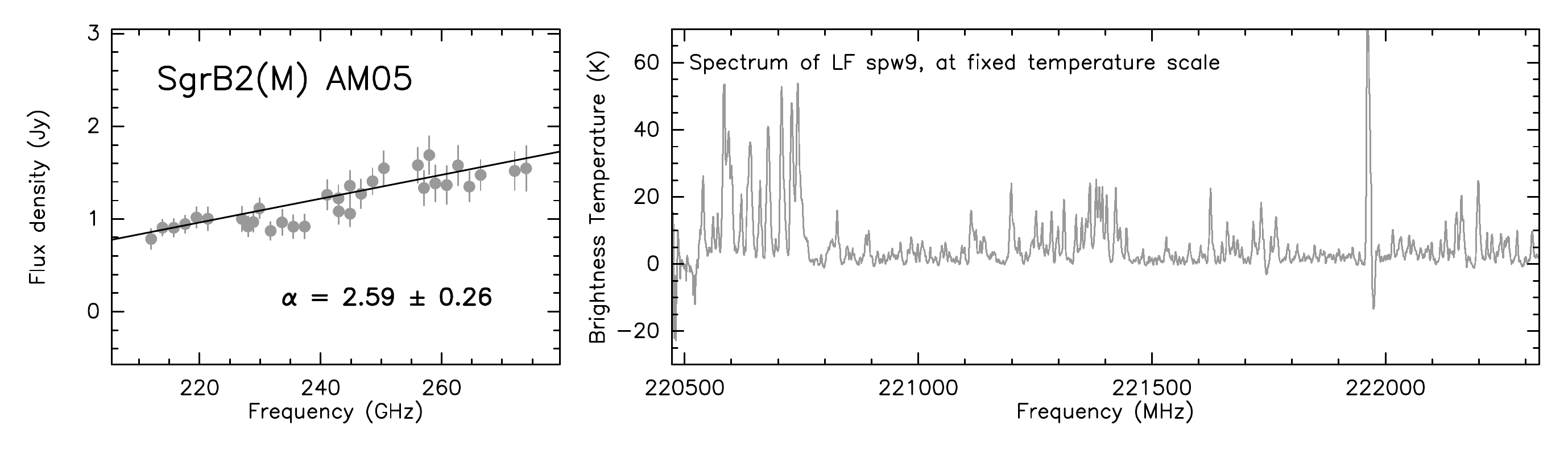} \\
\includegraphics[width=0.9\textwidth]{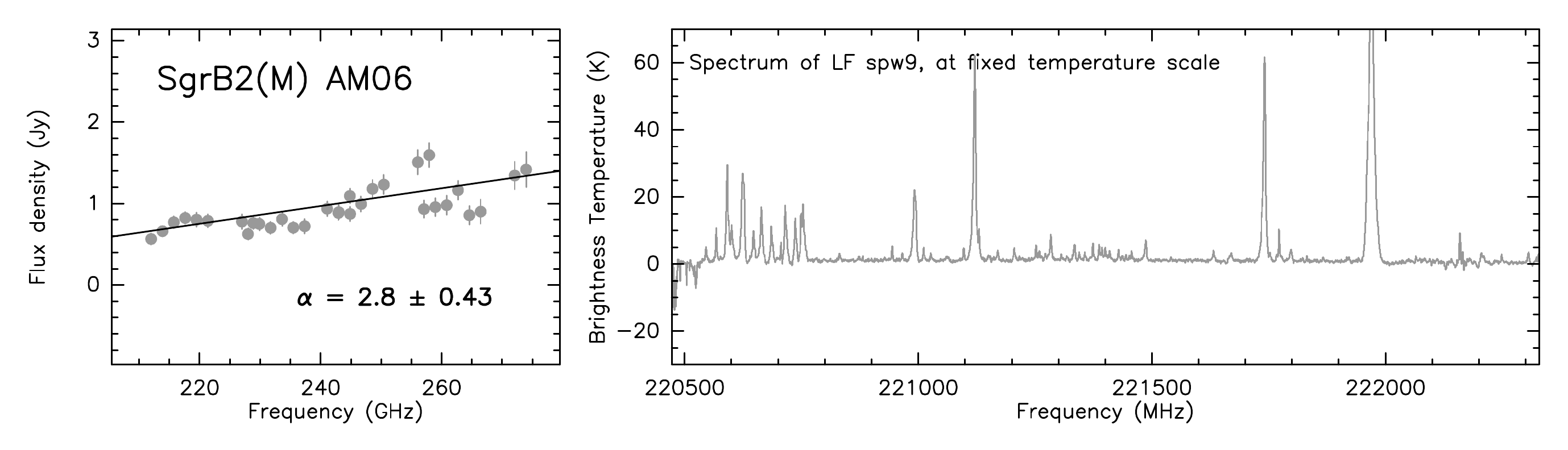} \\
\includegraphics[width=0.9\textwidth]{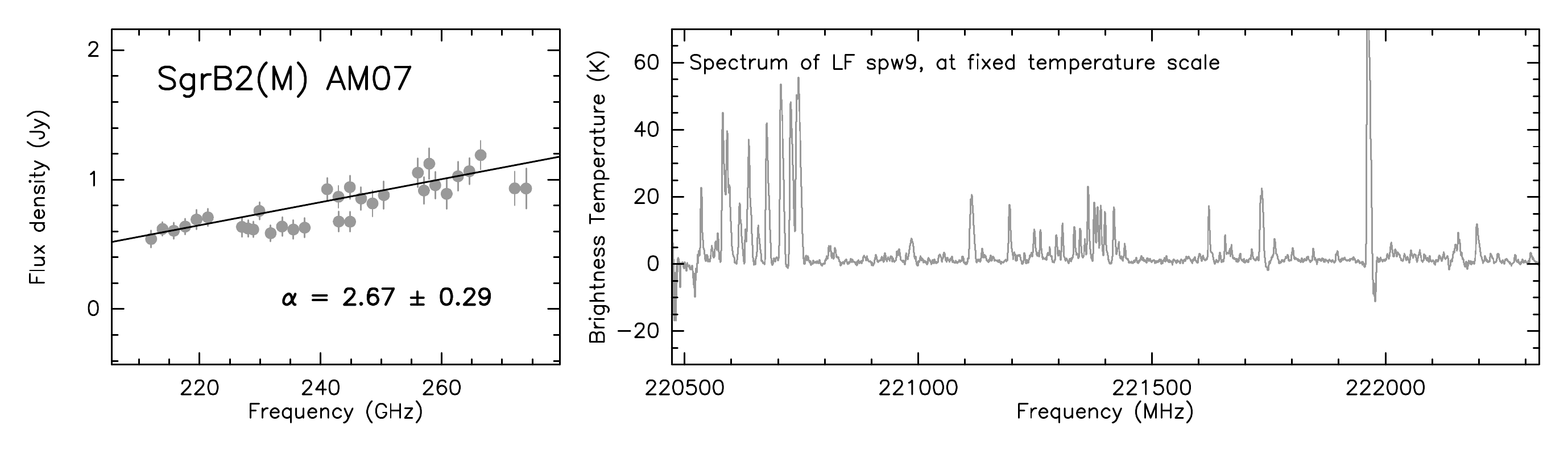} \\
\includegraphics[width=0.9\textwidth]{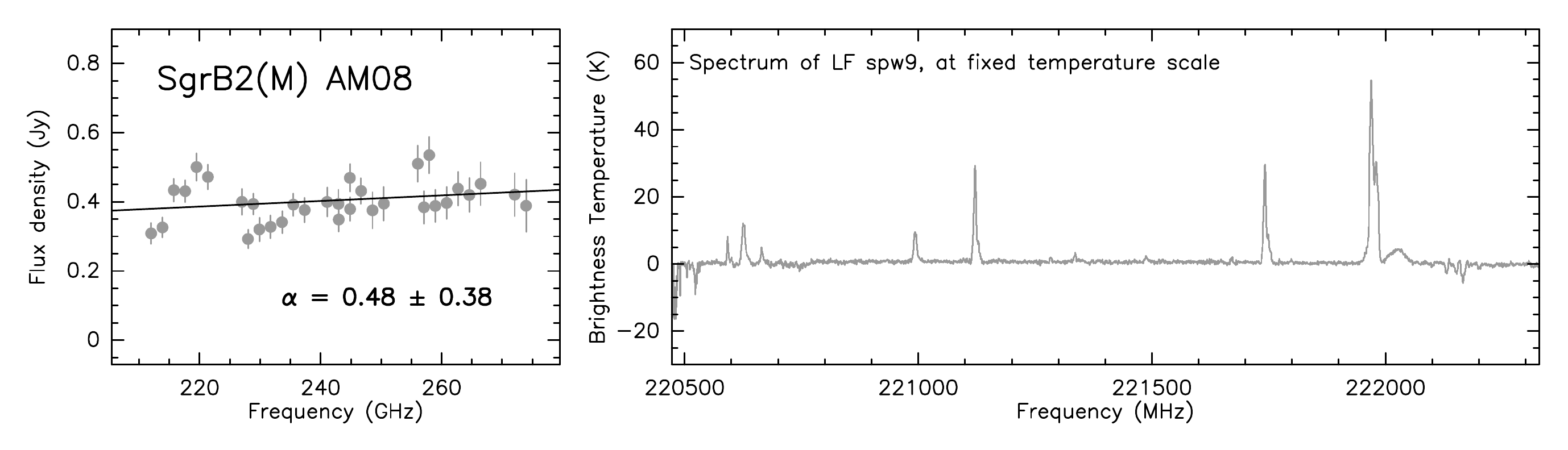} \\
\caption{\SgrB(M) mm-SEDs continued.}
\end{figure*}
\begin{figure*}[ht]
\ContinuedFloat
\centering
\includegraphics[width=0.9\textwidth]{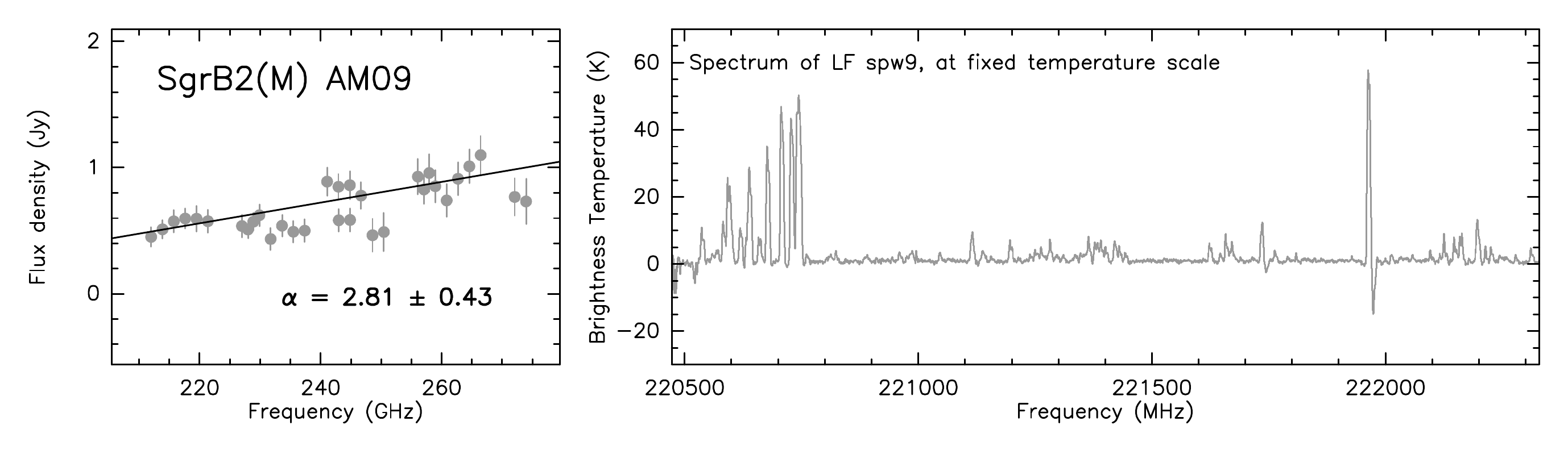} \\
\includegraphics[width=0.9\textwidth]{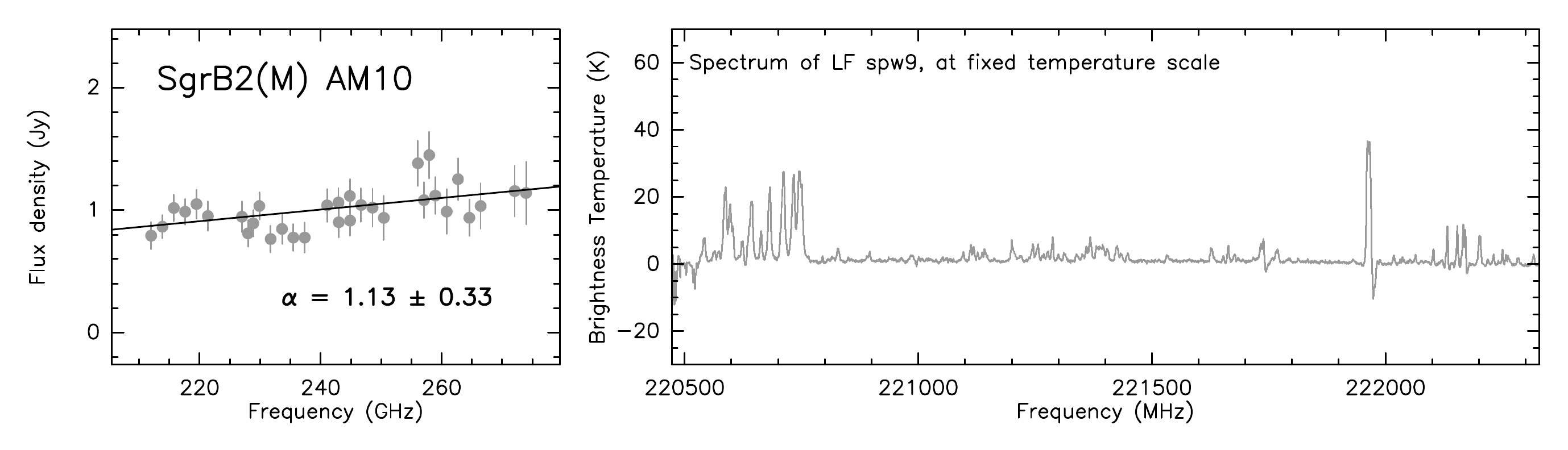} \\
\includegraphics[width=0.9\textwidth]{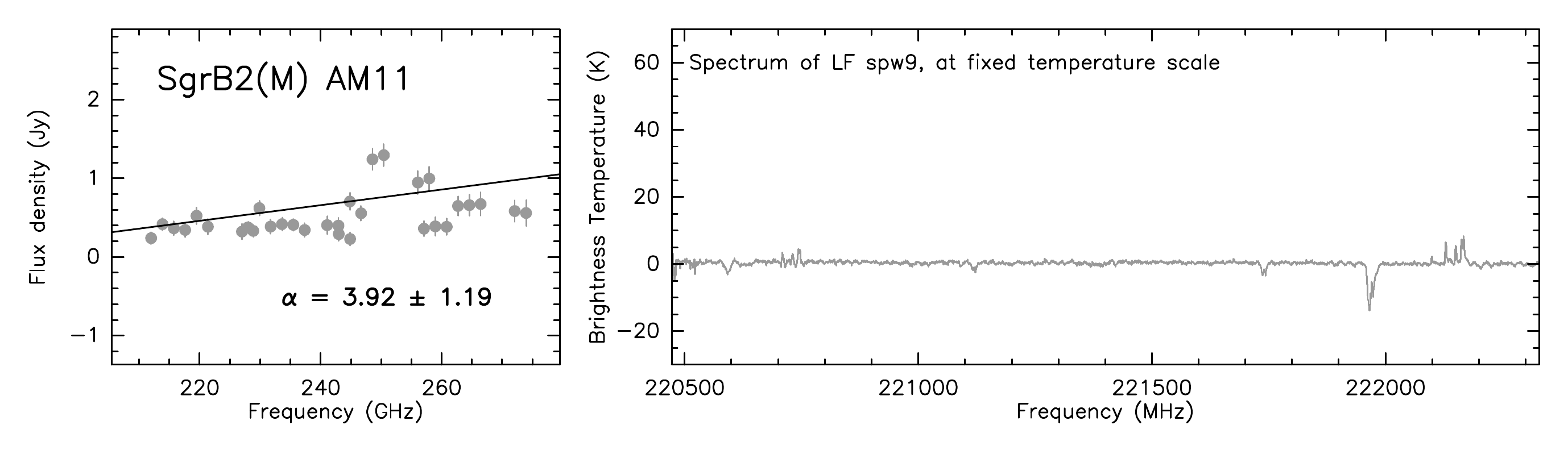} \\
\includegraphics[width=0.9\textwidth]{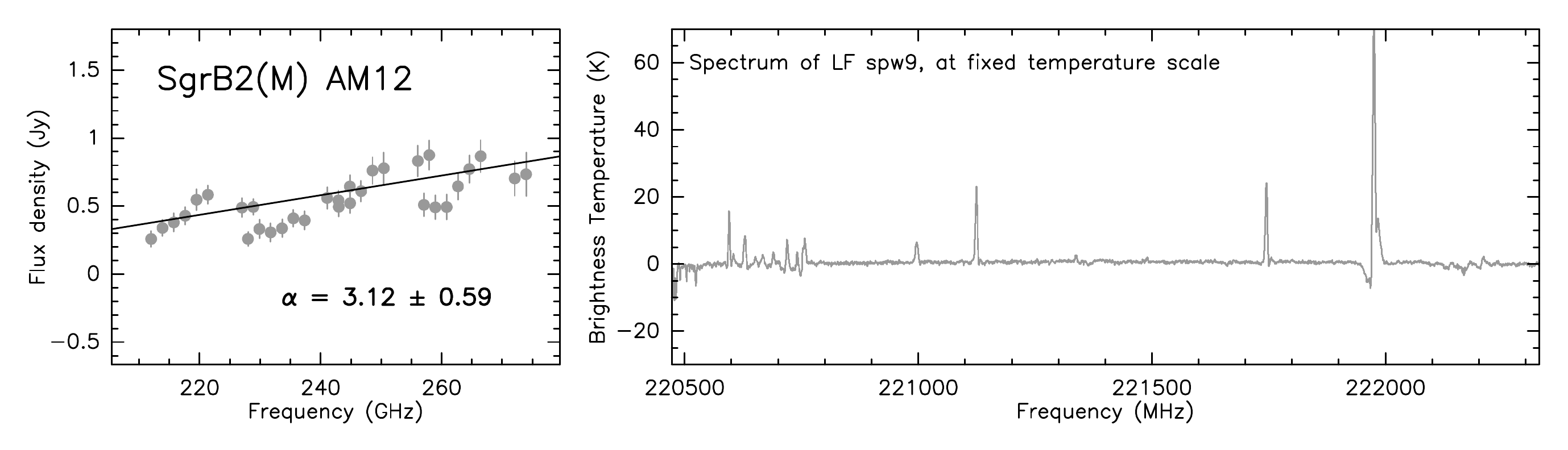} \\
\includegraphics[width=0.9\textwidth]{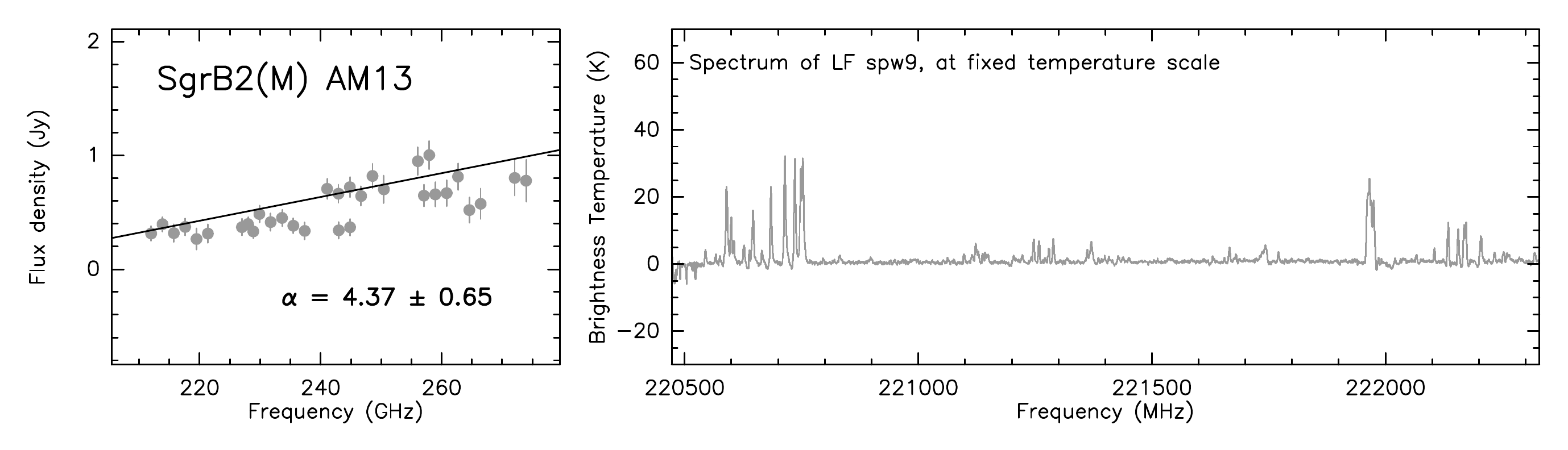} \\
\caption{\SgrB(M) mm-SEDs continued.}
\end{figure*}
\begin{figure*}[ht]
\ContinuedFloat
\centering
\includegraphics[width=0.9\textwidth]{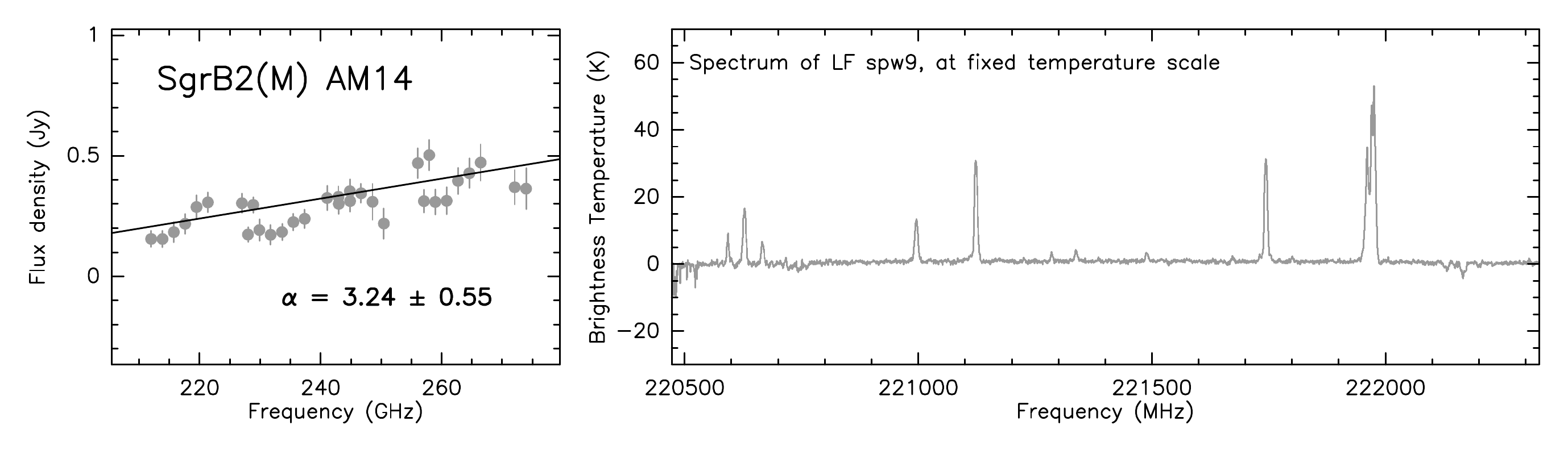} \\
\includegraphics[width=0.9\textwidth]{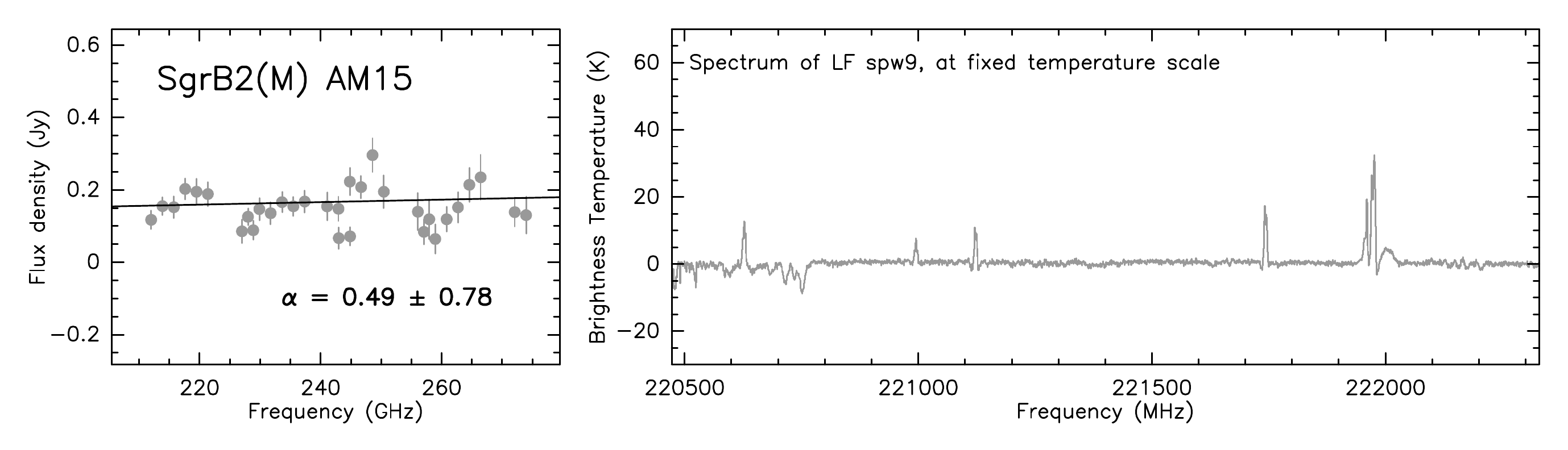} \\
\includegraphics[width=0.9\textwidth]{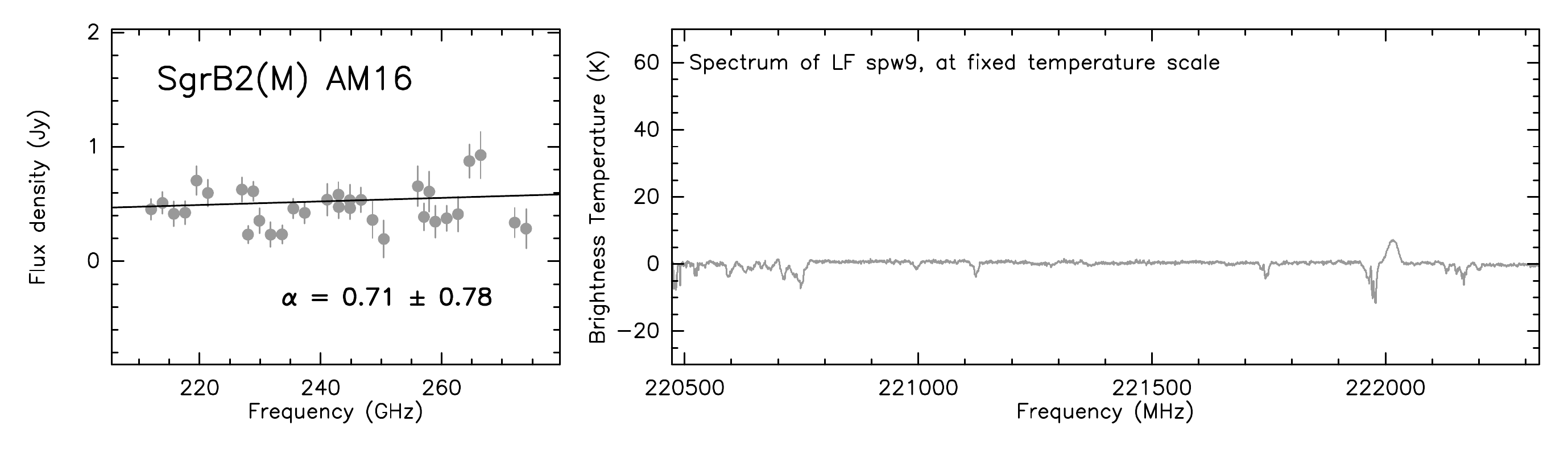} \\
\includegraphics[width=0.9\textwidth]{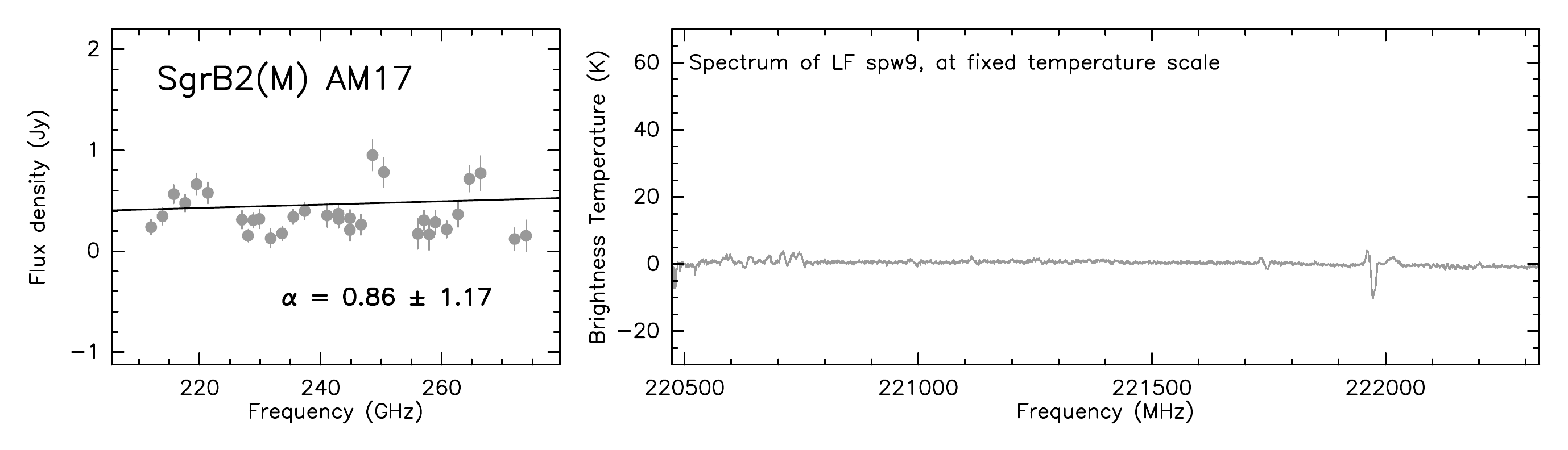} \\
\includegraphics[width=0.9\textwidth]{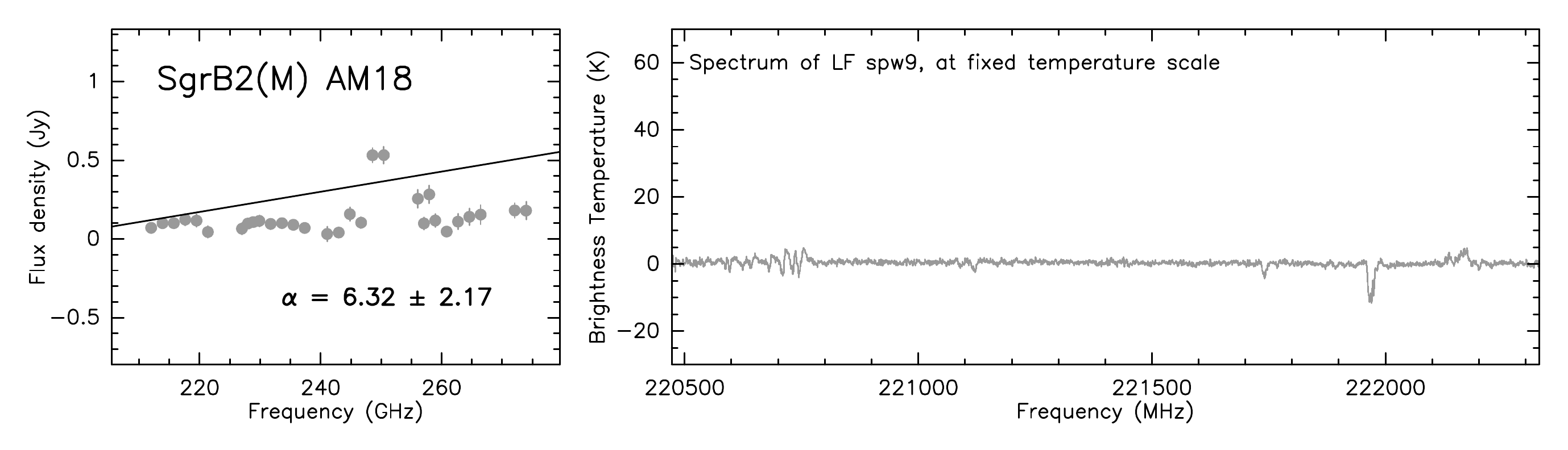} \\
\caption{\SgrB(M) mm-SEDs continued.}
\end{figure*}
\begin{figure*}[ht]
\ContinuedFloat
\centering
\includegraphics[width=0.9\textwidth]{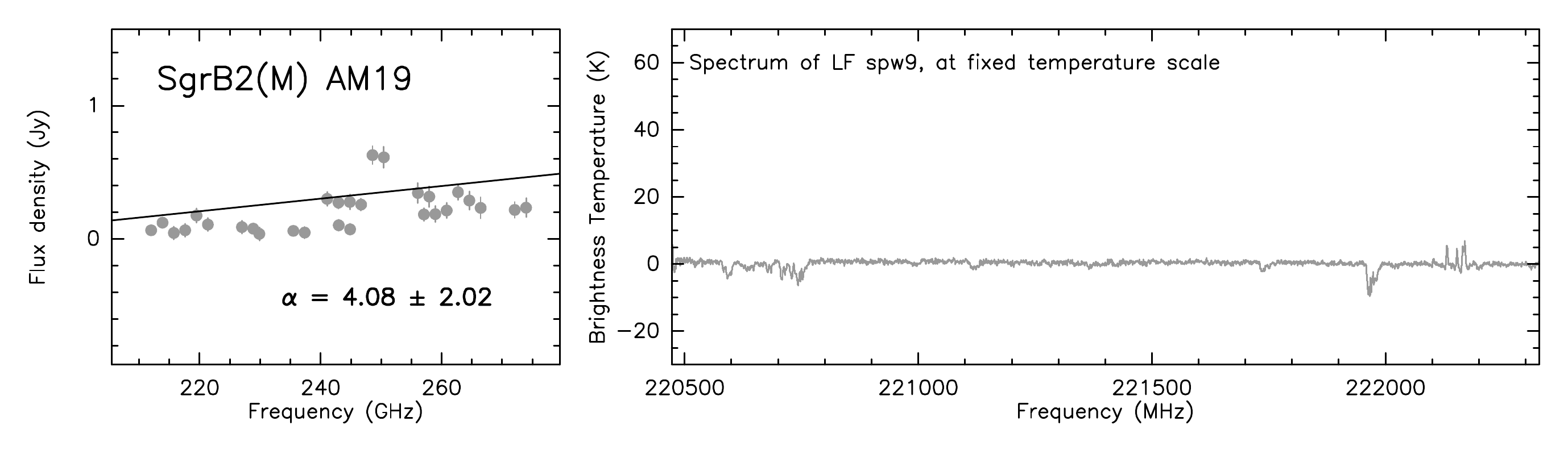} \\
\includegraphics[width=0.9\textwidth]{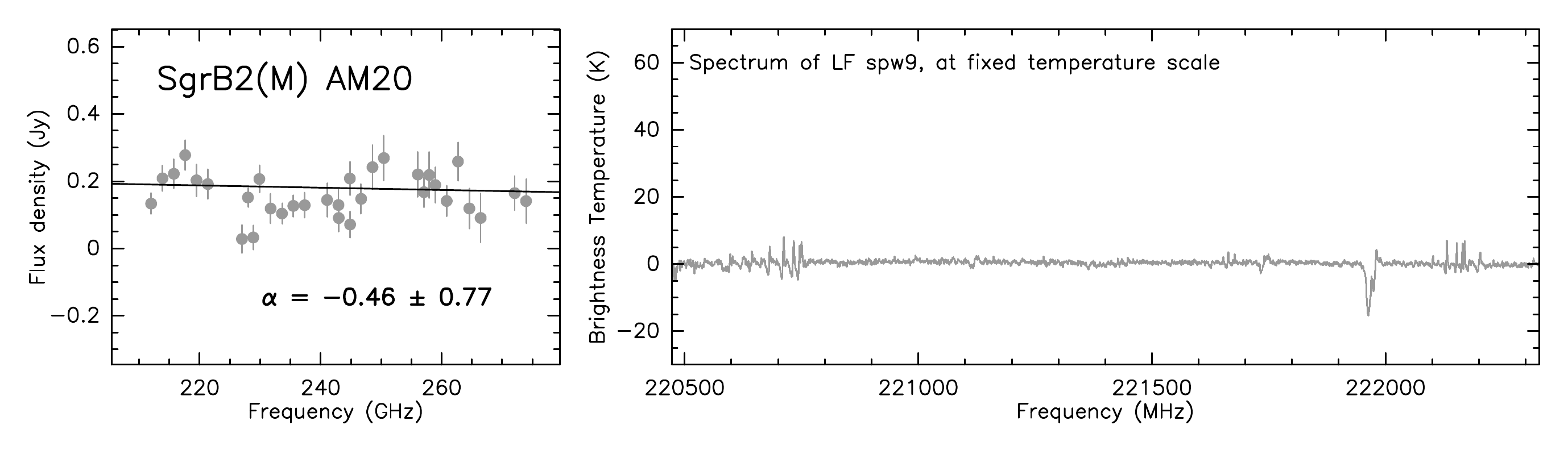} \\
\includegraphics[width=0.9\textwidth]{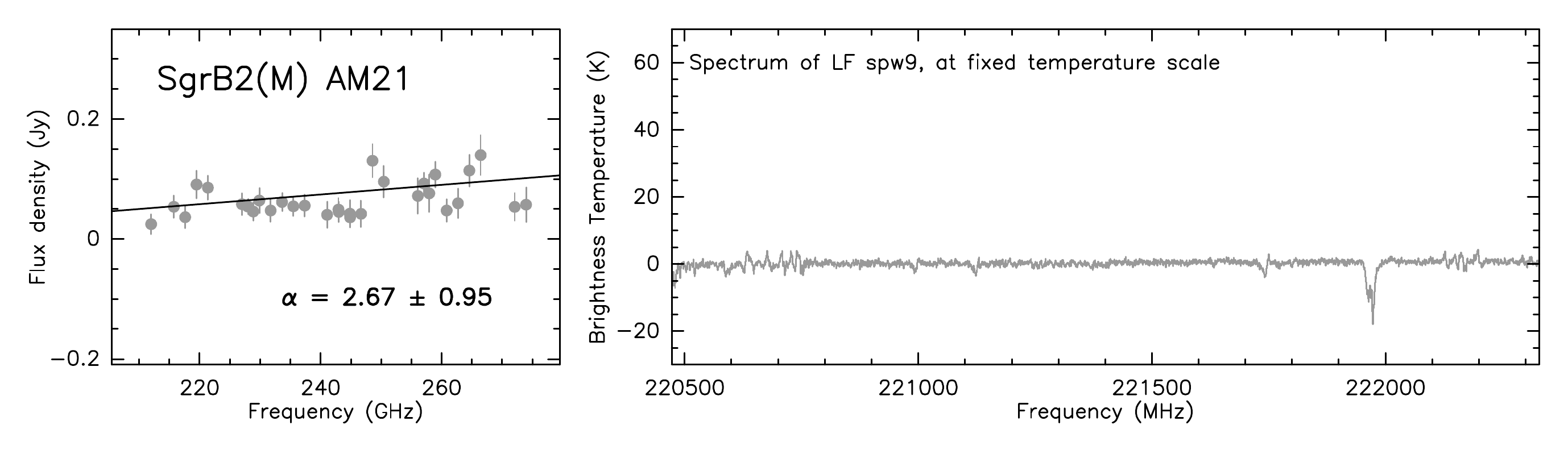} \\
\includegraphics[width=0.9\textwidth]{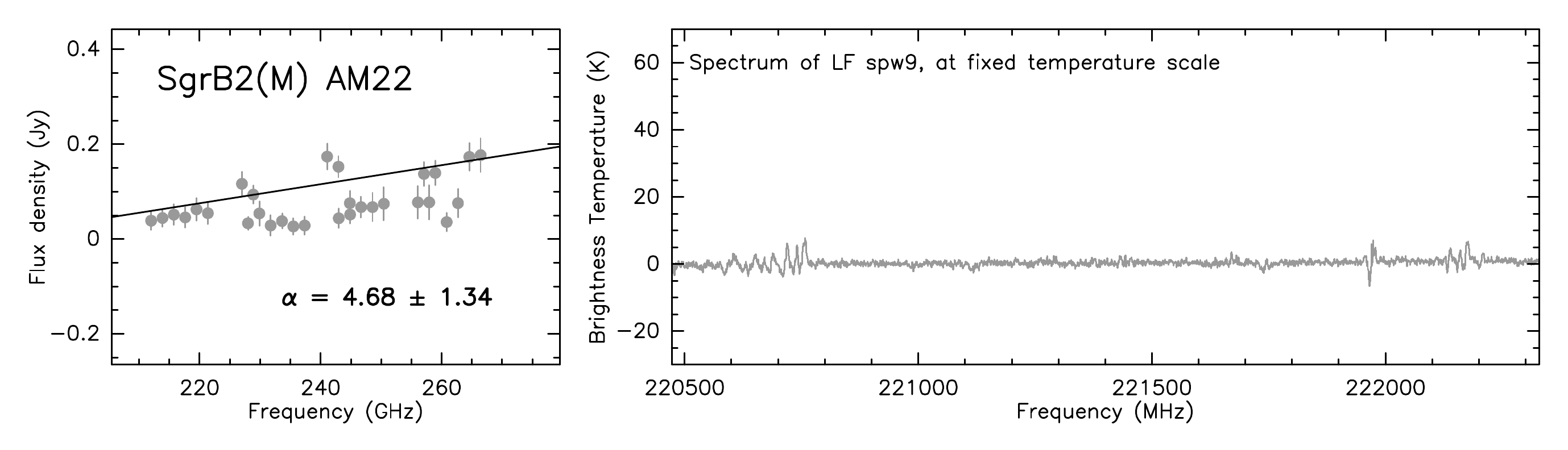} \\
\includegraphics[width=0.9\textwidth]{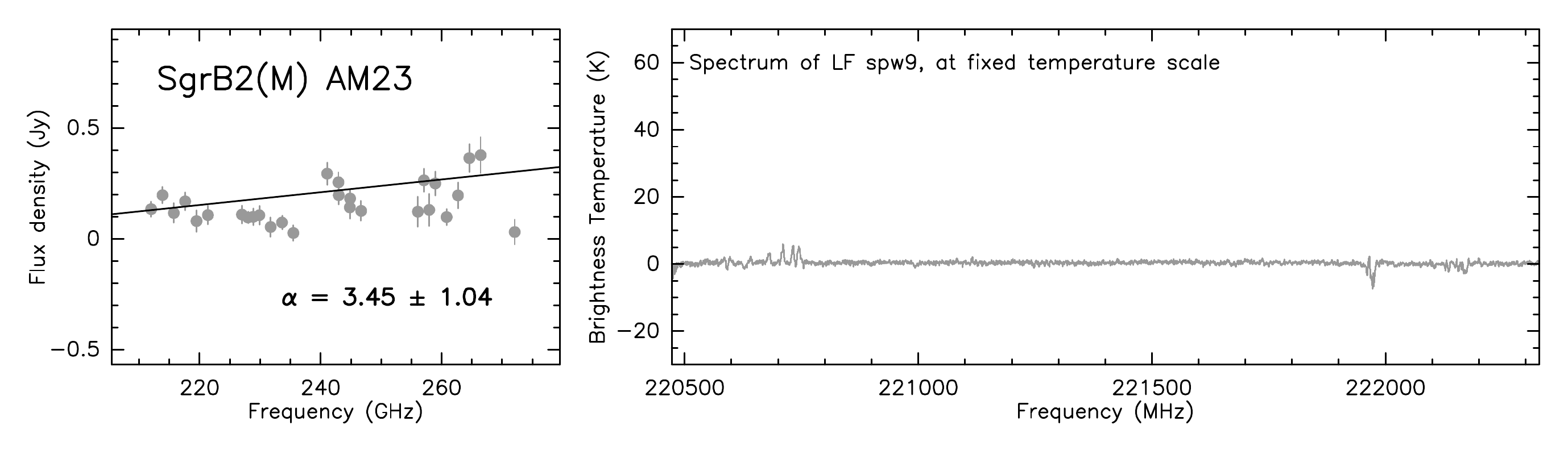} \\
\caption{\SgrB(M) mm-SEDs continued.}
\end{figure*}
\begin{figure*}[ht]
\ContinuedFloat
\centering
\includegraphics[width=0.9\textwidth]{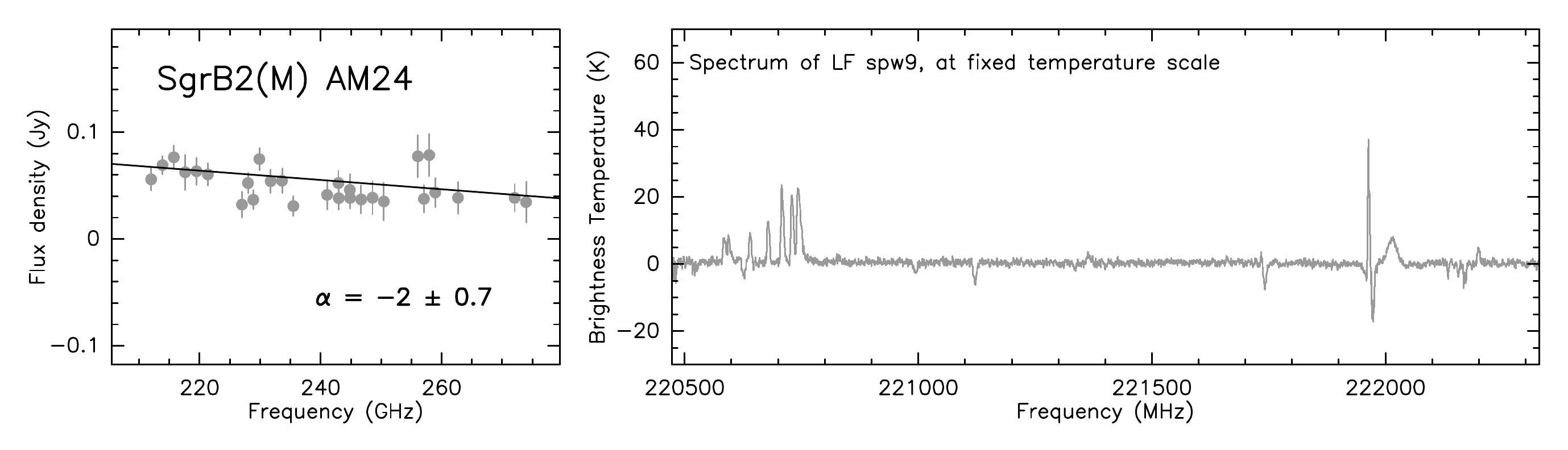} \\
\includegraphics[width=0.9\textwidth]{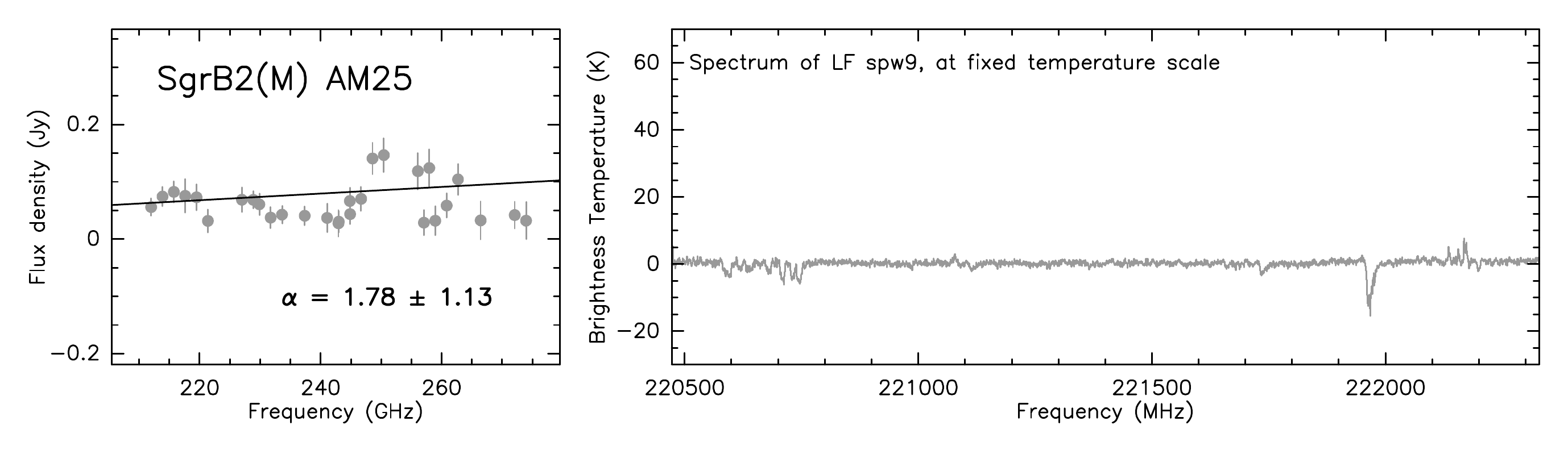} \\
\includegraphics[width=0.9\textwidth]{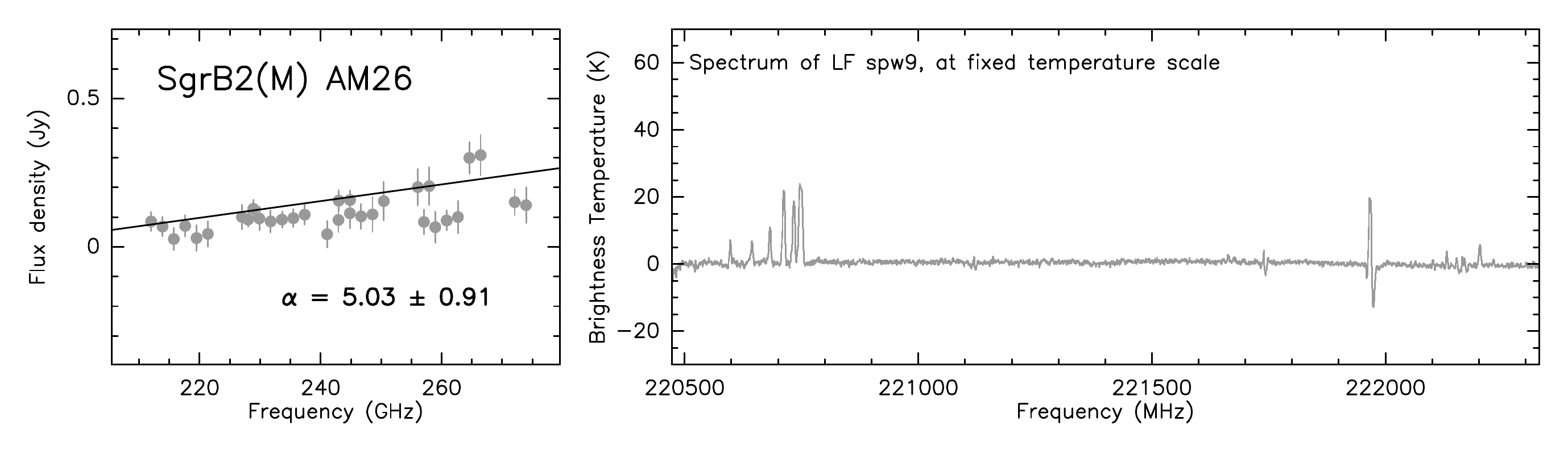} \\
\includegraphics[width=0.9\textwidth]{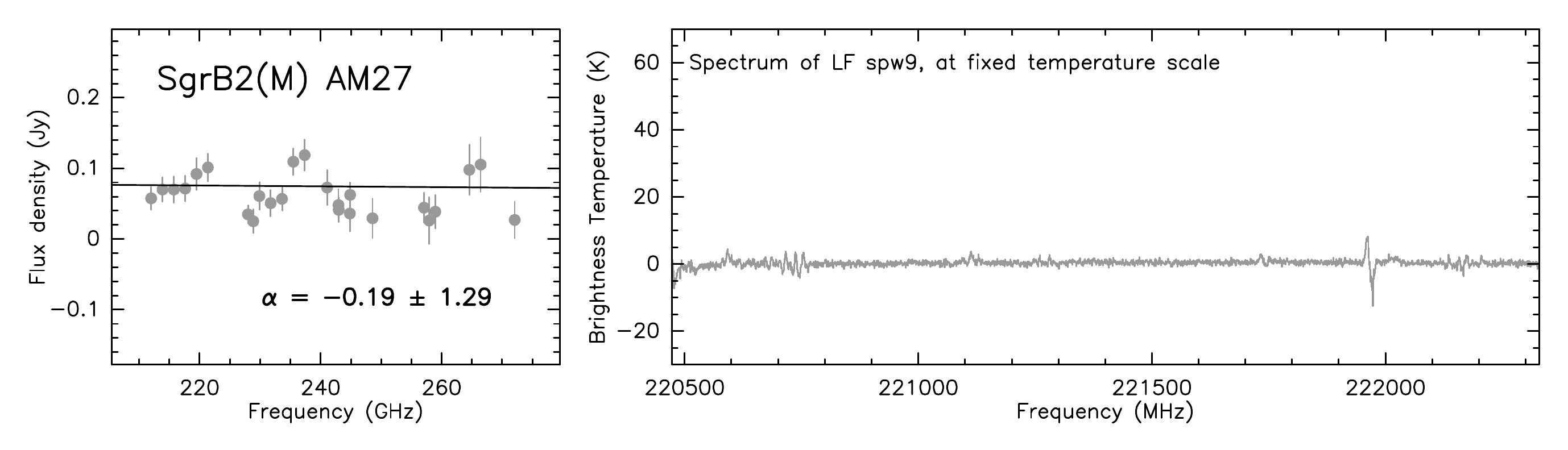} \\
\caption{\SgrB(M) mm-SEDs continued.}
\end{figure*}

\end{appendix}
\end{document}